\documentclass[11pt,twoside,a4paper]{article}
\usepackage[american]{babel}
\usepackage{bm}
\usepackage{amsfonts}
\usepackage{graphicx}
\usepackage{amsmath}
\usepackage{fancyhdr}
\usepackage{a4wide}
\usepackage{siunitx} 
\usepackage{setspace}
\usepackage[utf8]{inputenc}

\usepackage{amsthm}

\newcommand{\disp}{\displaystyle}

\usepackage{color}

\definecolor{applegreen}{rgb}{0.0, 0.5, 0.0}

\newcommand{\zs}{\sigma}

\newcommand{\zb}{\beta}
\newcommand{\zg}{\gamma}
\newcommand{\zl}{\lambda}
\newcommand{\zm}{\mu}

\newcommand{\zo}{\omega}
\newcommand{\zr}{\rho}
\newcommand{\za}{\alpha}
\newcommand{\zy}{\psi}

\newcommand{\zh}{\eta}

\newcommand{\zO}{\Omega}

\newcommand{\dif}{\; \textrm d }

\newcommand{\dpp}[2]{\frac{  \partial #1 }{  \partial #2   } }
\theoremstyle{definition}

 \usepackage{movie15}
\usepackage{hyperref}
\usepackage{booktabs}

\usepackage{cite}
\begin{document}

\begin{center}
\Large{\textbf{A unified 4D phase-space framework for two-level  quantum dynamics: open-source library. 
		%\\
%4D phase-space evolution of two-level quantum gases: beyond specific physical implementations.
}}\\
	%\\[1cm]
	\small{O. Morandi}	\\
\vskip0.5cm
	\textit{\textsf University of Florence  \\
		Department of Mathematics and Informatics U. Dini.\\
Firenze, Italy.}	
\vskip0.5cm
	\textit{omar.morandi@unifi.it}
\end{center}

\begin{center}
\begin{minipage}[h]{0.8\textwidth}
\section*{\textsf{Abstract}}
 \small
\textsf{We present a numerical scheme for simulating the 2D quantum dynamics of a two-level particle gas with internal degrees of freedom such as spin, pseudo-spin, or a two-band electronic structure. 
We adopt the Wigner formulation of quantum mechanics  consisting of a 4D phase-space representation of the quantum dynamics. 
The numerical scheme is based on a spectral splitting method applied to the integro-differential Wigner-Weyl formulation of the dynamics.
The computational architecture of our method is independent of specific physical implementations, resulting in broad applicability.
We illustrate the versatility of our approach by simulating dynamical systems relevant to nanomaterials science, cold atom physics, interacting gases, spintronics, and topological superconductors.}

\section*{\textsf{Code summary}}
\begin{description}
%	\item[Program Title] 
	\item[Program files ] \url{https://doi.org/10.5281/zenodo.19468814} .
%	\item[Licensing provisions]
	\item[Programming language] MATLAB version.
	\item[Supplementary material] Detailed documentation with examples and benchmarking runs \url{https://doi.org/10.5281/zenodo.19468814} .
\end{description}
 
\end{minipage}
\end{center}
%\vspace{0.5cm}
%\textit{\textsc{Keywords:}} Insert.
\vspace{1cm}
\normalsize

\section{Introduction}

The concept of quantum phase-space was introduced by Wigner in 1932 to study thermal equilibrium states in quantum mechanics. Central to this approach was the definition of a pseudo-distribution function arising from an isometric bilinear transformation of the particle wave function \cite{Wigner1932}. The Wigner framework allows quantum dynamics to be naturally interpreted as the result of some nonlocal quantum mechanisms acting on a particle system, that would otherwise evolve as a classical gas. Despite being fully equivalent to the standard quantum mechanical description of particle motion, the Wigner evolution equation may be formally represented as a perturbation applied to the classical Liouville equation in the classical phase-space. The Wigner formalism has also proven to be an excellent tool for investigating the classical limit of the quantum dynamics of interacting systems in a rigorous way \cite{Markowich_89,Markowich_Ringhofer,Gerard_91,Lions_93,Sparber_03,Pulvirenti_06,Gat_14,Figalli_14,
	Morandi_26_arXiv,Morandi_KRM_25}. General introductions to the Wigner formalism and its mathematical aspects and applications can be found in several reports \cite{Heller1976,Tatarskii1983,Lee_95,Shabtay_98,Schleich_01,ZachosEtAl05,Case2008,Polkovnikov_10,De_Gosson_17}. 

\noindent
Since its introduction, the Wigner formalism has become a popular and well-established framework in various research areas, including solid-state physics, nanomaterials, quantum optics, and cold atoms  \cite{Bordone_99,Demeio_02,Unlu_04,Dragoman_05,Sverdlov_05,Revzen_06,Zhu_09,Weinbub_18,Camiola_21,Wen_22}. 
Motivated by the potential offered by the quantum phase-space approach for simulating quantum dynamics in various contexts, several numerical schemes for solving the Wigner equation have been developed over the last few decades. These can be classified into two groups: deterministic algorithms, based on the numerical discretization of the Wigner equation, and stochastic algorithms, based on Monte Carlo methods. The latter exploit the analogy between classical transport and the Wigner formulation of quantum mechanics, extending numerical protocols originally developed for the classical Liouville or Boltzmann equations to the quantum phase-space.

\noindent
Due to the intrinsic complexity of phase-space methods arising from the use of twice the number of variables compared to wave-function descriptions based on the Schr\"odinger equation, the practical application of the Wigner description to complex structures has faced some criticism. 
One of the first  deterministic methods was proposed by Ringhofer in the 1990s \cite{Ringhofer1990}, consisting of a Galerkin spectral finite-difference scheme. A few years later,  popular schemes were proposed by Frensley \cite{Frensley} and Ravaioli et al. \cite{Ravaioli} for the numerical solution of both time-dependent and stationary states for a gas of particles in a single-band structure under the effective mass approximation. In particular, Frensley's algorithm combined an upwind finite-difference scheme for the discretization of the classical kinetic term $v\cdot \nabla_x f$ with a stepwise Galerkin approximation for the pseudodifferential operator. In this approach, the generator of the dynamics was represented as a super-operator where space and momentum were treated as a single variable.  Because this method requires large, typically ill-conditioned sparse matrices, its applications are restricted to small systems, and scaling to larger systems or higher dimensions remains challenging. 

\noindent
A seminal theoretical study concerning the accuracy and numerical convergence of algorithms based on a splitting strategy was performed by Arnold and Ringhofer \cite{ArnoldRinghofer1996}. Following these theoretical results, splitting algorithms for solving the Wigner dynamics became popular.

\noindent
One of the prototypical configurations repeatedly used as a benchmark to test the ability of a quantum simulator to accurately describe systems with strong quantum signatures is the Resonant Tunneling Diode (RTD) (see, e.g., \cite{Ravaioli,Kluksdahl_89,Buot_90,Buot_00,Shifren_03}). The RTD is a single-band heterostructure consisting of several nanometric layers with different chemical compositions but nearly identical atomic spacing. The material composition is chosen such that the conduction band energy produces a double-barrier potential structure, where the barrier height is sufficient to confine a few resonant states.

\noindent
Utilizing the Weighted Essentially Non-Oscillatory (WENO) scheme, originally developed for hyperbolic equations in kinetic theory, Dorda and Sch\"urrer demonstrated that the Wigner description of quantum transport provides accurate results for the RTD. The Wigner formalism reproduces the resonant current and the shape of the negative differential resistance that characterizes RTD devices \cite{DordaSchuerrer2015}. It should be noted that the RTD problem in the Wigner framework had also been previously addressed using Monte Carlo methods by Shifren et al. \cite{Shifren_03}, but the method was unable to provide the same degree of accuracy.

As an alternative to deterministic schemes, stochastic Monte Carlo methods reconstruct the phase-space Wigner pseudo-distribution function through sampling techniques. In these models, a set of test particles is generated, and the stochastic model reproduces the statistical evolution of the ensemble. These particles interact in a way that mimics hard-sphere collisions in real particles.
Several such schemes have been proposed by various groups \cite{Shifren_03,Nedjalkov_11,Muscato_16,DimovNedjalkovSellierSelberherr2015,SellierNedjalkovDimov2015,Weinbub_18}. The primary differences between these schemes lie in how the negative part of the Wigner distribution function, representing the signature of quantum dynamics, is reproduced by the underlying microscopic collisional stochastic dynamics of the fictitious particles. Recent versions of these Monte Carlo methods have been extended to include spin and magnetic effects in 2D systems \cite{Nedjalkov22}.
An alternative scheme, also based on a Monte Carlo approach but with a different conceptual basis, was proposed by Jacoboni's group, known as the Wigner path method \cite{JacoboniBertoniBordoneBrunetti2001,JacoboniBordone2004}. The Wigner path approach reformulates the Wigner problem in terms of a Dyson-like integral equation. The solution is then obtained through a perturbative approach reminiscent of the Feynman diagram techniques used in many-body dynamics.

Wigner function formalism provides also a well established tool in quantum optics and
numerical strategies have been designed which address optical problems \cite{Bastiaans_78,Wiseman1996,Dragoman_05,Dittrich_06,Zhang_09,Cuypers_11,Alonso_11,Sun_15,Lubk_15,Mout_18,Wuv_22,Morandi_JPA_24,Morandi_KRM_25}. 
 
 Numerical studies of Wigner dynamics are mostly restricted to 1D systems (2D in phase-space), with only a few examples developed for higher-dimensional spaces, typically limited to specific cases \cite{Xiong_23}. Moreover, the typical approach consists of focusing on a specific application or Hamiltonian system and developing numerical algorithms tailored to the resulting integro-differential Wigner equation \cite{ShaoLuCai2011,Cabrera_15,Chai_Morandi2015,Schulz_20}.
 In this paper, we adopt a different strategy. We present a numerical implementation of the 4D Wigner problem applicable to a general class of Hamiltonian systems. Our approach allows us to treat various quantum dynamical problems within a unified formalism and numerical algorithm. These problems are relevant to different areas of interest, ranging from solid-state physics and superconductors to cold atoms. This versatility is confirmed by a series of examples applying our model to different physical contexts.
 
 The paper is organized as follows. In Sec. \ref{Sec_Model}, we present the 4D Wigner model and specify the general class of problems that can be treated by our approach. Sec. \ref{Sec_Num} is devoted to clarifying certain aspects of the model's numerical implementation of the model. In Secs. \ref{Sec_ex_double_slit}-\ref{Sec_ex_graphene}, we provide a series of examples illustrating the capability of our 4D Wigner model to reproduce quantum dynamics in several physical contexts. 
 In Sec. \ref{Sec_ex_double_slit}, we consider, as a first textbook case, a simplified implementation of the double-slit experiment for electrons. In Sec. \ref{Sec_ex_Rashba}, we simulate the excitation of spin polarization in a particle gas within a Rashba semiconductor under an external magnetic field. In Sec. \ref{Sec_ex_atoms}, we discuss the steering of neutral cold atoms using optical lasers and describe the atom-atom scattering induced by dipolar interaction. In Sec. \ref{Sec_ex_BdG}, we show an example of Klein tunneling in a topological superconductor. Finally, in Sec. \ref{Sec_ex_graphene}, we investigate the excitation dynamics of charges in a graphene layer. Each case study stands as an autonomous application, allowing the reader to consult them individually.

\section{Model}\label{Sec_Model}

The aim of this work is to develop a numerical algorithm to solve the Wigner dynamics for systems with spin in 2D real space, resulting in a 4D phase-space. We discuss the implementation of a numerical solver based on a spectral splitting technique applied to the Wigner equation.
We consider a quantum system represented by a mixed state of particles with spin (or pseudospin, as discussed below) confined to a 2D layer. In a pure state, each particle is described by a wave function $\zy \in L^2 \left(\mathbb{R}^2;\mathbb{C}^2 \right)$. We denote the kernel of the density matrix by $\zr_{ij}(y,y') = \sum_k \zr_k \zy^k_i\left(y \right) \overline{\zy_j^k\left(y' \right)}$, where the probabilities $\rho_k \geq 0$ are normalized such that $\sum_k \rho_k = 1$. In the following, we refer to the spatial coordinates in the density matrix and to the position-momentum coordinates in the Wigner description as continuous degrees of freedom, while the indices denoting spin projections are referred to as discrete degrees of freedom.
We introduce the Wigner phase-space description of the quantum dynamics by applying a unitary transformation to the density matrix representation. First, we replace the spatial coordinates $(y,y')$ with the center-of-mass and relative coordinates. For convenience, we scale the relative coordinate by the Planck constant $(y,y')\rightarrow \left(x+\frac{\hbar \zh}{2} ,x-\frac{\hbar \zh}{2}\right)$. This change of variables corresponds to a counterclockwise rotation of the $(y, y')$ axes by an angle of $\pi/4$, followed by a rescaling. Finally, by taking the Fourier transform with respect to the relative coordinate, we obtain the  Wigner matrix function associated with the density matrix $\rho$
\begin{align}
	F_{ij}  =& \frac{1}{(2\pi)^2} \int \zr_{ij}\left(x+\frac{\hbar \zh}{2} ,x-\frac{\hbar \zh}{2} \right) e^{-ip\zh} \dif \zh \;. \label{def_wig_tras}
\end{align}
By this operation, we obtain a $2 \times 2$ Hermitian matrix, which we denote as the Wigner function $F\in L^2  \left(\mathbb{R}^2_x\times \mathbb{R}^2_p;\mathbb{C}^4 \right)$, whose elements are functions from the 4D phase-space to the complex plane. In the framework of the Wigner description of quantum dynamics, it is customary to refer to $x$ and $p$ as the position and momentum variables, respectively. 
We note that our definition of the Wigner matrix arises from the straightforward, component-wise application of the Wigner transform to the density matrix. Specifically, we do not apply any angular transformations associated with the spin degrees of freedom \cite{Agarwal_81,Varilly_89,Brif_98,Mukunda_04,Tilma_16,Klimov_17,Koczor_19,Kimov_22}. 

The evolution equation of the Wigner matrix can be obtained from the Schr\"odinger equation associated with the wave function $\zy$, $i\hbar \dpp{\zy}{t} = \mathcal{H}(x,-i\hbar\nabla ) \zy $, where $\mathcal{H}$ denotes a $2\times 2$ Hamiltonian symbol. The evolution equation for the Wigner matrix can be compactly formulated as a von Neumann-like equation within the Moyal algebra
\begin{align}
	i\hbar \dpp{F}{t} =  \left[\mathcal{H} ,F \right]_{\#}\doteq \mathcal{H} \# F -  F\# \mathcal{H} \;. \label{Wig_eq_comm}
\end{align}
The Moyal product, denoted by the symbol $\#$, is defined by the integral expression
\begin{align}
	\hspace{-30pt}	 \mathcal{H} \#  F
	=&    \frac{ 1}{(2\pi)^4}  \int_{\mathbb{R}^{8}}  \mathcal{H} \left(x -\frac{\hbar \zh}{2},p+ \frac{\hbar\zm}{2} \right) F \left(x',p' \right) e^{i\mu(x-  x')+i\zh(p -p') } \textrm{d}\mu \dif \zh   \dif x' \textrm{d} p'\;, \label{moyal_prod}
\end{align}
providing a clear indication of the nonlocality intrinsic to Wigner dynamics.
The implementation of a numerical algorithm to solve Eq. \eqref{Wig_eq_comm} in a 4D scenario for a generic Hamiltonian system is challenging. The main difficulties come from the presence of integrals over eight-dimensional space associated with the kernel of the dynamical generator \eqref{moyal_prod}. To mitigate these challenges, we restrict our analysis to a specific class of Hamiltonian systems that allows for a significant reduction in the computational effort required to solve Eq. \eqref{Wig_eq_comm}. In our model, we consider a class of Hamiltonian systems in which the total energy decouples into the sum of two symbols depending solely on the momentum variable or on the position variable. We assume
 \begin{align}
 	\mathcal{H} =\underbrace{\zl_0(p)\;\zl_0+\zl(p)\cdot \zs }_{\Lambda(p)} + \underbrace{u_0(x) \;\zs_0 +   u(x) \cdot \zs}_{U(x)}  \label{Ham_gen} 
 \end{align}
The first term, $\Lambda(p)$, describes the particle kinetic energy or the band structure associated with the material, and may account for the presence of external fields. Conversely, $U(x)$ describes the potential energy of the particle, which is typically expressed via space-dependent fields.
Both $\Lambda(p)$ and $U(x)$ are $2 \times 2$ Hermitian matrices depending on the particle momentum $p \in \mathbb{R}^2$ and the spatial position $x \in \mathbb{R}^2$, respectively. In our notation, we expand the matrices $U$ and $\Lambda$ in the basis of the Pauli matrices $\sigma_i$ (for $i=x,y,z$), where $\sigma_0$ denotes the identity matrix. By writing $U = (u_0, {u})$ and $\Lambda = (\lambda_0, {\lambda})$, where ${u} = (u_x, u_y, u_z)$ and ${\lambda} = (\lambda_x, \lambda_y, \lambda_z)$, we identify each matrix with its associated projection coefficients obtained via $u_i = \frac{1}{2} \text{Tr}(\sigma_i U)$, where Tr denotes the trace.
Under the assumption of Eq. \eqref{Ham_gen}, the Wigner evolution equation simplifies
\begin{align*}
	\dpp{F}{t} =&\phantom{+}    \frac{ 1}{ i \hbar (2\pi)^{2}}  \int_{\mathbb{R}^{4}} \left[U\left(x+ \frac{\hbar \zh}{2} \right)  F \left(x,p' \right) - F\left(x,p' \right)  U \left(x- \frac{\hbar\zh}{2}\right)\right]e^{-i\zh(p -p') }  \dif \zh  \dif p'\\
	&+    \frac{ 1}{i \hbar (2\pi)^{2}}  \int_{\mathbb{R}^{4}} \left[ \Lambda \left(p+ \frac{\hbar\zm}{2} \right)  F \left(x',p \right)  -F \left(x',p \right)  \Lambda\left(p- \frac{\hbar\zm}{2} \right)    \right] e^{i\mu(x-  x') } \dif \mu    \dif x'\;. 
\end{align*} 
As particular cases, Eq. \eqref{Ham_gen} encompasses several physically relevant Hamiltonians describing 2D systems. We provide a few examples covering diverse research fields, such as nanomaterials, solid-state physics, spintronics, ultracold atoms, and superconductors. 
Graphene is a 2D material composed of a single layer of carbon atoms arranged on the vertices of a honeycomb lattice, which has attracted enormous interest over the last few decades  \cite{Kane_05,Katsnelson_06,Sarma_11,Chen_12}. The graphene Hamiltonian for low-energy electrons shares a striking similarity with the Dirac Hamiltonian for relativistic particles. Consequently, the electron dynamic are formally equivalent to the evolution of massless Dirac fermions, where the speed of light is replaced by the Fermi velocity $v_F$.
 \begin{align}
		\mathcal{H}_{graph.} \doteq v_F \left( p_x  \zs_x +p_y  \zs_y \right) + U(x) \;\zs_0   =v_F\left(\begin{array}{cc}
		0  &  p_x+ i p_y \\
			p_x- i  p_y & 0
		\end{array}\right) + U (x)\zs_0, \label{Ham_grap}
	\end{align}
The potential $U$ represents an applied electric potential.   An example of the application of our formalism to the graphene Hamiltonian is presented in Sec. \ref{Sec_ex_graphene}
Another class of 2D systems that fits naturally within our framework is  two-band semiconductors. Depending on the material composition and lattice symmetries of the semiconductor crystal, the band structure may assume various forms that differ significantly from the simple parabolic electron-hole dispersion relation. A prominent example of such non-parabolicity is the $k \cdot p$ Hamiltonian, which provides a simple framework for describing the electronic properties of narrow-gap semiconductors. The effective Hamiltonian in this context is expressed as		
\begin{align}
		\mathcal{H}_{kp} \doteq\left(\begin{array}{cc}
			\frac{p^2}{2m}  & k \cdot p  \\
			k\cdot p & \frac{p^2}{2m}
		\end{array}\right) + U (x) \;,\label{Ham_kp}
	\end{align}
	where $k$ is a two-component vector characterizing the material. 
	As a third application of our model, we consider the evolution of a 2D electron gas with spin in the presence of external electric and magnetic fields, including spin-orbit effects represented by the Rashba effective field. In this case, the Hamiltonian is given by
 \begin{align}
		\mathcal{H}_{spin} =\left( \frac{p^2}{2m} + U(x)\right) \;\zs_0 + \left(p\wedge K - B \left( x \right) \right) \cdot \zs .  \label{Ham_spin}
\end{align}
Here, $U$ and $B$ denote the external electric and magnetic potentials, respectively. Similar to the $k \cdot p$ Hamiltonian, $K$ is a three-component vector characteristic of the material, commonly referred to as the Rashba vector. The application of our formalism to spintronic systems is discussed in Sec. \ref{Sec_ex_Rashba}. 
Finally, further applications of our formalism arise in the physics of superconductors, ultracold atoms, and optical lattices. A notable example is the Bogoliubov-de Gennes (BdG) Hamiltonian, which describes low-energy excitations in topological chiral superconductors. This model is of particular interest in the quest to detect Majorana zero modes
 \cite{Fu_09,Liang_11,Kitaev_06}
	\begin{align*}
		\mathcal{H}_{BdG} \doteq	\left(\begin{array}{cc}
			\frac{ p^2}{2m}-\mu  &  \Delta (i p_x-  p_y) \\
			\Delta (-i  p_x- p_y) & -\frac{ p^2}{2m} + \mu 
		\end{array}\right)  \;.
	\end{align*}
Here, $\mu $ is the chemical potential, $m$ the mass of the quasiparticle and $\Delta$ the superconductor pairing parameter. An example of application of our formalism to the BdG Hamiltonian is presented in Sec. \ref{Sec_ex_BdG}

To complete the description of the 2D system, we include a phenomenological relaxation mechanism toward local equilibrium, accounting for the interaction between the system and the surrounding  environment. We adopt a relaxation-time approximation (often referred to  as the BGK or Bhatnagar-Gross-Krook mechanism in kinetic theory). The final evolution equation for the Wigner matrix is expressed as follows
\begin{align*}
	 \dpp{F}{t} = \frac{1}{i\hbar}\left[\mathcal{H} ,F \right]_{\#} - \frac{F-F_{eq}}{\tau(x,p) } \;.
\end{align*}
The relaxation time $\tau$ may depend on both the momentum and position variables, with its specific expression determined by the relaxation mechanism under consideration. Here, $F_{eq}$ denotes the local equilibrium distribution. The determination of the most appropriate expression for the local distribution function in the Wigner model has been the subject of extensive research \cite{Markowich_90,Gardner_96,Degond_05,Romano_07,Hurst_14}.
Various models are available in which the local distribution is expressed as a nonlinear function of the moments of the dynamical solution $F$ and the external fields. In this general introduction to the splitting-spectral method for the 4D spinor Wigner matrix, we assume that the local equilibrium coincides with the classical local equilibrium distributions associated with Maxwell-Boltzmann (MB), Fermi-Dirac (FD), or Bose-Einstein (BE) statistics.
The equilibrium Wigner function is expressed in terms of the projector $	\Pi^\pm= \left| u^\pm \rangle \langle u^\pm \right| $,  onto the eigenspaces associated with the Hamitlonian eigenvalue problem $\mathcal{H} u^\pm=\zl^\pm u^\pm$. The local equilibrium Wigner matrix is given by
 \begin{align}
	F_{eq} =	\Pi^+ \; f_{eq} \left(\zl^+(x,p)\right)+ 	\Pi^-  \;  f_{eq} \left(\zl^-(x,p)\right) \;.\label{F_eq}
\end{align}
Depending on the particle statistics, we obtain the following local equilibrium distributions. MB: $ f_{eq} =  e^{-\frac{\zl-\mu}{ k_B T}  }$, FD: $f_{eq}  = \left[ 1+e^{\frac{\zl-\mu}{ k_B T} } \right]^{-1}$, BE: $f_{eq}  = \left[ 1-e^{\frac{\zl-\mu}{ k_B T} } \right]^{-1}$, 
where $\mu(x,t)$ is the local chemical potential, $k_B$ the Boltzmann constant, and $T(x,t)$ the local temperature.

\noindent
Due to the simple form of the Hamiltonian, the projectors $\Pi^\pm$ have an analytical representation. We write the Hamiltonian as $	\mathcal{H} (x,p)= d \cdot \zs + d_0 \zs_0$. The eigenvalues are $	\zl^{\pm}=d_0  \pm |d | $, with the eigenspace projectors given by %\cuc
\begin{align*}
	\Pi^\pm=&
	\frac{1}{2|d| }  \left(
		\begin{array}{cc}
			d_3\pm |d| & \pm (d_1-i d_2)  \\
	 \pm (d_1+i d_2)  & \frac{d_1^2 +d_2^2}{|d|\pm d_3} 
	 \end{array}\right)\;.
\end{align*}
We solve the Wigner problem in a rectangular domain for the spatial degrees of freedom, and in either $\mathbb{R}^2$ or a periodic first Brillouin zone for the momentum variables. Regarding the spatial boundaries, we consider both periodic and open systems. From a physical perspective, open systems represent 2D layers surrounded by ideal contacts. These act as infinite-capacity reservoirs that inject particles into the domain at equilibrium and absorb, without reflection, any particles leaving the sample. 
From a mathematical standpoint, such ideal contacts are implemented via transparent boundary conditions. In this approach, the incoming particle distribution is fixed on the subset of the phase-space boundary where the particle velocity is directed inward (i.e., opposite to the outward-pointing normal). We note that these open boundary conditions are of a classical nature and remain physically justified only if quantum effects are negligible near the boundaries.

\noindent
Regarding the final mathematical formulation of our model, we focus on the case of transparent boundary conditions (while the periodic case follows a similar structure with straightforward modifications)
\begin{align}
	\left\{ \begin{array}{l}
		\disp 	\dpp{F}{t} =\phantom{+}  \zg   \int_{\mathbb{R}^{4}} \left[U\left(x+ \frac{\hbar \zh}{2} \right)  F \left(x,p' \right) - F\left(x,p' \right)  U \left(x- \frac{\hbar\zh}{2}\right)\right]e^{-i\zh(p -p') }  \dif \zh  \dif p'\\[10pt]
		\phantom{\dpp{F}{t} =}\disp + \zg \int_{\mathbb{R}^{4}} \left[ \Lambda \left(p+ \frac{\hbar\zm}{2} \right)  F \left(x',p \right)  -F \left(x',p \right)  \Lambda\left(p- \frac{\hbar\zm}{2} \right)    \right] e^{i\mu(x-  x') } \dif \mu    \dif x' \\[10pt]
	\phantom{\dpp{F}{t} =} \disp 	- \frac{F-F_{eq}}{\tau(x,p) }	\hfill  \textrm{in } \zO \times \mathbb{R}_p^2 \times [0,T] \hspace{15pt} \\[10pt]
		\disp \left.F\right|_{t=0} = F_0\hfill \textrm{on } \partial \zO \times \mathbb{R}_{p}^2 \hspace{38pt} \\[10pt] 
		\disp \left.F\right|_{\partial\zO} = F_{in} \hfill \textrm{on }   \zO_{in} \times [0,T]
	\end{array}\right.\;. \label{wig_prob}
\end{align}
We have defined $\zg \doteq   \frac{ 1}{ i \hbar (2\pi)^{2}} $ and $\zO = [x_m,x_M]\times [y_m,y_M]$ is the spatial domain. The inflow conditions are applied to the domain $
	\zO_{in}	\doteq	\left\{ (x,p) \in \partial \zO \times \mathbb{R}_p^2\; : 
	\left.  \nabla_p \lambda_0 \right|_{\partial \zO } \cdot \widehat{n} < 0 
	\right\}$, where $ \widehat{n} $ denote the outgoing normal direction to the boundary and $ \nabla_p \lambda_0(p)$ represents the group velocity associated with the semi-classical motion.

\section{Numerical algorithm}\label{Sec_Num}

We discuss the algorithm implemented to obtain a numerical solution of the Wigner problem \eqref{wig_prob}. The approach is based on a splitting decomposition technique combined with spectral methods, which arise naturally from the constitutive role of the Fourier transform in the definition of the Wigner phase-space. Splitting schemes are based on the Trotter-Lie  semigroup decomposition formula and represent a popular and effective class of methods for the numerical approximation of Wigner dynamics \cite{ArnoldRinghofer1996}.
For simplicity, in this section, we neglect the relaxation mechanism, which can be included by straightforward extension of the splitting method discussed below. We approximate the evolution of the solution over a time step $\Delta $ using the expression $ 
F(t+\Delta) = \mathcal{S}^{\Delta/2} \mathcal{U}^{\Delta}  \mathcal{S}^{\Delta/2}   F(t)$, 
where the unitary operators $\mathcal{S}$ $, \mathcal{U}$ denoted as the streaming and field operators respectively, are the propagators associated with the following simplified problems 
\begin{align*} 
\mathcal{S}^{\Delta}  : \left\{
\begin{array}{ll}
\disp 	\dpp{F}{t} = 
\frac{ 1}{i \hbar (2\pi)^{2}}  \int_{\mathbb{R}^{4}} \left[ \Lambda \left(p+ \frac{\hbar\zm}{2} \right)  F \left(x',p' \right)  -F \left(x',p' \right)  \Lambda\left(p- \frac{\hbar\zm}{2} \right)    \right] e^{i\mu(x-  x') } \dif \mu    \dif x' &   \\
\left.F\right|_{t=0} = F_0
\end{array} \right. \;, 
\end{align*}
and
\begin{align} 
 \mathcal{U}^{\Delta}  : \left\{
 \begin{array}{ll}
\disp  \dpp{F}{t} =   \frac{ 1}{ i \hbar (2\pi)^{2}}  \int_{\mathbb{R}^{2d}} \left[U\left(x+ \frac{\hbar \zh}{2} \right)  F \left(x,p' \right) - F\left(x,p' \right)  U \left(x- \frac{\hbar\zh}{2}\right)\right]e^{-i\zh(p -p') }  \dif \zh  \dif p' & \\
  \left.F\right|_{t=0} = F_0
 \end{array} \right. \;. \label{split_sch_field_op}
\end{align}
In both problems, the time spans the interval $t\in (0,\Delta] $. 
To illustrate our method, we discuss in detail the numerical algorithm that provides the solution associated with the field propagator. The case for the streaming operator proceeds similarly.
The problem \eqref{split_sch_field_op} admits an analytical solution in the Fourier space. We apply the Fourier transform with respect to the momentum variable, and we denote In both problems the time spans the interval $t\in (0,\Delta] $. 
To illustrate our method, we discuss in detail the numerical algorithm that provides the solution associated with the field propagator. The case for the streaming operator proceeds similarly.
The problem \eqref{split_sch_field_op} admits an analytical solution in the Fourier space. We apply the Fourier transform with respect to the momentum variable, and we denote the transformed Wigner function by
$\widehat{F}(x,\zh)\doteq  \mathcal{F}_{p \rightarrow \zh }(F(x,\cdot))  =\frac{1}{(2\pi)^2}\int F\left(x,p \right) e^{-i  p\zh } \dif p$. We obtain
\begin{align*}
	\dpp{\widehat{F}}{t} = 
	\frac{1}{i \hbar }	\left[ U  \left({x}+\frac{  \hbar \zh}{2}  \right) \widehat{F} ({x},\zh)- \widehat{F} ({x},\zh) U \left(x-\frac{ \hbar \zh}{2} \right)\right]   \;.
\end{align*}
The solution is given by $
\widehat{F}(x,\zh,t+\Delta) = 
e^{-\frac{i}{\hbar} U  \left({x}+\frac{  \hbar \zh}{2}  \right)\Delta } \widehat{F} (x,\zh,t) e^{\frac{i}{\hbar} U \left(x-\frac{ \hbar \zh}{2} \right)\Delta }$. 
The exponential terms can be evaluated explicitly using 
$
\exp(i U \Delta) = e^{i u_0 \Delta} \left[ \cos(\Delta |u|) I + i \sin(\Delta |u|) \frac{u \cdot \vec{\sigma}}{|u|} \right]
$,  
where we have expressed the matrix $U=(u,u_0)$ in terms of the Pauli components $u_i=\frac{1}{2}\textrm{Tr}(\zs_i U)$. We obtain   
\begin{align*}
	\left( \mathcal{U}^{\Delta}  	\widehat{F}\right) (x,p )	=& \mathcal{F}^{-1}_{\zh \rightarrow p} \left[
	e^{- \frac{i}{\hbar} \left[u_0  \left({x}+\frac{  \hbar \zh}{2}  \right) -  u_0  \left({x}-\frac{  \hbar \zh}{2}  \right)\right] \Delta}
	M^{+}(x,\zh) \widehat{F} (x,\zh) M^{-}(x,\zh)  \right] \;,
\end{align*}
where we have defined 
\begin{align*}
	M^{\pm}(u;x,\zh) \doteq & 
	\cos\left(\frac{\Delta}{\hbar} \left|u \left({x}\pm\frac{  \hbar \zh}{2}  \right) \right| \right) I \mp  i \sin\left(\frac{\Delta}{\hbar} \left|u \left({x}\pm \frac{  \hbar \zh}{2}  \right) \right| \right) \frac{u \left({x}\pm\frac{  \hbar \zh}{2}  \right)  \cdot \vec{\sigma}}{\left|u \left({x}\pm \frac{  \hbar \zh}{2}  \right) \right|}  \; . 
\end{align*}
%\subsubsection*{Streaming propagator $ \mathcal{S}^{\Delta} $}
The same procedure can be applied t0o the streaming problem, and we obtain the solution in Fourier space with the difference that the solution is now  expressed in terms of the Fourier transform of the Wigner matrix with respect to the position coordinates. 
 $\widehat{F}(x,\zh)\doteq  \mathcal{F}_{p \rightarrow \zh }(F(x,\cdot))  =\frac{1}{(2\pi)^2}\int F\left(x,p \right) e^{-i  p\zh } \dif p$. We obtain
\begin{align*}
	\dpp{\widehat{F}}{t} = 
\frac{1}{i \hbar }	\left[ U  \left({x}+\frac{  \hbar \zh}{2}  \right) \widehat{F} ({x},\zh)- \widehat{F} ({x},\zh) U \left(x-\frac{ \hbar \zh}{2} \right)\right]   \;.
\end{align*}
The solution is given by $
	\widehat{F}(x,\zh,t+\Delta) = 
	e^{-\frac{i}{\hbar} U  \left({x}+\frac{  \hbar \zh}{2}  \right)\Delta } \widehat{F} (x,\zh,t) e^{\frac{i}{\hbar} U \left(x-\frac{ \hbar \zh}{2} \right)\Delta }$. 
The exponential terms can be evaluated explicitly, by using 
$
\exp(i U \Delta) = e^{i u_0 \Delta} \left[ \cos(\Delta |u|) I + i \sin(\Delta |u|) \frac{u \cdot \vec{\sigma}}{|u|} \right]
$,  
where we have expressed the matrix $U=(u,u_0)$ in terms of the Pauli components $u_i=\frac{1}{2}\textrm{Tr}(\zs_i U)$. We obtain   
\begin{align*}
	\left( \mathcal{U}^{\Delta}  	\widehat{F}\right) (x,p )	=& \mathcal{F}^{-1}_{\zh \rightarrow p} \left[
	e^{- \frac{i}{\hbar} \left[u_0  \left({x}+\frac{  \hbar \zh}{2}  \right) -  u_0  \left({x}-\frac{  \hbar \zh}{2}  \right)\right] \Delta}
 M^{+}(x,\zh) \widehat{F} (x,\zh) M^{-}(x,\zh)  \right] \;,
\end{align*}
where we have defined 
\begin{align*}
	M^{\pm}(u;x,\zh) \doteq & 
	 \cos\left(\frac{\Delta}{\hbar} \left|u \left({x}\pm\frac{  \hbar \zh}{2}  \right) \right| \right) I \mp  i \sin\left(\frac{\Delta}{\hbar} \left|u \left({x}\pm \frac{  \hbar \zh}{2}  \right) \right| \right) \frac{u \left({x}\pm\frac{  \hbar \zh}{2}  \right)  \cdot \vec{\sigma}}{\left|u \left({x}\pm \frac{  \hbar \zh}{2}  \right) \right|}  \; . 
\end{align*}
The same procedure can be applied to the streaming problem and we obtain the solution in Fourier space with the difference that now the solution is expressed in terms of the Fourier transform of the Wigner matrix with respect to the position coordinates $\widehat{F}(\mu,p)\doteq  \mathcal{F}_{x \rightarrow \mu }(F(\cdot,p))  =\frac{1}{(2\pi)^2}\int F\left(x,p \right) e^{-i  x\mu } \dif x$.   %\cuc
%solve
%\begin{align*} 
%	\dpp{F}{t} =& 
%	 \frac{ 1}{i \hbar (2\pi)^{2}}  \int_{\mathbb{R}^{4}} \left[ \Lambda \left(p+ \frac{\ze\zm}{2} \right)  F \left(x',p' \right)  -F \left(x',p' \right)  \Lambda\left(p- \frac{\ze\zm}{2} \right)    \right] e^{i\mu(x-  x') } \dif \mu    \dif x' \;.
%\end{align*}
%\begin{align*}
%	\dpp{\widehat{F}}{t} = \frac{ 1}{i \hbar } 
%	\left[ \Lambda \left(p+\frac{  \hbar \mu}{2}  \right) \widehat{F} (\mu,p)- \widehat{F} (\mu,p) \Lambda \left(p-\frac{ \hbar \mu}{2} \right)\right]   \;.
%\end{align*}
%The solution is 
%\begin{align*}
%	\widehat{F}(\mu,p,t+\Delta) = 
%	e^{-\frac{ i}{\hbar }  \Lambda \left(p+\frac{  \hbar \mu}{2}  \right)\Delta } \widehat{F} (\mu,p,t) e^{\frac{ i}{ \hbar }    \Lambda \left(p-\frac{ \hbar\mu}{2} \right)\Delta }  \;.
%\end{align*}
%\begin{align*}
%	\widehat{F}(\mu,p,t+\Delta) =& 
%	e^{- \frac{i}{\hbar} \left[\zl_0  \left(p+\frac{  \hbar \mu}{2}  \right) -  \zl_0  \left(p-\frac{  \hbar \mu}{2}  \right)\right] \Delta}
%	G^{+}(\mu,p) \widehat{F} (\mu,p,t) G^{-}(\mu,p)  \;,
%\end{align*}
The solution of the streaming propagator is obtained as 
\begin{align*}
	\left( \mathcal{S}^{\Delta}  	\widehat{F}\right) (x,p ) =&  \mathcal{F}^{-1}_{\mu \rightarrow x} \left[
	e^{- \frac{i}{\hbar} \left[\zl_0  \left(p+\frac{  \hbar \mu}{2}  \right) -  \zl_0  \left(p-\frac{  \hbar \mu}{2}  \right)\right] \Delta}
	M^{+}(\zl;\mu,p) \widehat{F} (\mu,p) M^{-}(\zl;\mu,p)  \right] \;.
\end{align*}
%where
%\begin{align*}
%	G^{\pm}(p,\mu) =& 
%	\cos\left(\frac{\Delta}{\hbar} \left|\zl \left(p\pm\frac{  \hbar \mu}{2}  \right) \right| \right) I \mp  i \sin\left(\frac{\Delta}{\hbar} \left|\zl \left(p\pm \frac{  \hbar \mu}{2}  \right) \right| \right) \frac{\zl \left(p \pm\frac{  \hbar \mu}{2}  \right)  \cdot \vec{\sigma}}{\left|\zl \left(p \pm \frac{  \hbar \mu}{2}  \right) \right|}  \; . 
%\end{align*}
%

We describe in detail the strategy adopted for the numerical discretization of the 4D phase-space and its associated Fourier space. 
We employ a uniform mesh for both the direct and Fourier spaces. The phase-space consists of $N_x \times N_y$ points in position and $N_{p_x} \times N_{p_y}$ in momentum. We discuss the connection between the discretization size in the momentum $p$ and its Fourier conjugate variable $\eta$. Similar considerations apply to the $x-\mu$ pair. The 2D momentum space $[p_{x}^m,\; p_{x}^M] \times [p_{y}^m, \; p_{y}^M]$ is sampled by a uniform grid with step sizes $\Delta_{p_x}$ and $\Delta_{p_y}$, where $\Delta_{p_x} = (p_x^M - p_x^m) / (N_{p_x} - 1)$ (and similarly for the $p_y$ direction). For simplicity, we now fix the momentum along the $y$-direction and focus solely on the $p_x$ axis. In this view, we are formally working with a 1D problem defined on  the interval $[p^m, p^M]$, where we drop the $x$ subscript to simplify the notation.
The discrete  points correspond to the momenta $p_i=p_m+(i-1)\Delta_p$. We denote by $\zh$ the Fourier conjugate axis $\zh \in [-\Delta_\zh N_p/2,\Delta_\zh N_p/2]$, symmetric with respect to the origin. Since the Fourier space is periodic, in practical implementations the last point is omitted, and the discretized points belong to the interval  $ [-\Delta_\zh N_p/2,\Delta_\zh\left( N_p/2-1\right)]$. 
For the $\zh$ axis, we take the same number of points $N_p$ as in the momentum axis, resulting a discretization size  $\Delta_\zh=\frac{2\pi}{(p_M-p_m)} =\frac{2\pi}{\Delta_p (N_p-1)}$, and $\zh \in [-\frac{\pi}{\Delta_p} ,\frac{\pi}{\Delta_p}  - \Delta_\zh ]$.  This constraint, relating the size of the momentum interval to the discretization of the Fourier variable, is constitutive of the Fourier series and must be satisfied regardless the numerical scheme adopted whenever one applies the direct and inverse Fourier transforms on equal footing.  In this case, the associated sampled distribution $f_k$ is assumed to be periodic in both $p$ and the dual variable $\eta$.  We denote the discrete Fourier transform (DFT) of the sequence $f_k$ by $\widehat{f}_r \doteq \widehat{f}(p_r) = 1/N \sum_{k=0}^{N-1} f_k e^{i p_r \Delta_{\eta} k}$, with $L_p$ and $L_{\eta}$ representing the periods in the direct and dual variables, respectively. We have $ \widehat{f}(p_r+L_p) =1/N \sum_{k=0}^{N-1} f_k e^{i p_r \Delta_\zh k }  e^{i L_p \Delta_\zh k }$. Periodicity is ensured only if $L_p \Delta_\zh k =n2\pi $ for some integer $n$ and for any integer $k$. This implies the fundamental relation $L_p \Delta_{\eta} = 2\pi$, as this condition ensures that the exponent remains an integer multiple of $2\pi$ for any $k \in \mathbb{N}$.
As remarked by Frensley in \cite{Frensley}, the efficient implementation of a numerical solver for Wigner dynamics benefits from adopting an additional constraint relating the discretization size in position $\Delta_x$ to the discretization size of the momentum Fourier variable $\Delta_{\eta}$. This requirement arises from the definition of the dynamical generator, which is expressed via the pseudodifferential operator in Eq. \eqref{wig_prob}. Integral operators, such as
$
\int_{\mathbb{R}^{4}}  U\left(x\pm \frac{\hbar \zh}{2} \right)  F \left(x,p' \right)  e^{-i\zh(p -p') }  \dif \zh  \dif p'$ 
 require sampling the potential matrix at the discrete points $x_{i,j} = i\Delta_x \pm j \frac{\hbar \Delta_{\eta}}{2}$. 
 However, the structure of the Wigner equation suggests that it is particularly convenient to enforce the relation $\Delta_{\eta} = \frac{2\Delta_x}{\hbar}$. Combining this with the fundamental Fourier relation $\Delta_{\eta} = \frac{2\pi}{L_p}$, we obtain the consistency condition 
$\Delta_x=\hbar \pi/(p_M-p_m)=\hbar\pi/\Delta_p/(N_p-1)$. 
 Although the latter is a natural discretization choice, there is no fundamental reason why it should always hold, and different choices may be more appropriate depending on the physical context. 
In order to optimize the numerical algorithm, it is natural to impose the following constraints on the discretization sizes of both the physical variables and their conjugate Fourier coordinates 
$\Delta_\zh  = [N_{\zh}] \frac{2 \Delta_x}{\hbar} $, where $[N_{\eta}]$ denotes either an integer or the reciprocal of an integer. Similar considerations applied to the space-dual Fourier pair $x-\mu$ (see Eq. \eqref{wig_prob}) lead to $  \Delta_\mu  = [N_{\mu}] \frac{2 \Delta_p}{\hbar}$. 
Ideally, the factors $[N]$ should be as close to unity as possible (noting that in the original Frensley algorithm $[N_{\eta}]$ was set exactly to one) otherwise, the precision of the algorithm may degrade.
Using $\Delta_\zh=\frac{2\pi}{P} $ and  $\Delta_\mu=\frac{2\pi}{L}$, where we indicate $P\doteq p_M-p_m $ and $L\doteq x_M-x_m $, we obtain 
\begin{align}
	 \frac{\pi  \hbar}{ PL}	 = \frac{[N_{\zh}]}{ N_x-1}  = \frac{[N_{\mu}]}{N_p-1}\;.\label{discr_int_ratio}
\end{align}
%It is important that the interval spanned by the Fourier variable $\zh$ be considerable larger than $\Delta_x$ and, in the same way, that the $\mu$ axes be considerable larger than $\Delta_p$, otherwise there will be "blind" areas in space or in momentum, over which the potential or the band structure may changes without that the numerical scheme notice them. We impose 
%			\begin{align*}
%			\Delta_x  \ll  \frac{\hbar\Delta_\zh}{2 } \frac{N_p}{2}\\
%			\Delta_p \ll   \frac{\hbar\Delta_\mu}{2 }\frac{N_x}{2}
%		\end{align*}
%
We note that in the classical limit, it is generally not possible to ensure that $[N_\eta]$ and $[N_\mu]$ remain close to unity. As $\hbar \to 0$, the relationship $\Delta_{\eta} = [N_{\eta}] \frac{2 \Delta_x}{\hbar}$ implies that for fixed grid spacings $\Delta_x$ and $\Delta_{\eta}$, the scaling factor $[N_{\eta}]$ must scale proportionally to $\hbar$. Consequently, maintaining $[N_{\eta}] \approx 1$ would require an excessively fine spatial discretization, leading to a significant increase in computational cost. The application of our algorithm in the semi-classical regime is discussed in the following Section.
\subsection{Semi-classical regime}
Similarly to the previous section, we neglect the relaxation mechanism, which is formally unaffected when considering the classical limit of the dynamics, and can be added without modifications to the final dynamical equation. 
The classical limit of the Wigner equation has been extensively investigated both from the theoretical side and in its application to particle dynamics \cite{Gerard_91,Lions_93,Sparber_03,Gat_14}. Here, we limit ourselves to formal considerations. We assume that all the quantities are smooth enough to justify the differentiability and the convergence of integral operators.  We expand the Wigner equation \eqref{wig_prob} with respect to the Plank constant $\hbar$ up to the first order, obtaining 
\begin{align}
	\dpp{F}{t} =&   \frac{ 1}{ i \hbar }  \left[ \left(U +\Lambda  \right) , F  \right] +  \frac{ 1}{ 2}       \left\{\nabla_x U , \nabla_p F  \right\}    +  \frac{ 1}{  2}   \left\{ \nabla_p \Lambda   ,   F \right\}   + o(\hbar)\;, \label{wig_prb_cl_lim}
\end{align} 
where the curly brackets denote anticommutators. 
The classical limit of the dynamics for particles with spin displays several specific features that lack a counterpart in the standard spinless case. 
Unlike the scalar Wigner function, where the semiclassical limit leads directly to the classical Liouville equation, the multi-band (or spinfull) case must account for the nontrivial topology of the Bloch bundle. This results in the emergence of geometric phases, such as the Berry curvature, which acts as an effective magnetic field in phase space, and the presence of interband transitions that remain relevant even as $\hbar$ vanishes.
In the formal limit $\hbar \rightarrow 0$, the problem in Eq. \eqref{wig_prb_cl_lim} is, in general, ill-posed. Indeed, a solution exists only if the initial data belong to the kernel of the commutator, i.e., $[U(x) + \Lambda(p), F(x, p)] = 0$. In fact, setting $\hbar = 0$, leads to two distinct equations: one algebraic, representing the constraint that the solution must reside within the kernel of the commutator, and a second one describing the dynamics restricted to such a submanifold. The algebraic constraint forces the Wigner distribution to be diagonal in the basis of the local energy eigenstates. If the initial datum does not belong to the kernel of the commutator with the total energy $U + \Lambda $, the system exhibits infinitely rapid oscillations that prevent the existence of a strong classical limit.
The analysis of the critical behavior of the solution for small values of the Planck constant, along with the necessary modifications to the design of the numerical scheme, will be the subject of future work.
Here, we limit ourselves to straightforward modifications of the numerical scheme that become relevant for small $\hbar$. Up to the first order in $\hbar$, the field and streaming groups become 
\begin{align} 
	\mathcal{U}^{\Delta}  : \left\{
	\begin{array}{ll}
		\disp  	\dpp{F}{t} =  \frac{ 1}{ i \hbar }  \left[ U\left(x\right), F \left(x,p \right)   \right]+	    \frac{1 }{2} \left\{\nabla_x U , \nabla_p   F \left(x,p' \right)    \right\}  & t \in (0,\Delta] \\
		\left.F\right|_{t=0} = F_0
	\end{array} \right. \label{split_sch_field_op_CL}
\end{align}
and
\begin{align} 
	\mathcal{S}^{\Delta}  : \left\{
	\begin{array}{ll}
		\disp 		\dpp{F}{t} =  \frac{ 1}{ i \hbar }  \left[ \Lambda \left(p\right), F \left(x,p \right)   \right]+ \frac{ 1}{  2}   \left\{ \nabla_p \Lambda   ,   F \left(x',p \right) \right\}  & t \in (0,\Delta] \\
		\left.F\right|_{t=0} = F_0
	\end{array} \right.  \label{split_sch_drift_op_CL}
\end{align}
whose solution are
\begin{align*}
\left( \mathcal{U}^{\Delta} F\right) (x,p) 	= & \mathcal{F}^{-1}_{\zh \rightarrow p} \left[
	e^{-  i \zh \nabla_x u_0  \Delta}
	C^{+}(u;x,\zh) \widehat{F} (x,\zh) C^{-}(u;x,\zh)  \right] \\
		\left( \mathcal{S}^{\Delta} F\right) (x,p) 	= & \mathcal{F}^{-1}_{\mu \rightarrow x} \left[
	e^{-  i \mu \nabla_p \zl_0   \Delta}
	C^{+}(\zl; p,\mu) \widehat{F} (p,\mu) C^{-}(\zl;p,\mu)  \right] \;,
\end{align*}
where we have defined 
\begin{align*}
	C^{\pm}(u;x,\zh) =& 
	\cos\left( \Delta \left| \frac{ u }{ \hbar}    \pm   \frac{ \zh}{2} \nabla_x u   \right| \right) I \mp  i \sin\left( \Delta \left|\frac{ u }{ \hbar}    \pm   \frac{ \zh}{2} \nabla_x u   \right| \right) \frac{ \left(\frac{ u }{ \hbar}    \pm   \frac{ \zh}{2} \nabla_x u   \right)  \cdot \vec{\sigma}}{\left|\frac{ u }{ \hbar}    \pm   \frac{ \zh}{2} \nabla_x u  \right|}  \; . 
\end{align*}
We note that we maintain the previous scheme based on the spectral representation. However, a significant difference from the full quantum case concerns the role played here by the conjugate variables $\eta$ and $\mu$. In this context, they no longer have a direct connection to the physical scales of the problem and are instead used solely to implement differential operators in Fourier space. Consequently, the constraints in Eq. \eqref{discr_int_ratio} are no longer relevant, and the scheme does not suffer from the increased numerical complexity or the computational cost for small $\hbar$ discussed above.
 
%whose solution is
%$\widehat{F}(\mu,p)=\mathcal{F}_x F$. In this case, we obtain the solution 
%%\begin{align*}
%%	\widehat{F}(x,p,t+\Delta) =&	\mathcal{S}^{\Delta} 	{F}(x,p,t) 	= \mathcal{F}^{-1}_{\mu \rightarrow x} \left[ 
%%	e^{-  i \mu \nabla_p \zl_0   \Delta}
%%	G^{+}(\mu,p) \widehat{F}_0 (\mu,p,t) G^{-}(\mu,p)\right]  \;.
%%\end{align*}
%\begin{align*}
%	\left( \mathcal{S}^{\Delta} F\right) (x,p) 	= & \mathcal{F}^{-1}_{\mu \rightarrow x} \left[
%	e^{-  i \mu \nabla_p \zl_0   \Delta}
%	C^{+}(\zl; p,\mu) \widehat{F} (p,\mu) C^{-}(\zl;p,\mu)  \right] \;.
%\end{align*}
%where
%\begin{align*}
%	G^{\pm}(p,\mu) =& 
%	\cos\left( \Delta \left| \frac{ \zl }{ \hbar}    \pm   \frac{ \mu}{2} \nabla_p \zl   \right| \right) I \mp  i \sin\left( \Delta \left|\frac{ \zl }{ \hbar}    \pm   \frac{ \mu}{2} \nabla_p \zl   \right| \right) \frac{ \left(\frac{ \zl }{ \hbar}    \pm   \frac{ \mu}{2} \nabla_p \zl   \right)  \cdot \vec{\sigma}}{\left|\frac{ \zl }{ \hbar}    \pm   \frac{ \mu}{2} \nabla_p \zl  \right|} \; . 
%\end{align*}

%%%%%%%%%%%%%%%%%%%%%%%%%
%%%%%%%%%%%%%%%%%%%%%%%%%

\section{Test cases}
We discuss some applications of our numerical simulator to different physical systems, showing the versatility and the spectrum of potential applications of the 4D quantum phase-space dynamics.
\subsection{Double-slit experiment}\label{Sec_ex_double_slit}

\begin{figure}[!h]
	\begin{center}
		\includegraphics[width=0.4\textwidth]{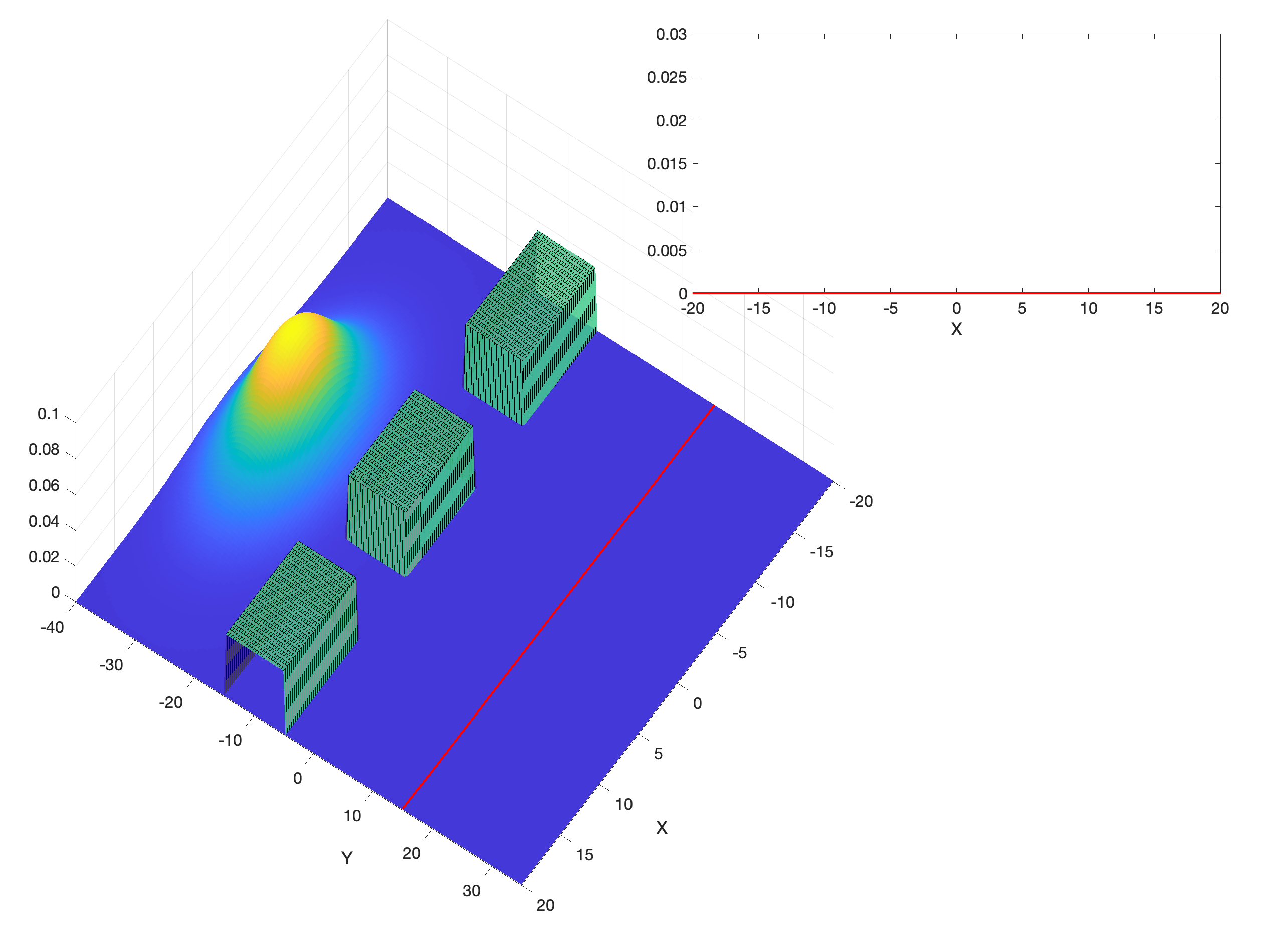} 	
		\includegraphics[width=0.4\textwidth]{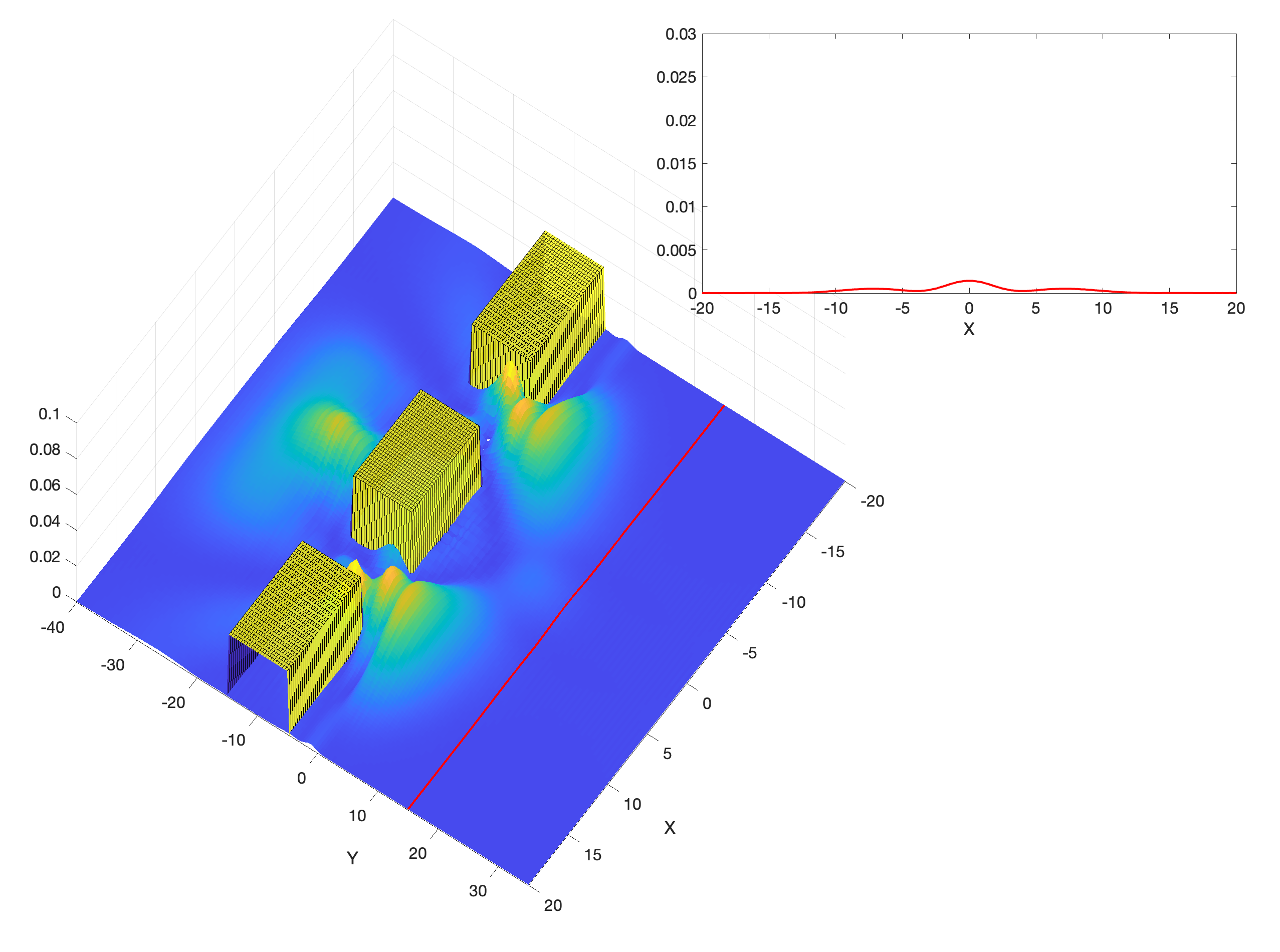}\\ 	
		\includegraphics[width=0.4\textwidth]{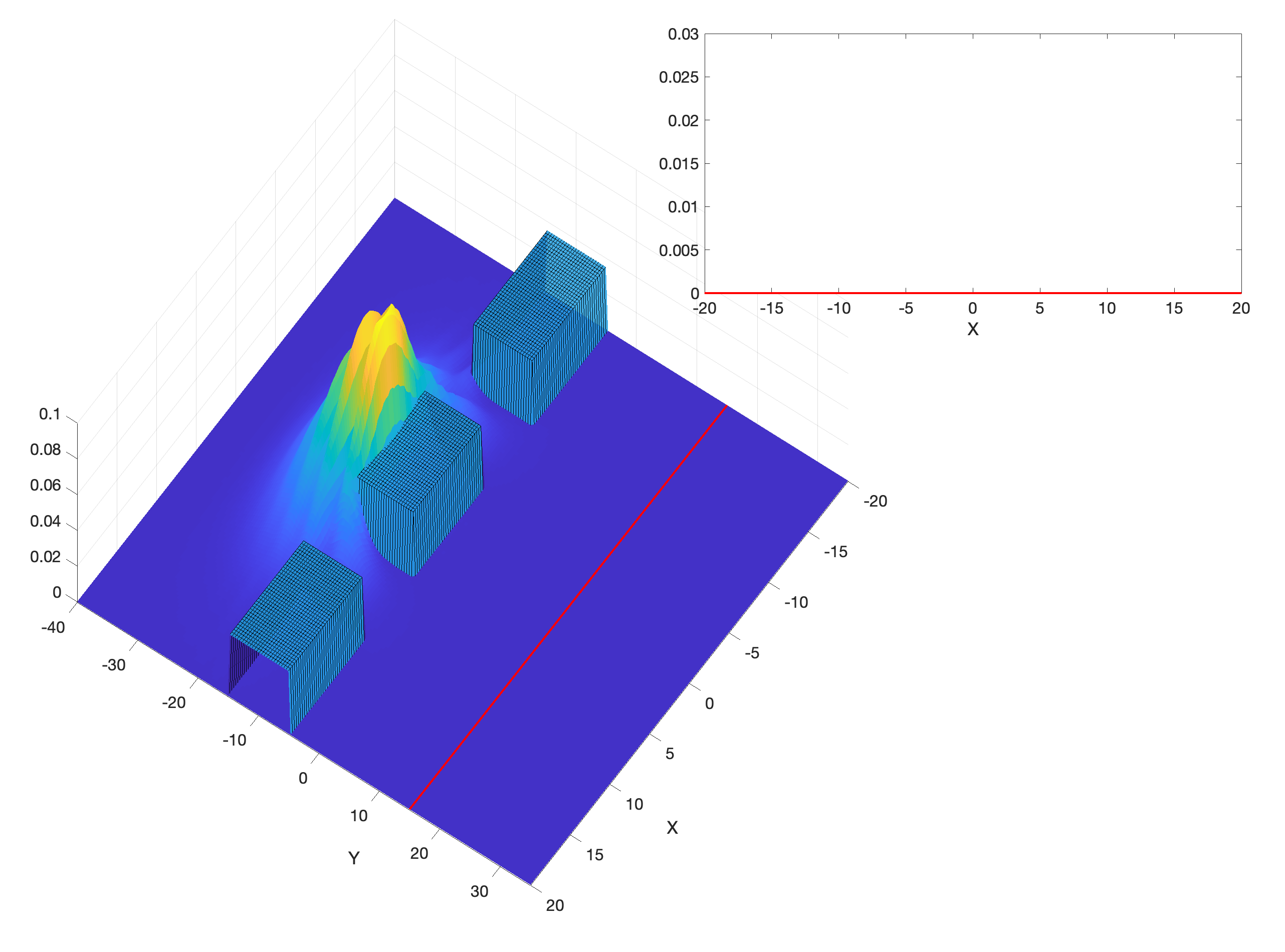} 	
		\includegraphics[width=0.4\textwidth]{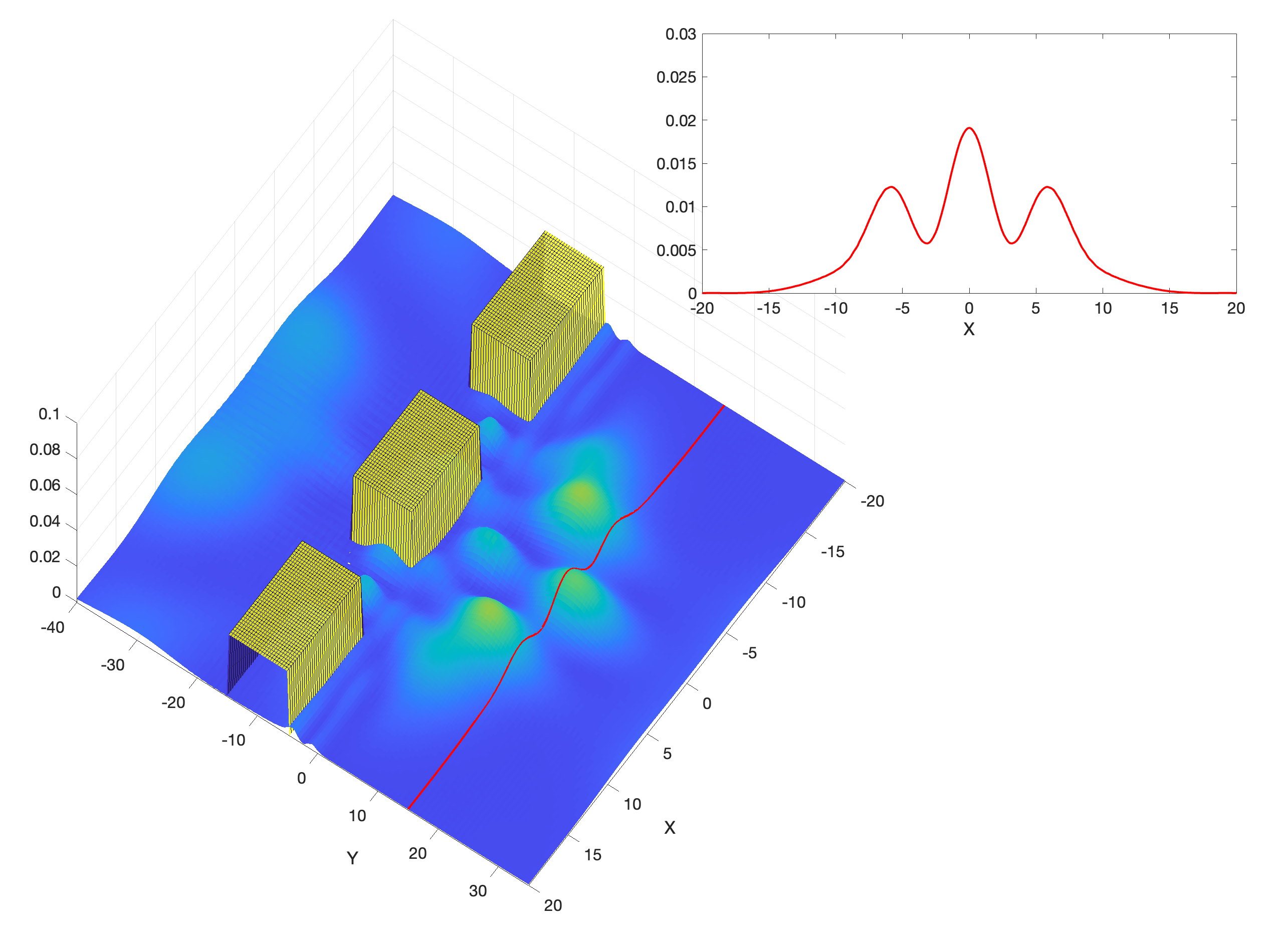}\\
		\includegraphics[width=0.4\textwidth]{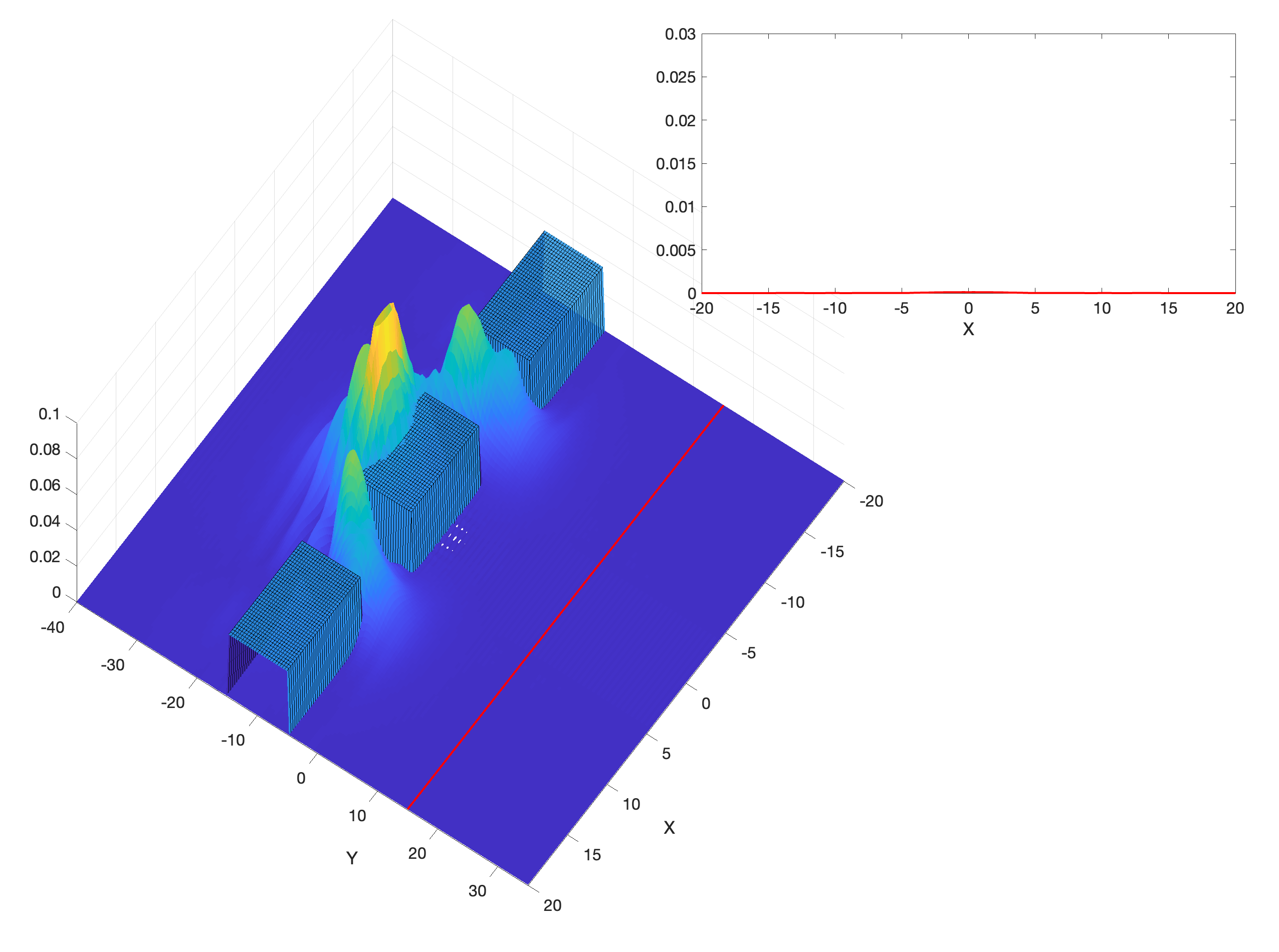} 	
		\includegraphics[width=0.4\textwidth]{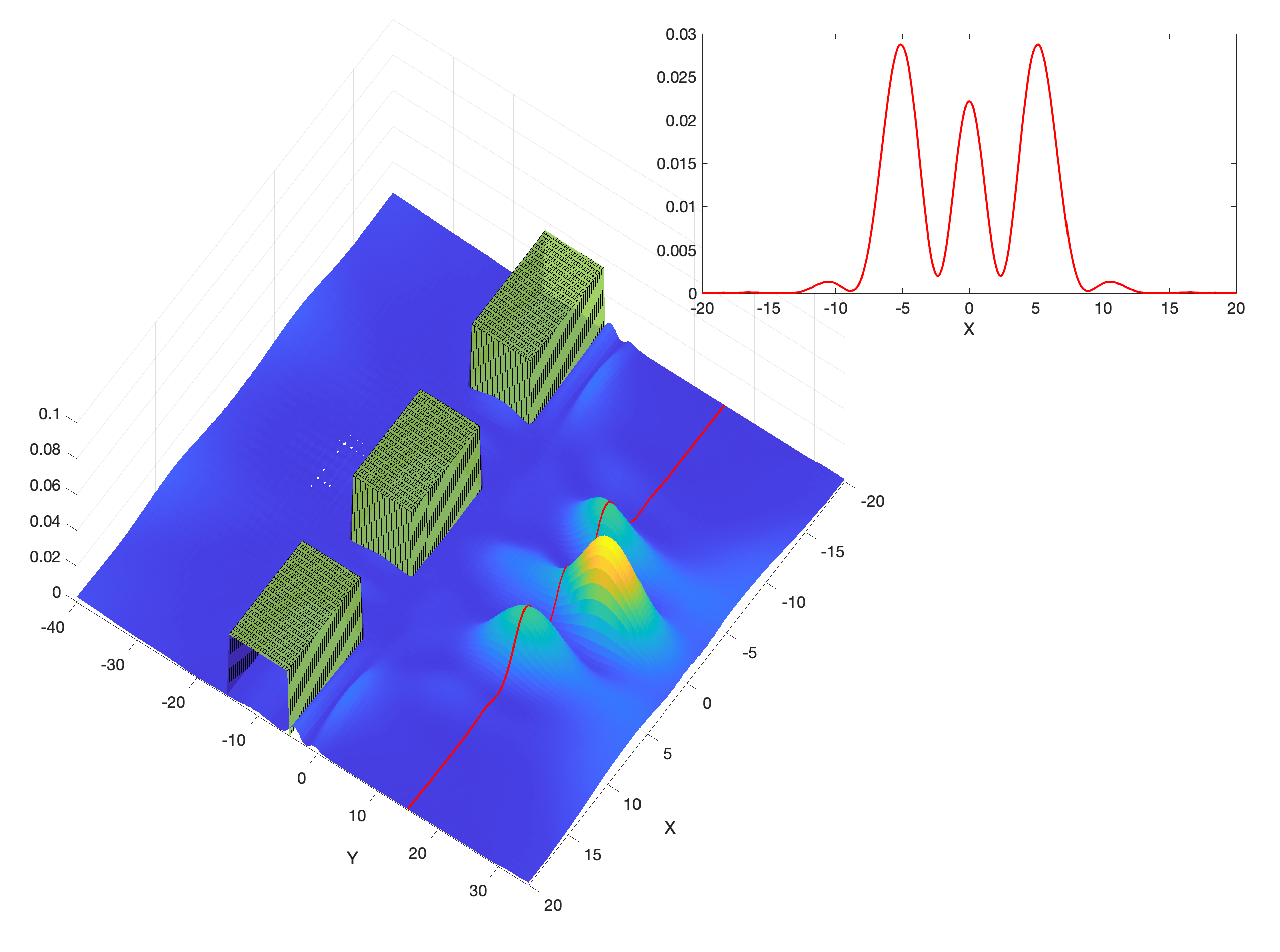}
		%\\	%\href{run:./movie/movie_double_slit.mp4}{$\circ$}
		\caption{\href{https://youtu.be/FGD3cQkYtL0}{Movie (YouTube)}. Double slit experiment. The panels correspond to the times $t=0,\;95,\;190,\;285,\;380,\;475\; \qty{}{\femto\sec}$ (from top to bottom, from left to right). }\label{fig_double_split} 
	\end{center}
\end{figure}

As a first example, we simulate a setup realizing a double-slit experiment under idealized conditions. We consider a single spinless electron, represented by a minimum uncertainty wave packet, impinging upon a potential wall containing two identical narrow slits. The initial Wigner function is given by
\begin{align*}
F (x,p)=	\frac{1}{(\hbar \pi)^2  } e^{-\frac{2\left(x - \overline{x} \right)^2}{\Delta_x^2} -\frac{2\left(y-\overline{y}  \right)^2}{\Delta_y^2} -\frac{\Delta_x^2}{2\hbar^2} \left(p_x-\overline{p_x}\right)^2-\frac{\Delta_y^2}{2\hbar^2} \left(p_y-\overline{p_y}\right)^2} \zs_0  \;,
\end{align*}
which represents the Wigner transform of the minimum-uncertainty packet (see Appendix \ref{App_Wig_from_Sch}).
The parameters used in the simulations are reported in Tab. \ref{tab_IC_double_slit} (concerning the initial condition and the numerical algorithm) and in Tab. \ref{tab_well_double_slit} (regarding the geometry of the double-slit potential).
\begin{table}[b!]
	\centering
	\begin{tabular}{lll}
		Description &Symbol & Value	\\	\midrule
		$x-y$ variances: & ($\Delta_x,\Delta_y$)& $(10,10)$\;$\qty{}{\nano\meter}$ \\
		Mean position : & ($\overline{x},\overline{y}$)& $(0,-30)$\;$\qty{}{\nano\meter}$ \\
		Mean momentum : & ($\overline{p_x},\overline{p_y}$)&  $(0,1)$\;$\hbar \qty{}{\nano\meter^{-1}}$ \\
		\midrule
		Points in position space  &$(N_x,N_y)$&$(160,180)$\\
		Points in momentum space &$(N_{p_x},N_{p_y})$&$(160,180)$\\
	Time step & $\Delta_t$ &\qty{0.86}{\femto\sec}		
	\end{tabular}\\
	\caption{Parameters used in our simulations. The upper lines refer to the initial condition of the electron, represented by a minimum uncertainly packet. Lower lines indicate some relevant numerical parameters. }\label{tab_IC_double_slit}
	\end{table}
\begin{table}[b!]
\begin{minipage}{0.45\textwidth}\vskip 20pt
	\includegraphics[width=0.9\textwidth]{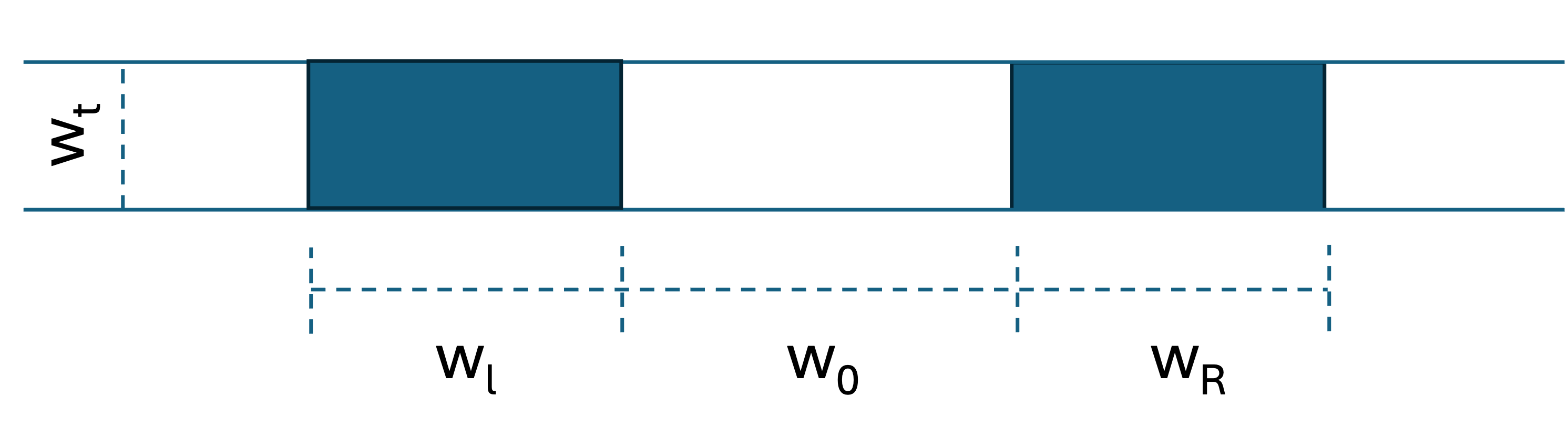} 	
\end{minipage}	
\begin{minipage}{0.45\textwidth}
	\begin{tabular}{ll}
	Potential wall & Value	\\	\hline
	$w_L=w_R$ & $ 6$\;$\qty{}{\nano\meter}$ \\
	$w_0$  & $9$\;$\qty{}{\nano\meter}$ \\
	$w_t$ &  $5$\;$\qty{}{\nano\meter}$ \\
	Height of the barriers &$\qty{4e-2}{eV}$
\end{tabular}
\end{minipage}\\	
\caption{Parameters characterizing the double well potential used in the simulation: $w_l$ and $w_r$ represent the thickness of the left and right slit, respectively, $w_0$ the distance between the slits and $w_t$ the barrier width. }\label{tab_well_double_slit}
\end{table}
	The result of the simulation is represented in Fig. \ref{fig_double_split}. We depict the evolution of the particle density $n(x,y,t)=	\frac{1}{2} \textrm{Tr} 
	\int_{\mathbb{R}^2_p}	F \dif p $ at different times. 
%\notaf{Sim in Wig\_4D\_10 case\_double\_slit }
The minimum uncertainly packet moves toward the potential barrier and impacts the wall orthogonally. The quantum probability density splits into two waves of equal amplitude. To demonstrate the single-particle interference pattern, the two components of the quantum density emerging from the slits are focused toward an ideal screen located behind the wall. This is achieved by a harmonic potential (not depicted in the figure) situated in the half-plane $y>0$. The harmonic potential is defined as $U_h(x,y)= \Theta (y) V_h x^2 $, where $\Theta$ is the Heaviside step function and we set $V_h = \qty{1.5 e-2}{eV.nm^{-2}} $. The interference pattern, arising from the coherent superposition of the two components of the initial Gaussian beam split by the slits, is observed on a screen located at a distance of $\qty{20}{\nano\meter}$ behind the double-slit potential. The position of the screen is indicated by a red curve, which is modulated by the electron density profile. The resulting interference pattern is depicted in the panel located in the top-right corner of each subfigure.

\subsection{Particle gas with spin in magnetic and Rasbha fields}\label{Sec_ex_Rashba}

Spin-orbit coupling is an intrinsic property of quantum particles and gives rise to unique phenomena in certain condensed matter materials. Bychkov and Rashba proposed a simple model of spin–orbit coupling that accounts for electron spin polarization in two-dimensional semiconductors with structural inversion asymmetry. In such materials, carriers experience a net momentum-dependent effective magnetic field, resulting in spin-dependent velocity corrections. These can be modeled by an effective interaction, known as the Rashba Hamiltonian, which shares the same form as the conventional spin-orbit interaction.
The great potential of spin-active transistors for spintronics, based on the Rashba effect where non-equilibrium magnetic polarization can be controlled via electrostatic potentials, has stimulated growing research interest.
Spintronic devices integrated with conventional electronics offer the possibility of achieving electrical control over spin polarization by varying the voltage of a nearby metallic gate electrode. 
%Such devices are of critical importance for modern quantum information technology, as quantum bits encoded in spin degrees of freedom can be manipulated electrically. 
A comprehensive introduction to this field can be found in \cite{Manchon_15}.

\begin{figure}[!h]
	\begin{center}
			\includegraphics[width=0.45\textwidth]{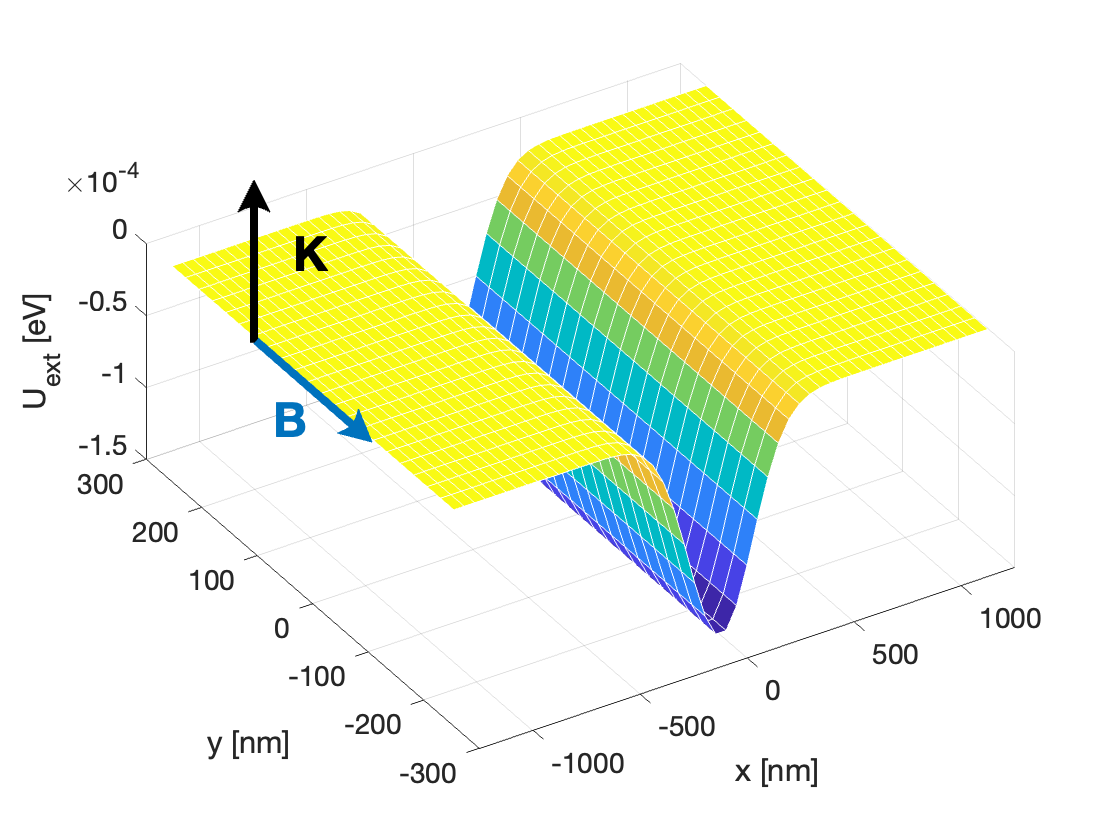} 	
		\includegraphics[width=0.45\textwidth]{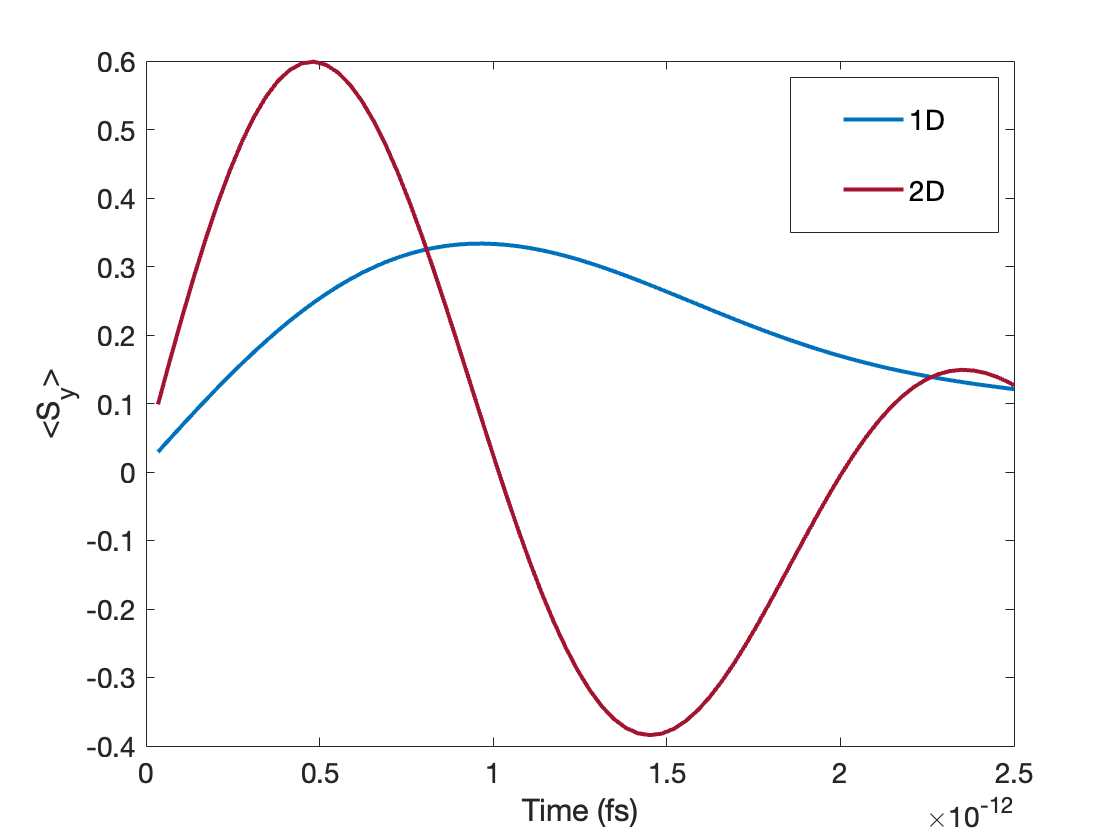} 	
		\caption{ Left panel: Representation of the uniform magnetic field $B$, the uniform Rashba coefficient $K$ and the external electric potential. Right panel: Time evolution of the spin polarization for the 1D (blue curve) or the 2D (red curve) model. }\label{fig_Rash} 
	\end{center}
\end{figure}

\begin{figure}[!h]
	\begin{center}
\boxed{	\includegraphics[width=0.48\textwidth]{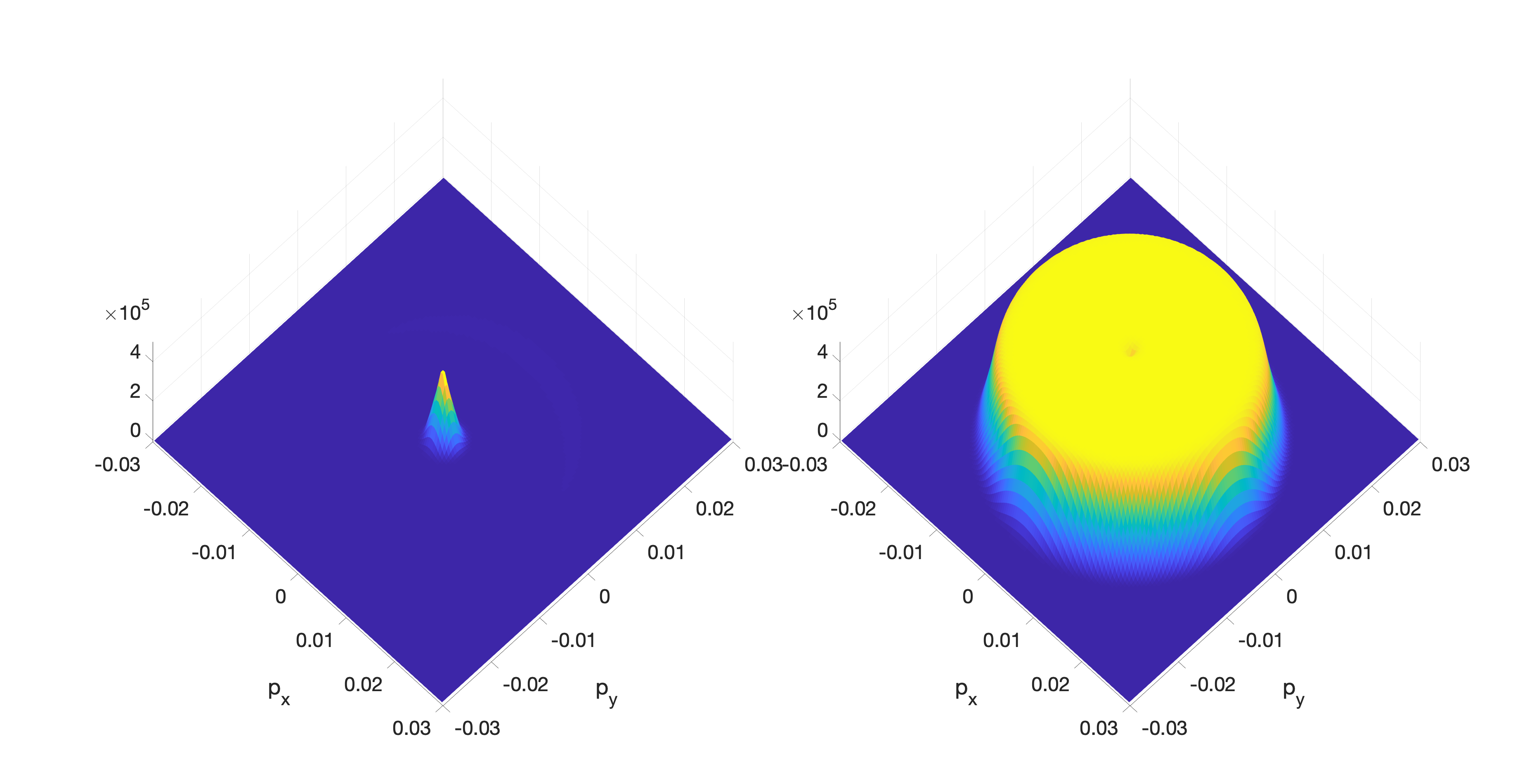} }\;\boxed{	\includegraphics[width=0.48\textwidth]{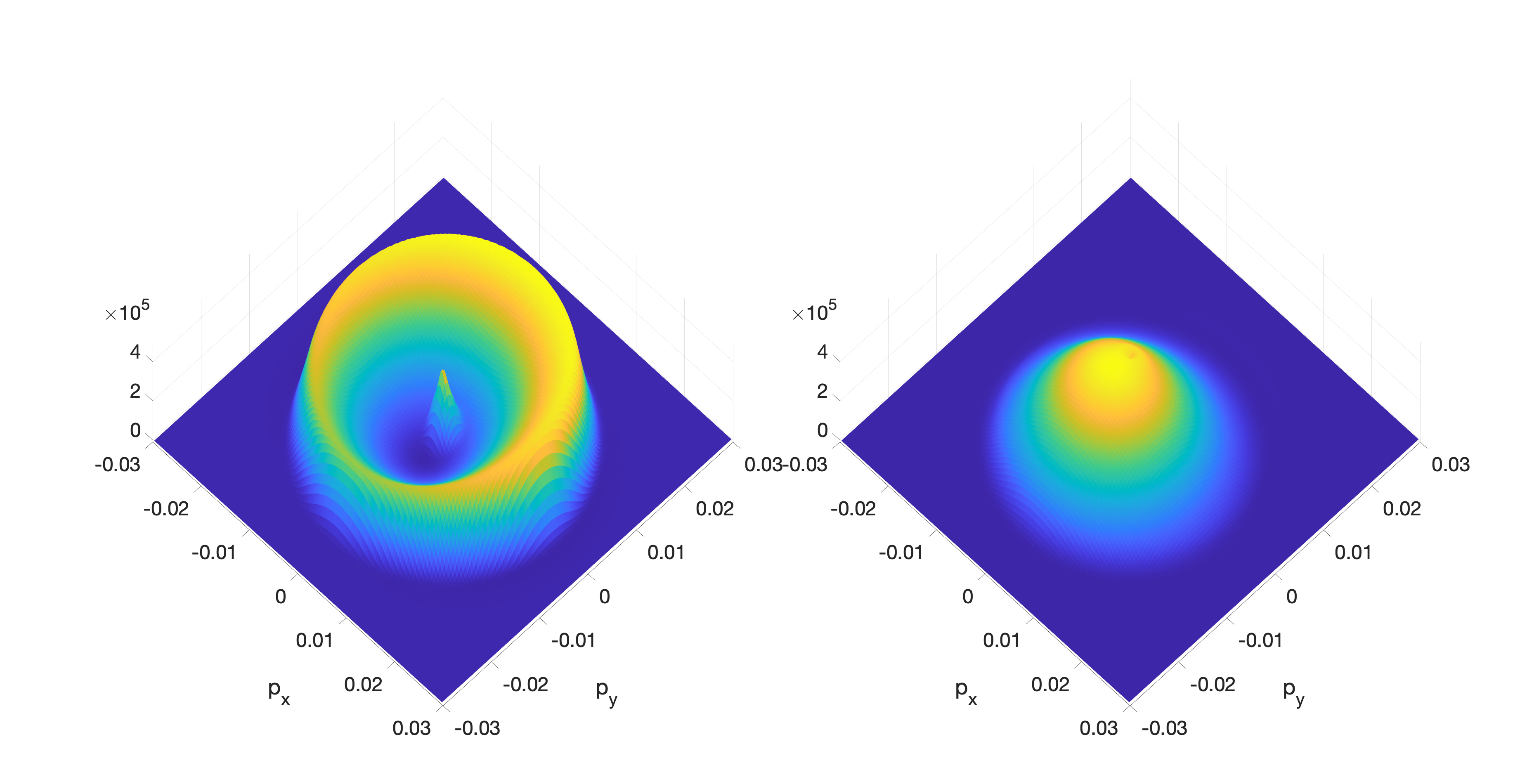}} 	\\	\boxed{\includegraphics[width=0.48\textwidth]{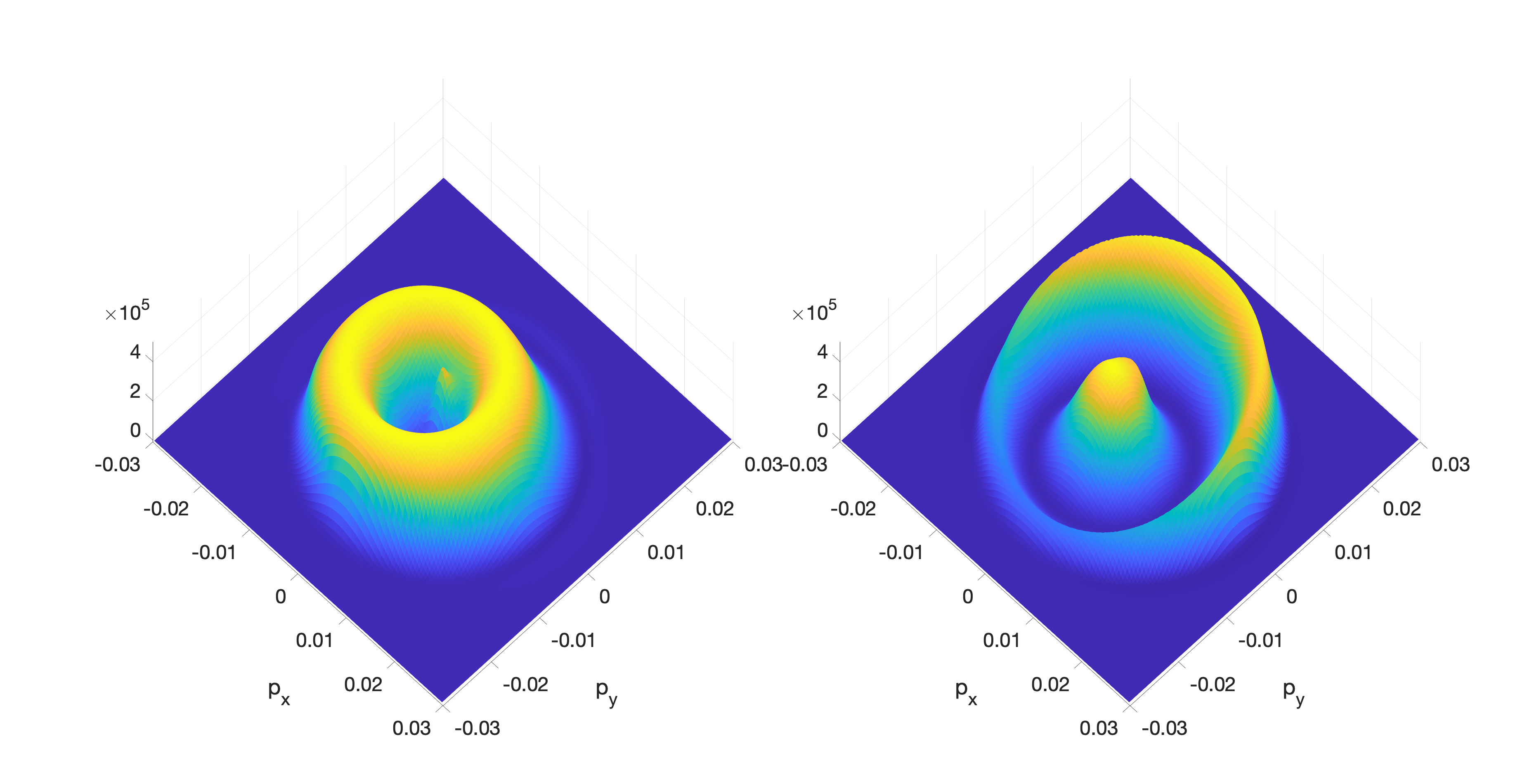} 	}\;\boxed{\includegraphics[width=0.48\textwidth]{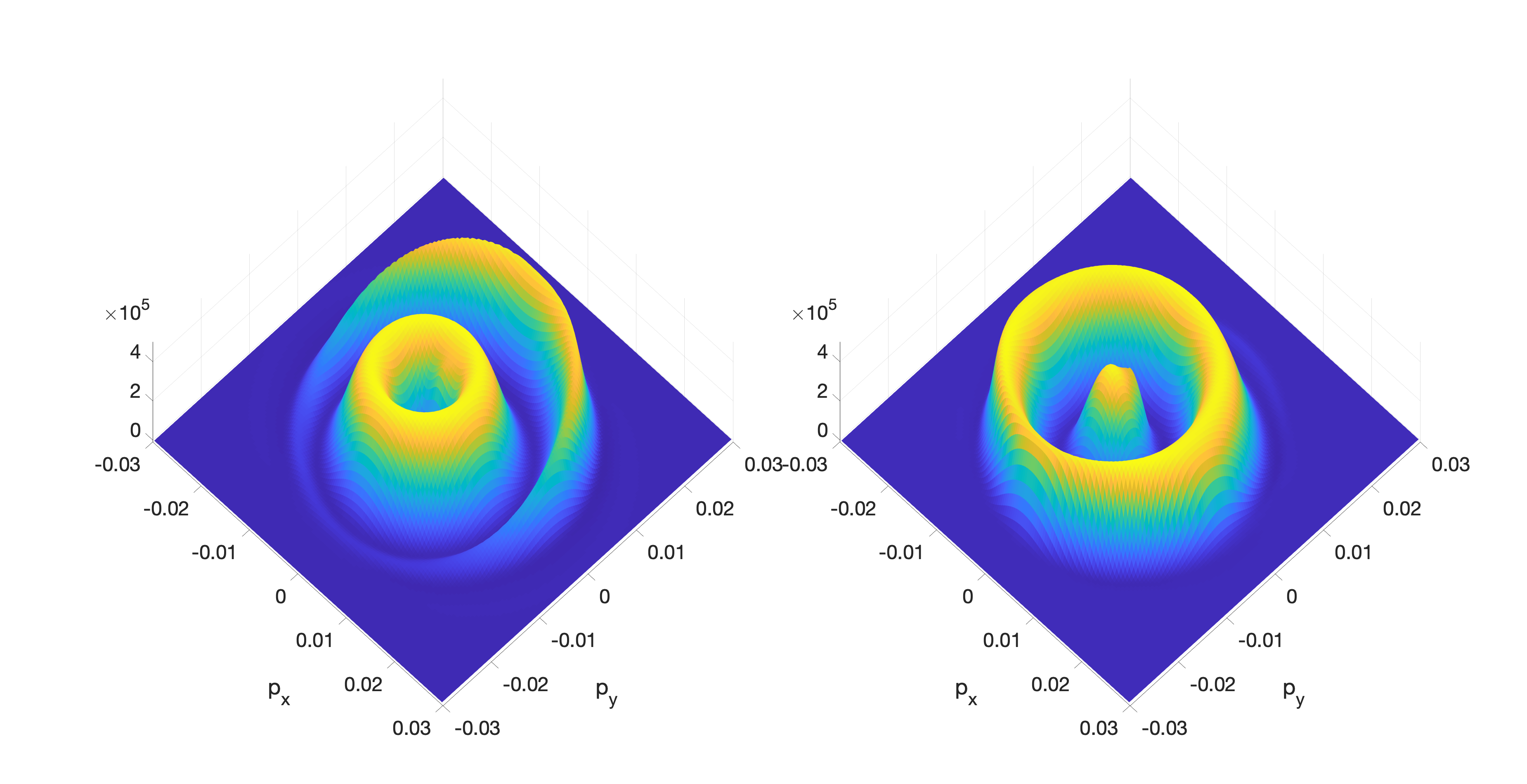}} 	%	\\
%	\href{run:./movie/movie_Rashba_01.mp4}{$\circ$}
 		\caption{\href{https://youtu.be/UsXecLWkXnY}{Movie (YouTube)}. Evolution of the momentum distribution for the spin up (left subfigure) and spin down (right subfigure) populations. The panels refer to the times $t=35,\;860,\;1700,\;260\; \qty{}{\femto\sec}$. }\label{fig_Rash_mom} 
	\end{center}
\end{figure}

We apply our model to simulate the spin evolution of a 2D electron gas within a semiconductor displaying the Rashba effect, in the presence of external electric and magnetic fields.
The total Hamiltonian is given by Eq. \eqref{Ham_spin}.
%\begin{align*}
%	\mathcal{H} =\left( \frac{p^2}{2m} + U \right) \;\zs_0 + \left(p\wedge K - B  \right) \cdot \zs   
%\end{align*}
Our test case is characterized by the following geometry: a uniform, constant magnetic field $B$  is oriented along the $y-$ direction. The Rashba coupling, characterized by the vector $K$, is  along the $z$ (out-of-plane) axis. The electric field $E =- \nabla U $ originates from a Gaussian potential well aligned along $x$, given by  $U=-U_0e^{-\frac{x^2}{2\zl_U^2}}$. 
The fields are represented in the left panel of Fig. \ref{fig_Rash}, and the parameters used in the simulation are described in Tab.  \ref{tab_coeff_Rashba} ($m_e$ denotes the electron mass) \cite{Cavaliere_19}.
 % We define $K=|K|\hat{i}$, with $|K| \doteq \sqrt{\frac{2 E_{SO}}{m_e m_{*}}}    \widehat{K}$, and  $B=E_{Z}  \hat{j}$. We use the following parameters: $E_{SO}  =0.25 \textrm{ m eV} $, $E_Z=0.2 \textrm{ m eV}$, with $U_0=0.15$ meV, $\zl_U=150 $ nm, chemical potential $\zm=0.0$. Effective mass $m_{*}=0.015$, Temperature $T=250 $ mK. 
 \begin{table}[!b]
 	\centering
 \begin{tabular}{lll}
 	Description & Symbol & Value	\\	\hline
 Spin-orbit energy	  & $E_{SO}$ &  $\qty{0.25}{\milli\electronvolt}$ \\
  Rashba coefficient &$ |K| $ & $  \sqrt{\frac{2 E_{SO}}{m_e m_{*}}}$  \\
   Magnetic field  &$ |B| $ & $  \qty{0.2}{\milli\electronvolt}$ \\
  Effective mass  &$ m_* $ & $ 0.015$\;$m_e$ \\
   Temperature  &$ T $ &  $\qty{250}{\milli \kelvin}$ \\
   Electric potential  &$ U_0 $ &  $\qty{0.15}{\milli \electronvolt}$ \\
     Gaussian spread  &$ \zl_U$ &  $\qty{150}{\nano \meter}$ \\
 \end{tabular} \label{tab_coeff_Rashba}
 \caption{Physical parameters used in the simulations.}
\end{table}
As an initial condition, we consider an unperturbed uniform gas described by the Fermi-Dirac distribution given by Eq. \eqref{F_eq}. Due to the presence of the Rashba coupling and the magnetic field, the electron gas exhibits a spontaneous net spin polarization at equilibrium.
We assume that the external potential $U$ is switched on instantaneously at $t=0^+$. In the right panel of Fig. \ref{fig_Rash}, we depict the evolution of the spin polarization along the $x$ direction, comparing it with the case of 1D solution (i. e. neglecting the $p_y$ component of the Rashba Hamiltonian). 
The vector nature of the Rashba interaction, defined by the cross product of momentum and the Rashba coefficient, imparts a distinctly two-dimensional behavior to the system. Our numerical findings confirm that a simplified 1D model, assuming uniformity along the y-axis, is insufficient to capture the full dynamics, regardless of the apparent symmetry of the applied fields.
To illustrate the complexity of the spin excitation, Fig. \ref{fig_Rash_mom} depicts the time evolution of the momentum distribution for the spin-up (left panel) and spin-down (right panel) populations. More precisely, the spin-up (down) distribution denotes the component of the particle density projected onto the $\Pi^+$ ($\Pi^+$) eigenspace associated with the higher (lower) energy branch.
%\notaf{Sim in Rashba\_01 e  Rashba\_01\_simpl ; plot Rashba\_plot\_res.m}

%\notaf{Sim in case\_Rashba\_01 $\backslash$ res\_sim\_01 comando make\_movie\_04}

\subsection{Steering neutral atoms by optical tweezers }\label{Sec_ex_atoms}

As a further application of our code, we consider the transport of ultracold neutral atoms along a planar trajectory using optical tweezer technology.
Two-dimensional arrays of optically controlled neutral atoms have emerged as promising platforms for implementing quantum computing and studying fundamental interactions between atom pairs \cite{Saffman_19, Henriet_20, Bluvstein_23}. Atoms at sub-microkelvin temperatures can be trapped in arrays of micron-scale optical traps, known as optical tweezers, whose configuration and position can be dynamically controlled \cite{Couvert_08, Chen_11,Rosi_13, Zhang_15,Manfredi_17, Gieseler_21, Hwang_23,Cicali_24,Morandi_24_SIAP, Morandi_25_PRA}. A review of recent advances in the manipulation of neutral atoms is provided in Ref. \cite{Browaeys_20}.

\noindent
While the transport of ultracold atoms in optical tweezers has been extensively studied assuming trajectories restricted to a line, recent progress in the parallel manipulation of multiple traps has enabled the implementation of unstructured 2D optical arrays with complex geometries. In this context, a fully 2D simulation of atom transport along nontrivial trajectories has become highly desirable.

\begin{figure}[!h]
	\begin{center}
			\boxed{	\includegraphics[width=0.48\textwidth]{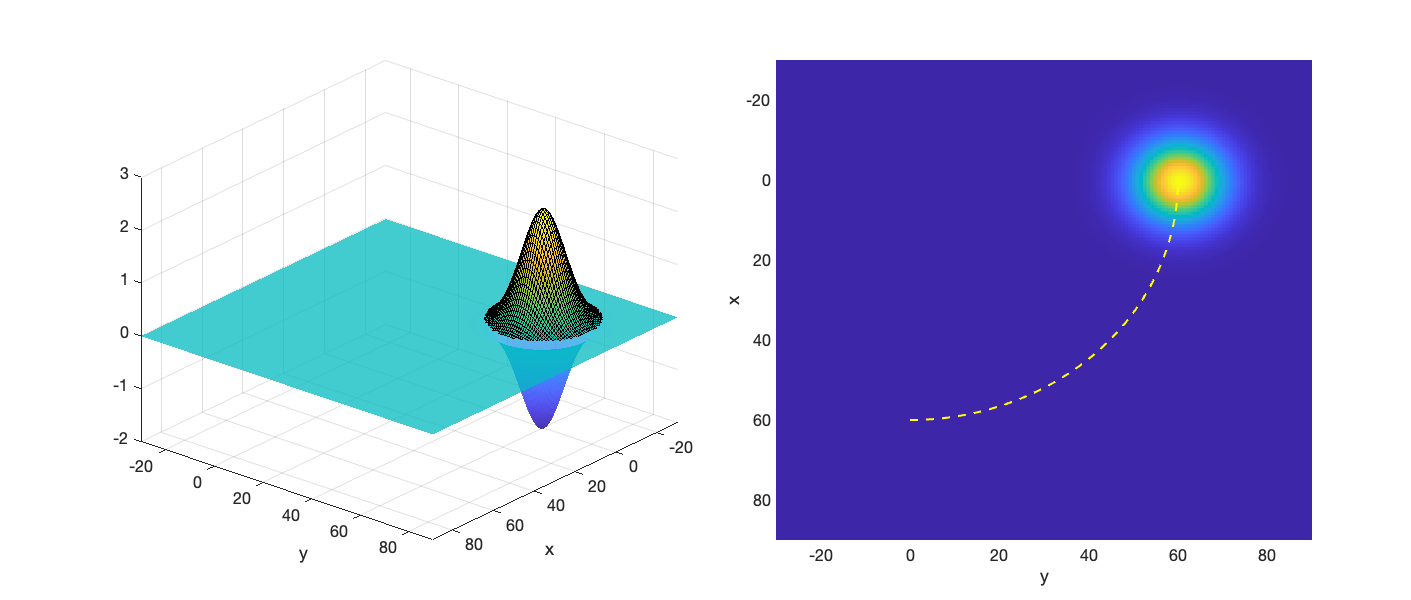}}\;\boxed{	\includegraphics[width=0.48\textwidth]{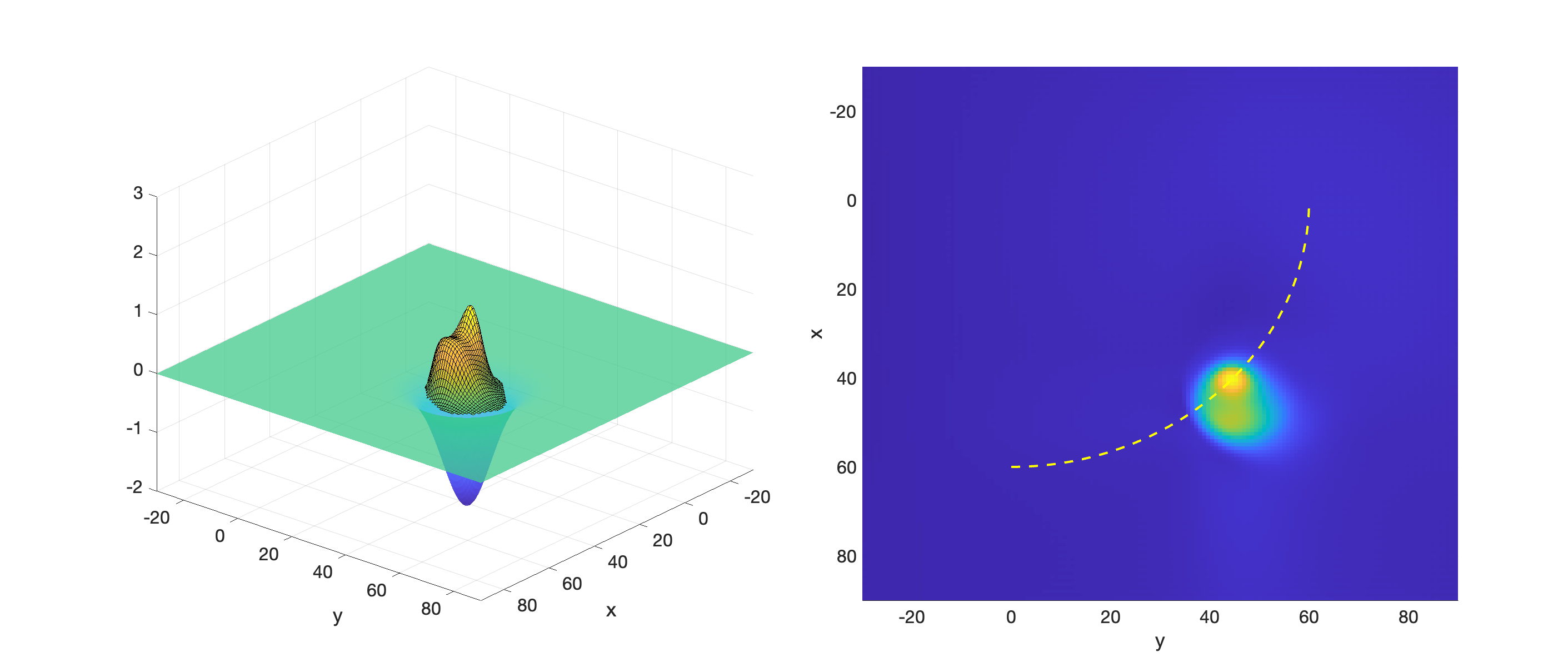}}\\
			\boxed{	\includegraphics[width=0.48\textwidth]{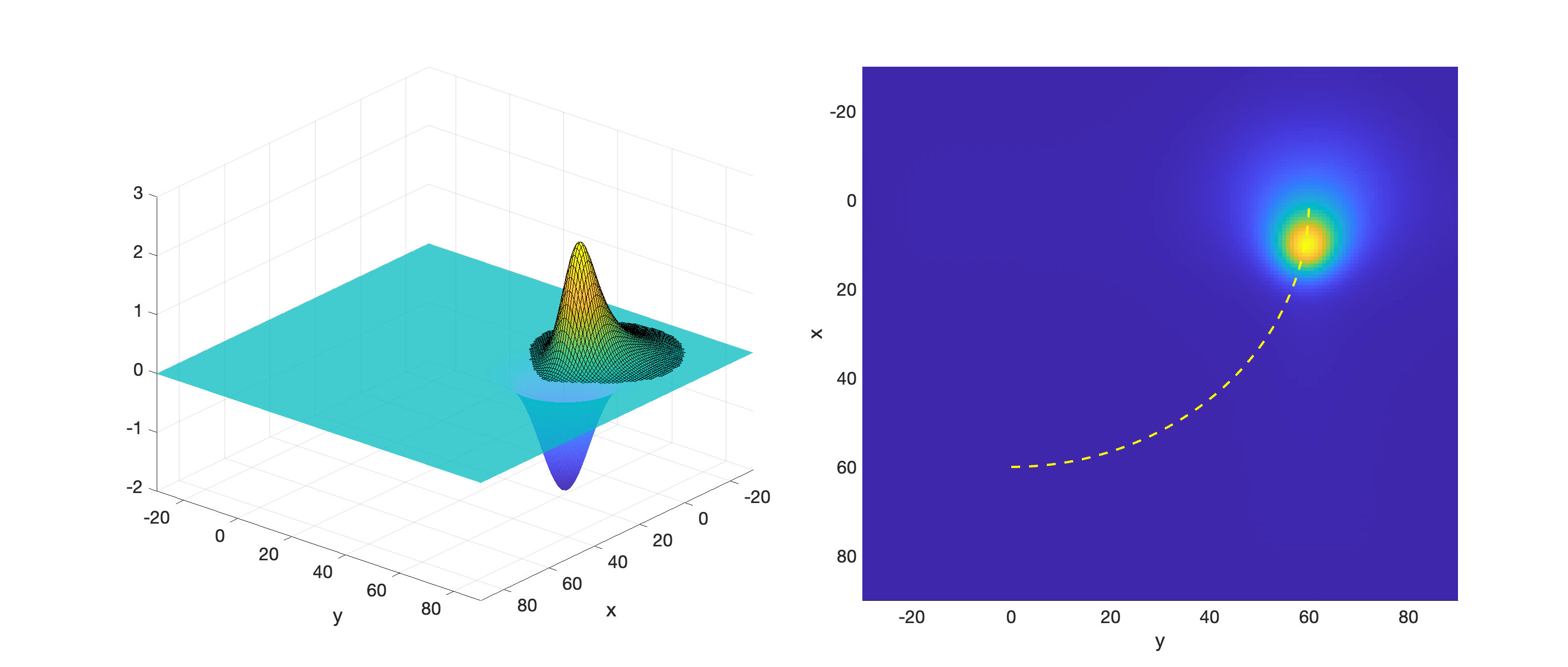}}\;\boxed{	\includegraphics[width=0.48\textwidth]{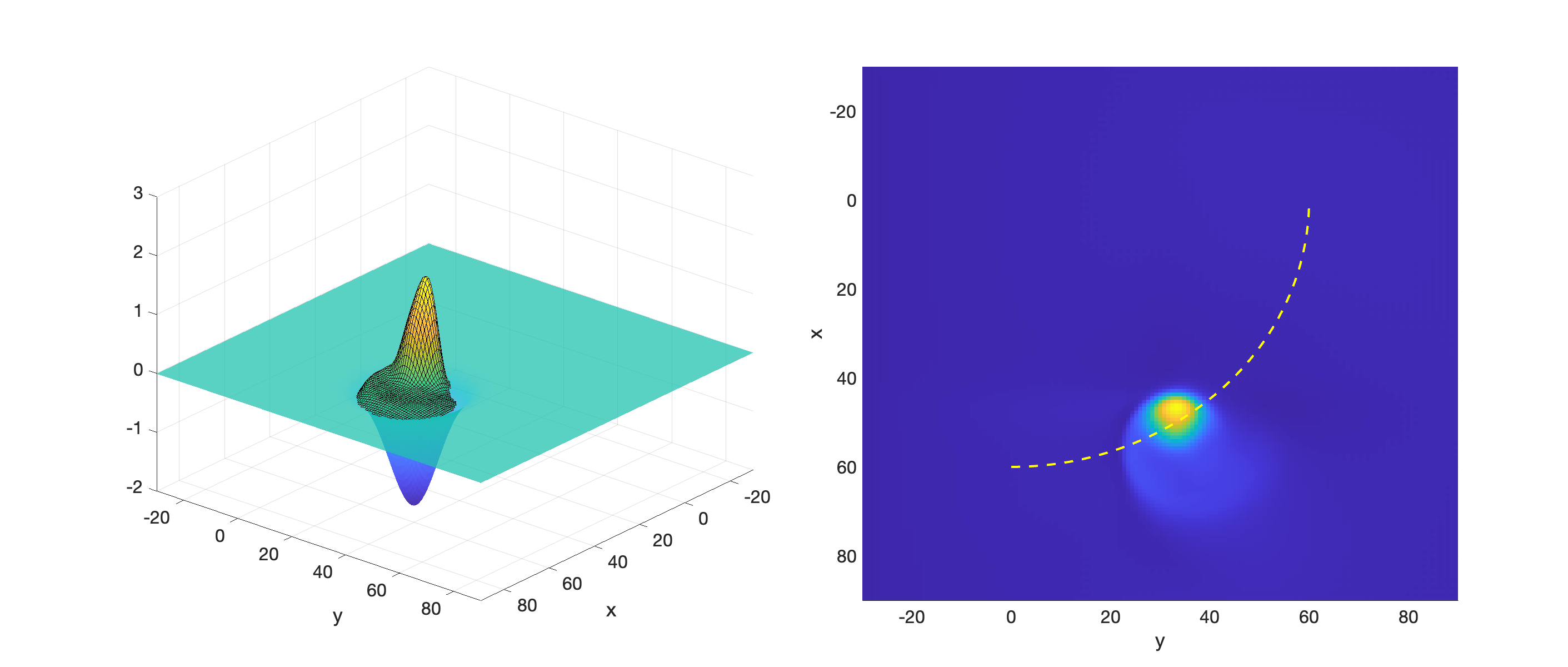}}\\
			\boxed{	\includegraphics[width=0.48\textwidth]{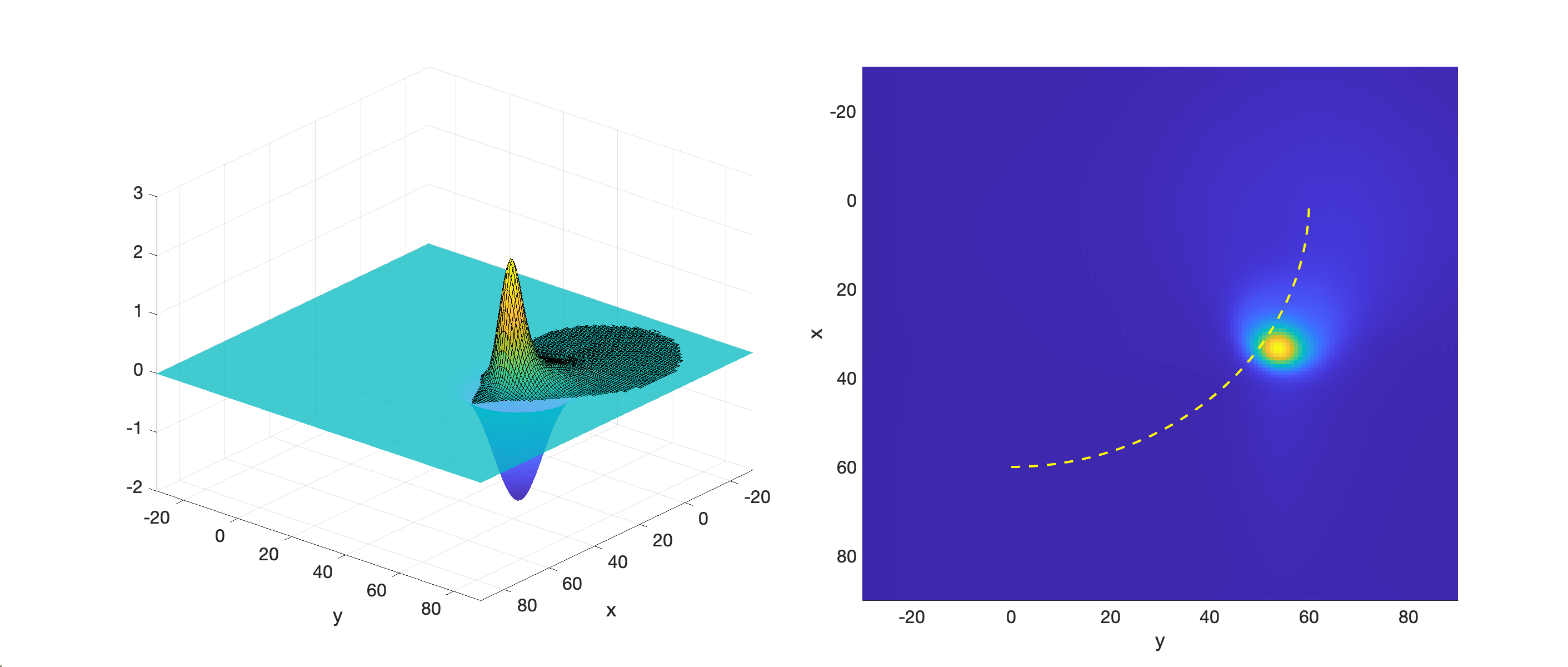}}\;\boxed{	\includegraphics[width=0.48\textwidth]{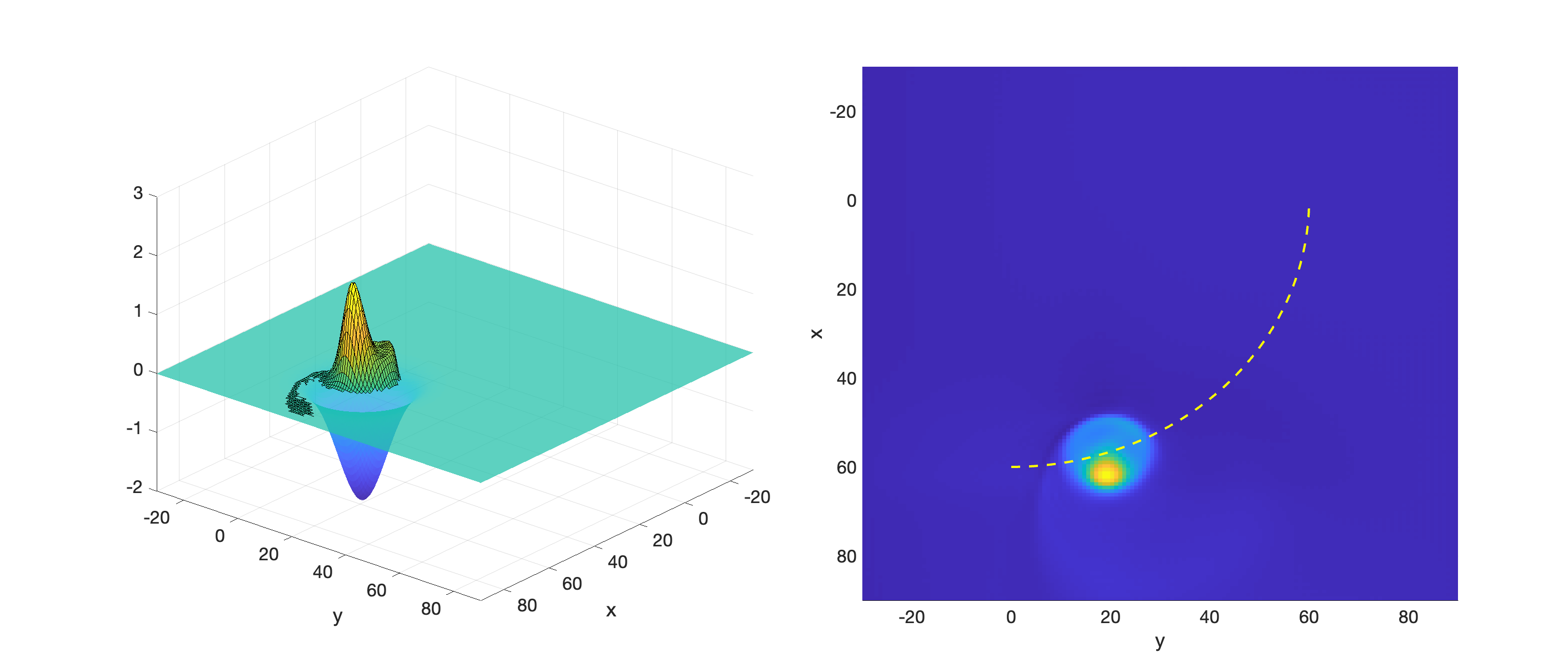}}
			%\\
			%\href{run:./movie/movie_tweez_2D.mp4}{$\circ$}
		\caption{\href{https://youtu.be/4NX1815MBSk}{Movie (YouTube)}. Atom transported by along a circle by the optical tweezer field. In the left side of the panels we depict the 3D view of the atom density and of the tweezer potential. Right side: top view of the atom density. The panels refer to the times $t=0,\;1.4,\;2.8,\;4.2,\;5.5,\;7\; \qty{}{\nano\sec}$ (from top to bottom, from left to right). The dashed yellow curve serves as a visual guide, depicting the trajectory of the optical tweezer's center.} \label{fig_one_atom_opt_tw} 
	\end{center}
\end{figure}

As a first application of the 4D Wigner transport model to optical trap technology, we simulate the steering of a single atom by an optical tweezer along a prescribed spatial trajectory. The optical trap is represented by a time-dependent, Gaussian-shaped potential
\begin{align}
U_{tw}(x,y,t) =& U_0\;   e^{- \frac{\left(x-x_t(t)\right)^2+\left(y-y_t(t)\right)^2}{\zs^2} }\;,\label{U_tw_Gauss2D}
\end{align}
where $(x_t(t),y_t(t))$ are the time-dependent coordinates of the trapping potential's center. The tweezer beam size $\zs$ and the potential depth $U_0$ are kept constant. We assume that the center of the tweezer travels along a quarter-circle arc, covering an angle of $\pi/2$ angle, with the trajectory defined as $(x_t(t),y_t(t))= R(\cos(\zo t ),\sin(\zo t ))$.  
The angular velocity $\zo$ is chosen to be sufficiently small to implement nearly-adiabatic transport.
Assuming a spinless atom and considering the typical experimental regimes of trap depth and temperature, the atom dynamics become essentially classical \cite{Morandi_25_PRA}. Consequently, the Wigner equation is replaced by the classical Liouville equation
 \begin{align*}
	&\disp	\dpp{f}{t} + \frac{p}{m} \cdot\nabla_r f -  \nabla_r U_{tw} \cdot \nabla_p f =    0 \;,
\end{align*}
where $f$ is a real scalar function. As the initial condition for the atom density, we assume a thermalized distribution localized inside the trap  
 \begin{align*}
	f^I (r,p)  =
	\frac{2\zs^2}{  \pi^2  }	\frac{1}{k_B T     } e^{-\frac{1}{k_B T}\frac{p_x^2+p_y^2}{2 m}-\frac{\left(x-x_t(t)\right)^2+\left(y-y_t(t)\right)^2}{\zs^2}  } \;,
\end{align*}
where $k_B$ denotes the Boltzmann constant and $T$ the temperature. We use the following parameters. Tweezer: $R= \qty{60}{\nano\meter}$, $\zo= \qty{180}{\mega\hertz}$, $\zs= \qty{10}{\nano\meter}$ $U_0=- \qty{0.1}{\kelvin}$. Initial condition: $T= \qty{30}{\milli\kelvin}$,  $m=10^4 \; m_e$. 
The results of the simulations are shown in Fig. \ref{fig_one_atom_opt_tw}. 
%\notaf{Sim in Wig\_4D\_10 case\_atoms\_CL\_nospin }
The atom density remains trapped within the moving optical tweezer and follows the prescribed trajectory. In the left panels of Fig. \ref{fig_one_atom_opt_tw}, we present a 3D visualization of the atom density alongside the moving trapping potential. To better track the atomic motion, the corresponding top-down views of the density are shown in the right panels. The dashed yellow curve serves as a visual guide, depicting the trajectory of the optical tweezer's center.
Despite the trapping, the transport process to the target position is accompanied by leakage. A significant portion of the probability density escapes the potential well, thereby reducing the fidelity (the probability of finding the atom within the trap at the conclusion of the protocol. Since experimental implementations typically demand high fidelities, optimal control techniques can be employed to design tweezer trajectories that minimize these losses \cite{Morandi_25_PRA, Cicali_24}. Our simulation platform provides a robust framework for testing and validating such optimal control protocols in 2D optical tweezer systems.

\begin{figure}[!h]
 	\begin{center}
 	\boxed{	\includegraphics[width=0.48\textwidth]{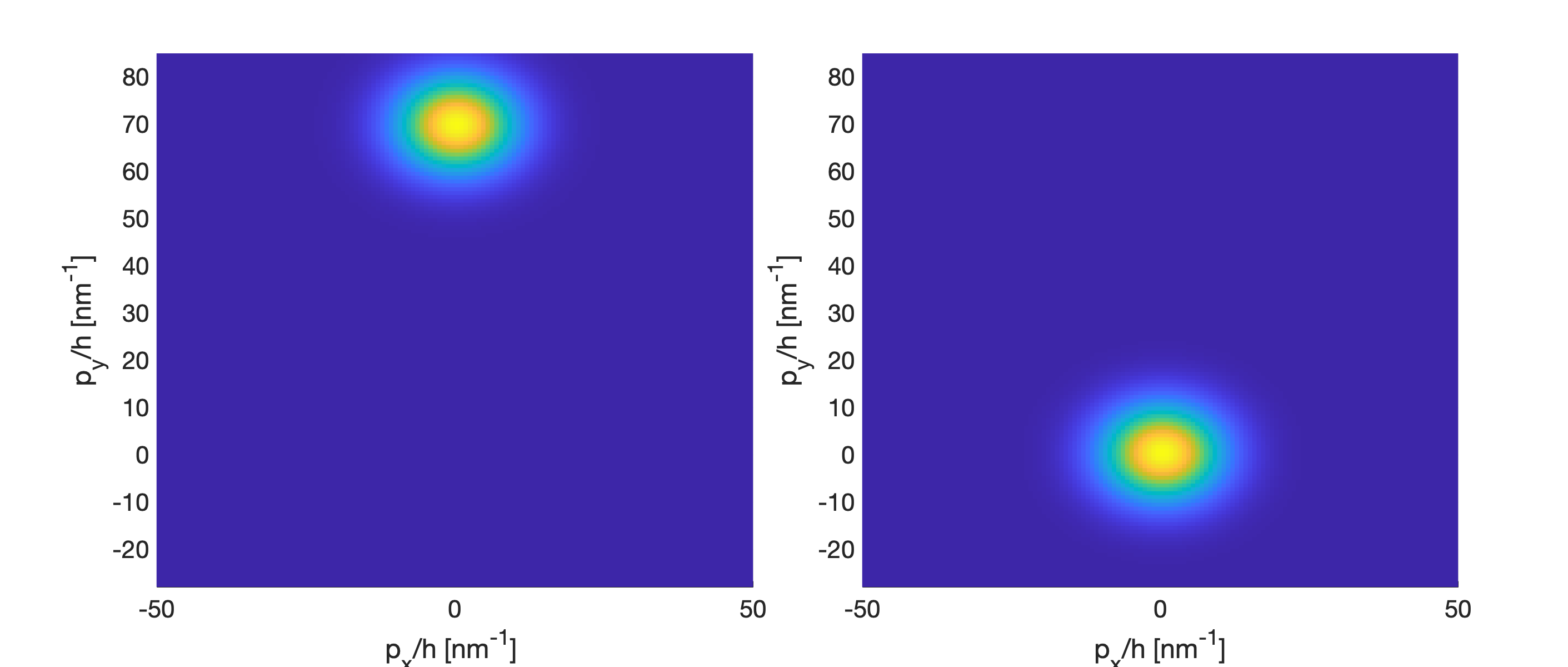}}
 	\boxed{	\includegraphics[width=0.48\textwidth]{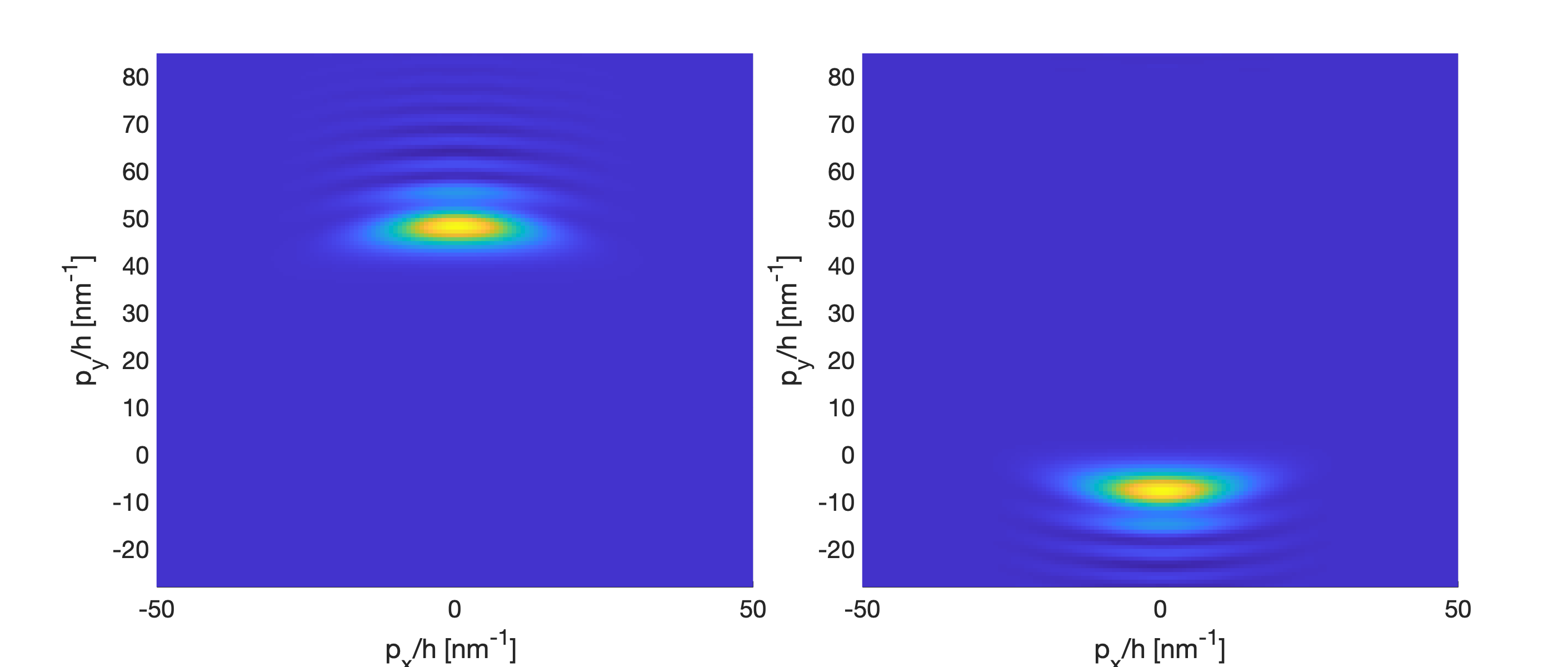}}\\
 	\boxed{	\includegraphics[width=0.48\textwidth]{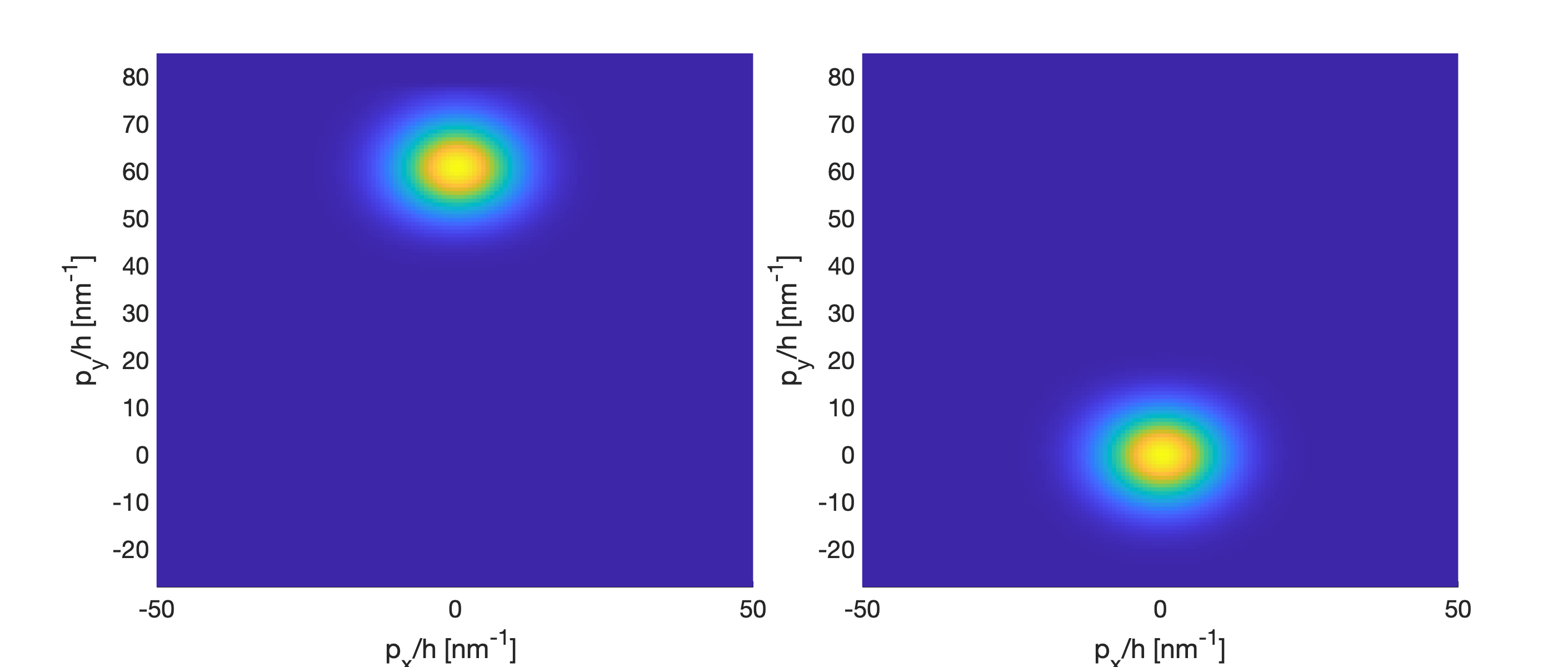}}
 	\boxed{	\includegraphics[width=0.48\textwidth]{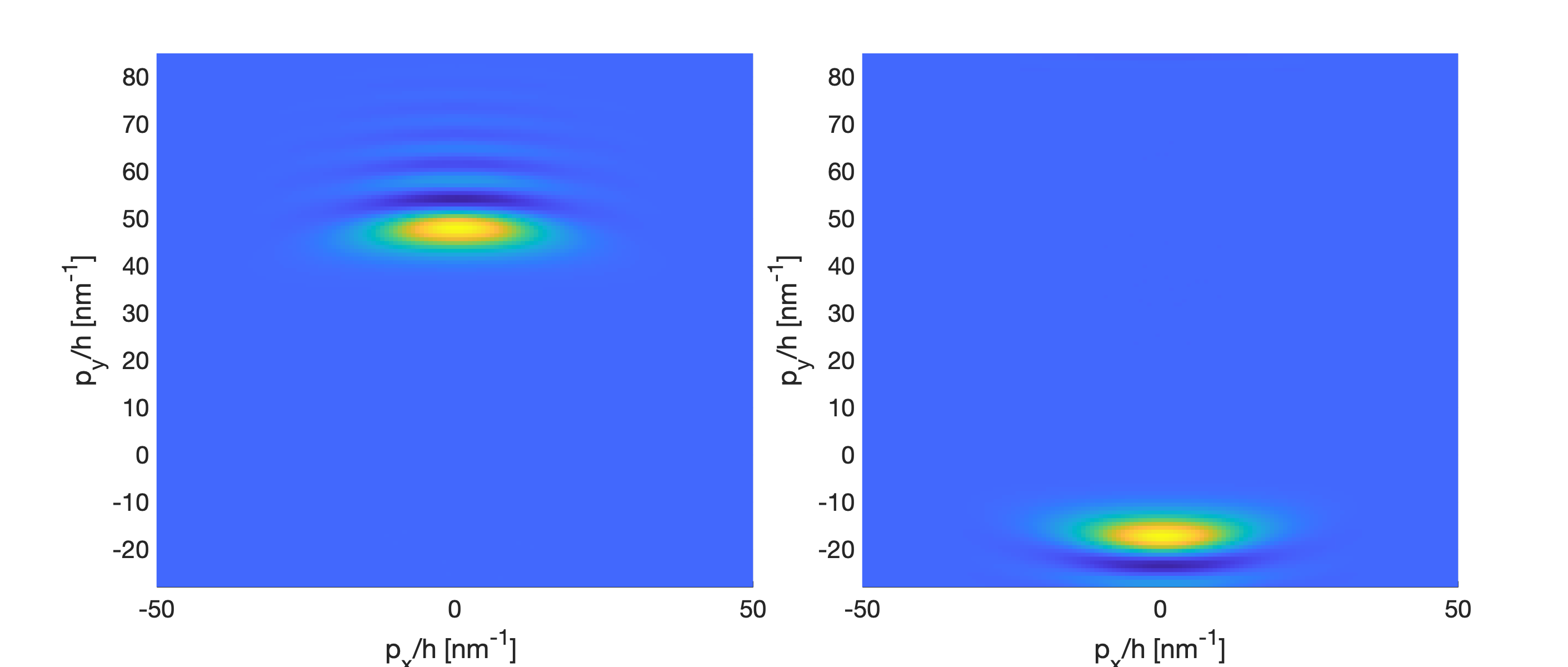}}\\
 	\boxed{	\includegraphics[width=0.48\textwidth]{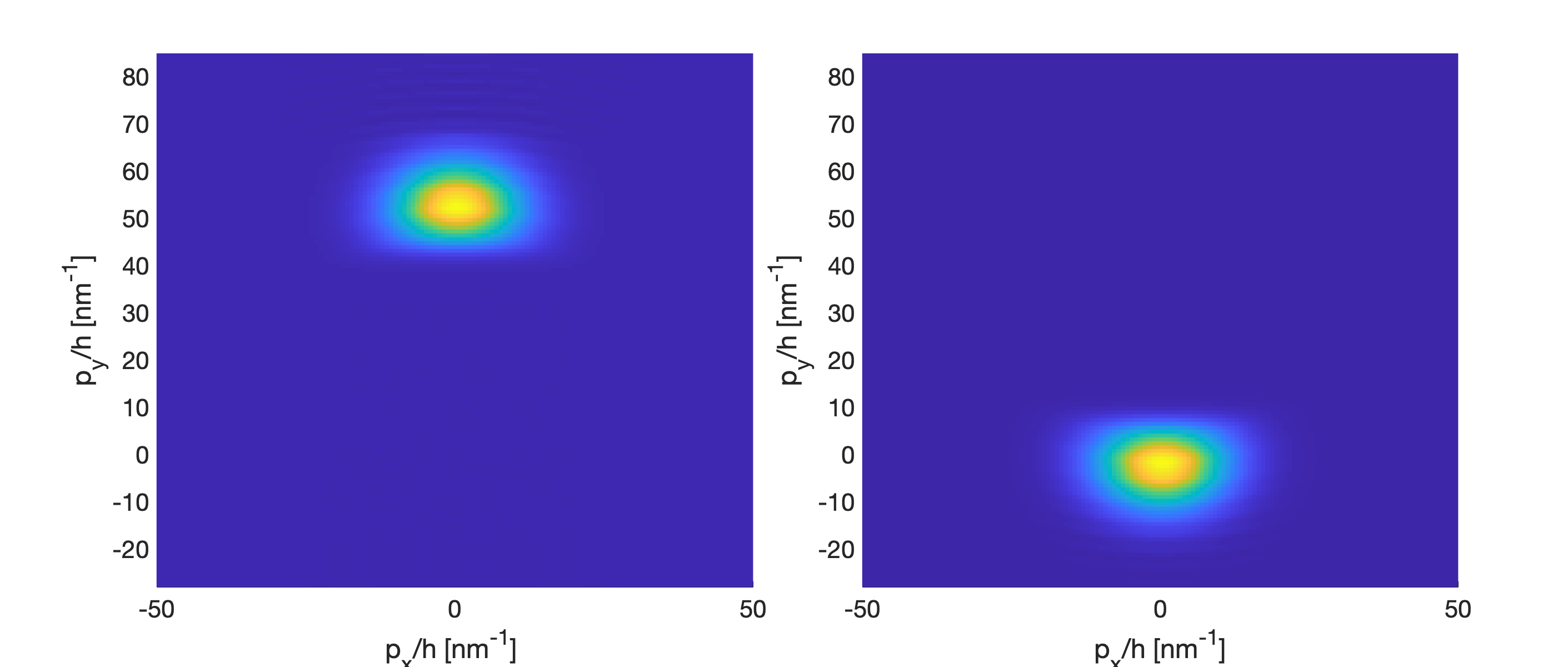}}
 	\boxed{	\includegraphics[width=0.48\textwidth]{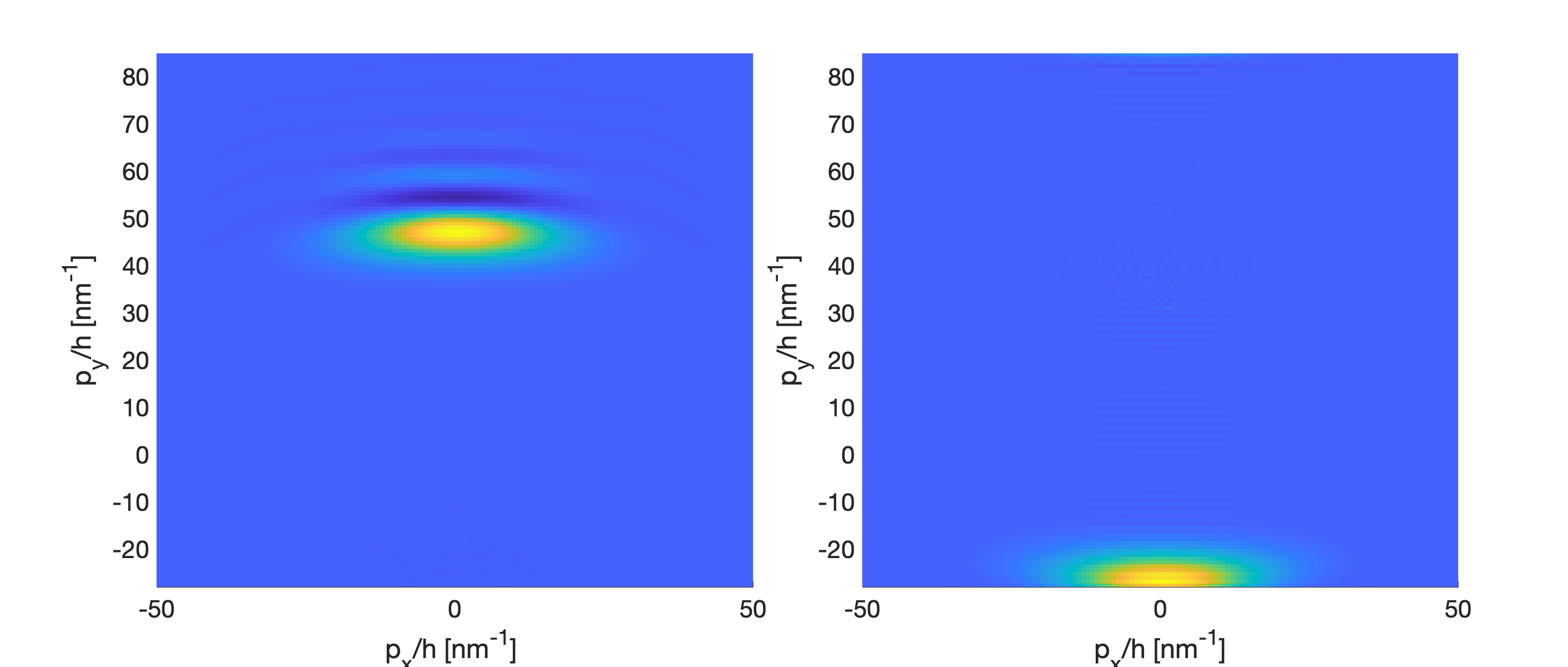}}	%\href{run:./movie/movie_atom_dip_int.mp4}{$\circ$}
 		\caption{\href{https://youtu.be/Y3QyPwGNpG0}{Movie (YouTube)}. Dipole-dipole interaction between two neutral atoms. Evolution of the atom density distribution for the two atoms. Left (right) subfigure depicts the initially moving (at rest) atom. The panels refer to the times $t=0,\;95,\;190,\;285,\;380,\;475\; \qty{}{\femto\sec}$ (from top to bottom, from left to right).  }\label{fig_atom_dip_int} 
 	\end{center}
 \end{figure}

As a second example related to ultracold atom technology, this time addressing quantum dynamics, we study the dipole–dipole interaction between two spinless neutral atoms. This case illustrates the use of the two-band Wigner system to simulate two distinct populations of particles interacting via a nonlinear effective Hamiltonian. For simplicity, we do not include optical tweezer fields in this example, meaning the atoms are unconfined. We model the dipole–dipole interaction between the atoms within the mean-field approximation \cite{Morandi_24_PLA,Morandi_26_PRA}. In our framework, each atom is described by a scalar Wigner distribution obtained by solving the Wigner equation
 \begin{align*}
	&\disp	\dpp{f^i}{t} + \frac{p}{m} \cdot\nabla_x f^i -  \Theta_{U_{dd}^i} [f^i]=    0  & i=1,2.
%	\label{Wig_KB_uni_01}
\end{align*}
The quantum character of the evolution is encoded in the presence of the pseudodifferential operator  $\Theta_{U}[ f ]  $ defined by %In dimensionless variables
\begin{equation*}
%	\label{theta_op_sca}
	\begin{aligned}
		\Theta_{U}[ f ]  (x,{p},t) 
		\doteq& \frac{1}{ i \hbar (2\pi)^2} \int_{\mathbb{R}^{4}} 
		\left[ U\left(x+ \frac{\hbar \zh}{2} \right)  - U\left(x- \frac{\hbar \zh}{2} \right)   \right] f ({x},{p}',t)   e^{-i\left({p} -{p}'\right)  {\zh}}\dif {p}'  \dif  \zh \;.
	\end{aligned}
\end{equation*} 
The dipole-dipole interaction is described by the following effective potential
\begin{align*}
%U_{dd}	=&  - \frac{\mu}{|r_{12}|^3}\left(3 (\mathbf{d}^1\cdot \widehat{\mathbf{r}})(\mathbf{d}^2\cdot \widehat{\mathbf{r}})-(\mathbf{d}^1\cdot \mathbf{d}^2) \right)\\
U_{dd}^i (r,t)	=&   \frac{\mu |d|^2}{| r - \langle r^{\overline {i}} \rangle |^3}\;.
\end{align*}
Here, $d $ is the dipole moment, $\mu$ the interaction strength, $r=(x,y)$ is the atom position, and we adopt the convention $\overline{1}=2,\overline{2}=1$. We assume that the dipole polarization of the atoms is orthogonal to the plane. The mean atom positions are obtained by averaging the corresponding Wigner distribution 
\begin{align*}
   \langle x^i\rangle\hat{i} +  \langle y^i\rangle\hat{j}  \doteq \frac{1}{\int _{\mathbb{R}^4}  f^{i} \dif p_x \dif p_y \dif x \dif y }	\int _{\mathbb{R}^4} \left(  x \;\hat{i}  + y\; \hat{j}  \right) f^{i} \dif p_x \dif p_y \dif x \dif y \;.
\end{align*}
This system can be described by our formalism, using the following diagonal two-band Hamiltonian
\begin{align*}
	\mathcal{H} =\frac{p^2}{2m}  \;\zs_0 +\left(\begin{array}{cc}
		U_{dd}^1 &  0 \\
		0 & U_{dd}^2
	\end{array}\right) \;.
\end{align*}
 The Hamiltonian matrix is always diagonal, consequently, the two populations remain decoupled, and no quantum superposition between them occurs. Initially, both atoms are represented by localized Gaussian wave packets. The centers of mass of the two atoms are aligned with the $y$-axis at a distance of $\qty{70}{\nano\meter}$. One atom, whose density is shown in the left subpanel of Fig. \ref{fig_atom_dip_int}, is at rest, while the second (represented in the right subpanel) travels with a momentum of $0.5\hbar\;\qty{}{\nano\meter^{-1}}$ toward the first. The results of the simulation are displayed in Fig. \ref{fig_atom_dip_int}. 
We depict the evolution of the density for the two atoms: the initially traveling atom is shown in the left subpanels, while the atom initially at rest is shown in the right subpanels. %
The dipole–dipole interaction mediates the transfer of kinetic energy from the first atom to the second. By the end of the simulation, the initially traveling atom has come to rest, and its momentum has been entirely transferred to the second atom, resulting in a complete exchange of momentum, characteristic of an elastic scattering process.
 We observe that the scattering mechanism exhibits a pronounced nonclassical signature, which is evident from the nonlocal character of the interaction and the presence of interference ripples in the two Wigner distributions.

% \notaf{Sim in Wig\_4D\_10 case\_atoms\_dip\_int }

 \subsection{Klein tunneling in topological superconductors }\label{Sec_ex_BdG}

Given the implications of quantum computing for future technology, the investigation of topological quantum systems has emerged as a central theme in condensed matter research over the last decade. 
The geometrical interpretation of a topologically nontrivial system is essential for modeling topological insulators, as well as for explaining topological protection and topological superconductivity.
A remarkable manifestation of topological superconductivity emerges in the form  of edge states in a bulk material. The electronic properties of these superconductors are explained in terms of Bogoliubov quasiparticles, which represent the coherent superposition of electrons and holes \cite{Kitaev_06,Fu_09,Liang_11,Beenakker_13}. Similarly, low-energy excitations in Bose-Einstein Condensates (BECs) at finite temperatures are also described in terms of Bogoliubov quasiparticles. The noncondensed atoms superpose coherently with the atoms belonging to the condensate \cite{Griffin_book,Choi_12,Mor_14PRA}.
 
\noindent
A simple effective model describing Bogoliubov quasiparticles in chiral topological superconductors is represented by the following Hamiltonian denoted as Bogoliubov de Gennes (BdG) Hamiltonian 
 \begin{align*}
  \mathcal{H}_{BdG} =\underbrace{\left(\begin{array}{cc}
 		\frac{p^2}{2m}-\mu  & -2 \Delta (p_y-ip_x) \\
 		- 2 \Delta (p_y+ip_x) & -\frac{p^2}{2m} + \mu 
 	\end{array}\right)}_{\doteq \Lambda_{BdG}} + U (x)\zs_0, \end{align*}
where $\Delta$ represents the superconducting paring parameter, $\mu$ is the chemical potential and $U$ some external potential.  The energy associated to a Bogoliubov quasiparticle is $\zl^\pm = U(x) \pm\sqrt{\left(\frac{p^2}{2m}-\mu \right)^2+ 4\Delta^2p^2 }$. 
\begin{figure}[!h]
 	\begin{center} 
 		\begin{align*}
 			\zl^+(p)  && \zl^-(p) 
 		\end{align*}
 		\includegraphics[width=0.9\textwidth]{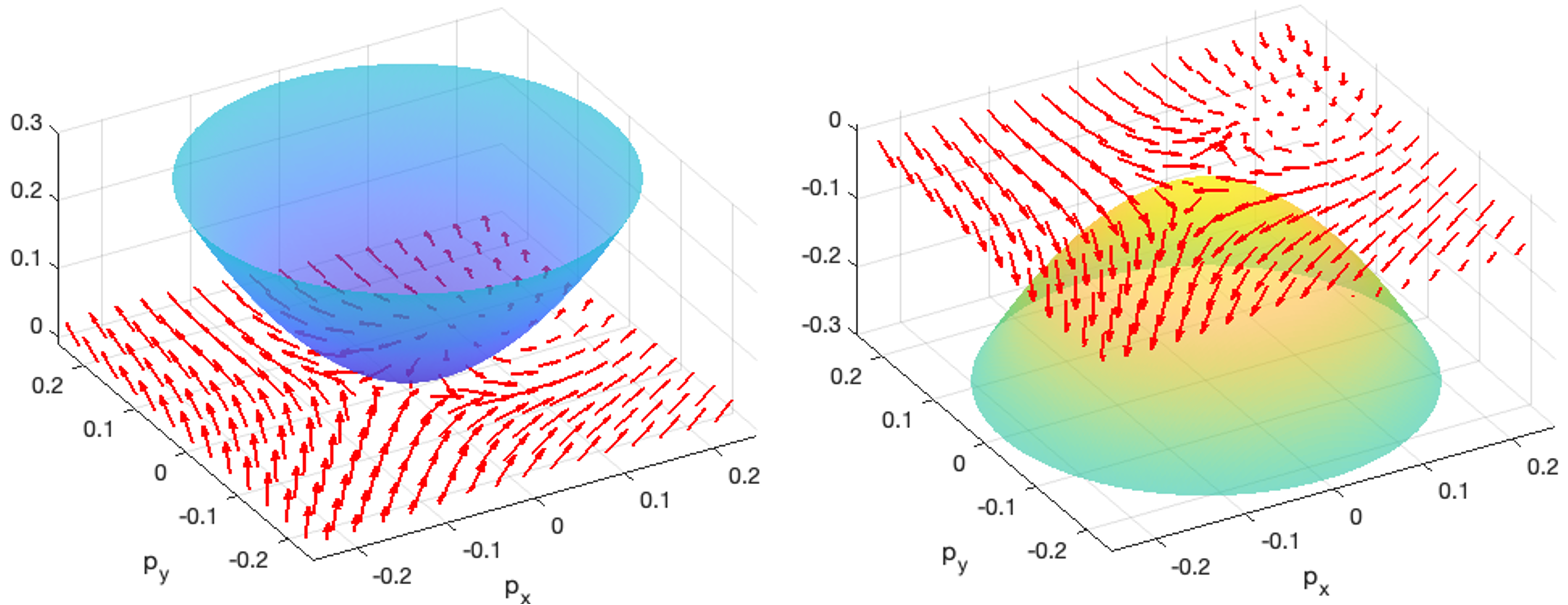} 	  	 
 		\caption{ Band structure and eigenvector direction on the Bloch sphere. }\label{fig_band_diag} 
 	\end{center}
 \end{figure}
The Bloch sphere representation offers a natural geometric framework for characterizing topological superconductors through their Hamiltonian eigenspaces.
The Bloch sphere represents the Hopf fibration, mapping the complex projective line $	\mathbb{P}^1( \mathbb{C}) $ onto the unit sphere $S^2$.  Algebraically, such a map is implemented by associating a spinor $\zy $ with the expectation values of the Pauli matrices $	 \mathbb{P}^1( \mathbb{C}) \ni\psi  \rightarrow  	\hat{n} \doteq  \langle \psi |\zs \psi \rangle  \in S^2 $. Explicitly, the Bloch vector is given by $(\hat{n}_x ,\hat{n}_y,\hat{n}_z ) = (2 \textrm{Re} \left(\overline{\psi_1} \psi_2  \right) ,  
 2 \textrm{Im} \left(\overline{\psi_1} \psi_2  \right),
 |\psi_1|^2-|\psi_2|^2)$. The  two eigenvectors of the BdG Hamiltonian, $	\Lambda_{BdG} \psi^\pm=\zl^\pm \psi^\pm $ are mapped to opposite vectors on the Bloch sphere, 
  i. e. $\psi^+ \rightarrow \hat{n}^+ $, $
 \psi^-\rightarrow \hat{n}^- = -\hat{n}^+$. 
 
\begin{figure}[!h]
	\begin{center} 	\includegraphics[width=0.8\textwidth]{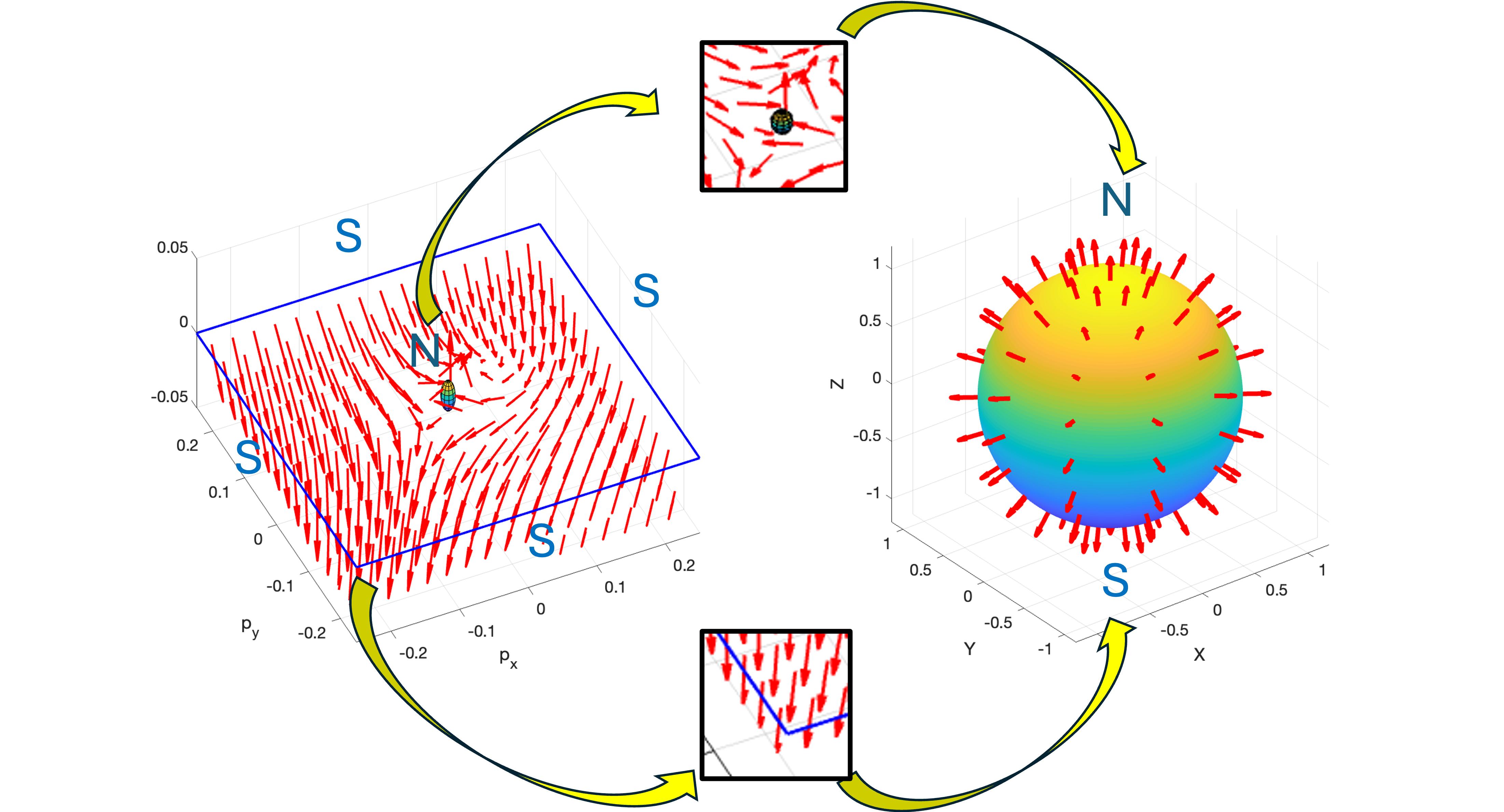} 
	 	\caption{ Representation of the Bloch sphere topology of the BdG Hamiltonian. The plot illustrates the compactification of the momentum plane onto the unit sphere. The Bloch vectors are mapped to the sphere via inverse stereographic projection, effectively realizing the hedgehog map. }\label{fig_rep_top} 
	\end{center}
\end{figure}

 \noindent
 The band structure of the BdG Hamiltonian and the orientation of the associated Bloch vectors are depicted in Fig. \ref{fig_band_diag}. The left panel corresponds to the lower-energy states, while the right panel shows the higher-energy states. The blue and green paraboloids represent the quasiparticle energy as a function of momentum. The Bloch vectors, indicated by red arrows, exhibit a complex spin texture, reflecting the underlying topology of the superconducting phase. To illustrate this in more detail, Fig. \ref{fig_rep_top} shows the inverse stereographic projection of the Bloch vectors onto a sphere. We discuss the geometrical structure of the upper band, similar considerations apply to the lower band.
Noting that $\lim_{p\rightarrow \infty} \hat{n} = -  \hat{z}$, the Bloch vector remains well-defined over the entire momentum space, allowing us to compactify the  $p_x-p_y$ plane into a sphere.  Accordingly, we map the origin of the momentum plane to the North Pole and the point at infinity to the South Pole of the sphere. Such a mapping from momentum space to the Bloch sphere is topologically equivalent to the well-known  ``hedgehog" map of a sphere onto itself, as illustrated in the right panel of Fig. \ref{fig_rep_top}. 
The ``nontriviality" of the map is quantified by the topological charge $Q= \frac{1}{4\pi}\int_{\mathbb{R}^2} \hat{n} \cdot \left( \partial_{p_x} \hat{n} \wedge \partial_{p_x}\hat{n} \right) \dif p_x \dif p_y $, which associates the values of $+1$ and $-1$ to the upper and lower bands, respectively. For simplicity, hereafter we will refer to upper (lower) band as the states with topological charge one (minus one). Accordingly,  the bands represent two topologically distinct configurations of the quasiparticle gas. 

\noindent
From a classical viewpoint, a particle may belong to only one band, and transitions to the second band are allowed only through inelastic scattering. In contrast, within the quantum mechanical framework, a single-particle state can exist as a coherent superposition of two bands.
We apply the two-band 4D Wigner model to investigate this point, studying the dynamics of quantum tunneling between two configurations with opposite topological charges. This can be understood in terms of Klein tunneling induced by an external field, which describes the coherent transfer of a population across two different band branches separated by a finite energy gap.

 \begin{figure}[!h]
	\begin{center} 	
		\begin{center}
			\includegraphics[width=0.6\textwidth]{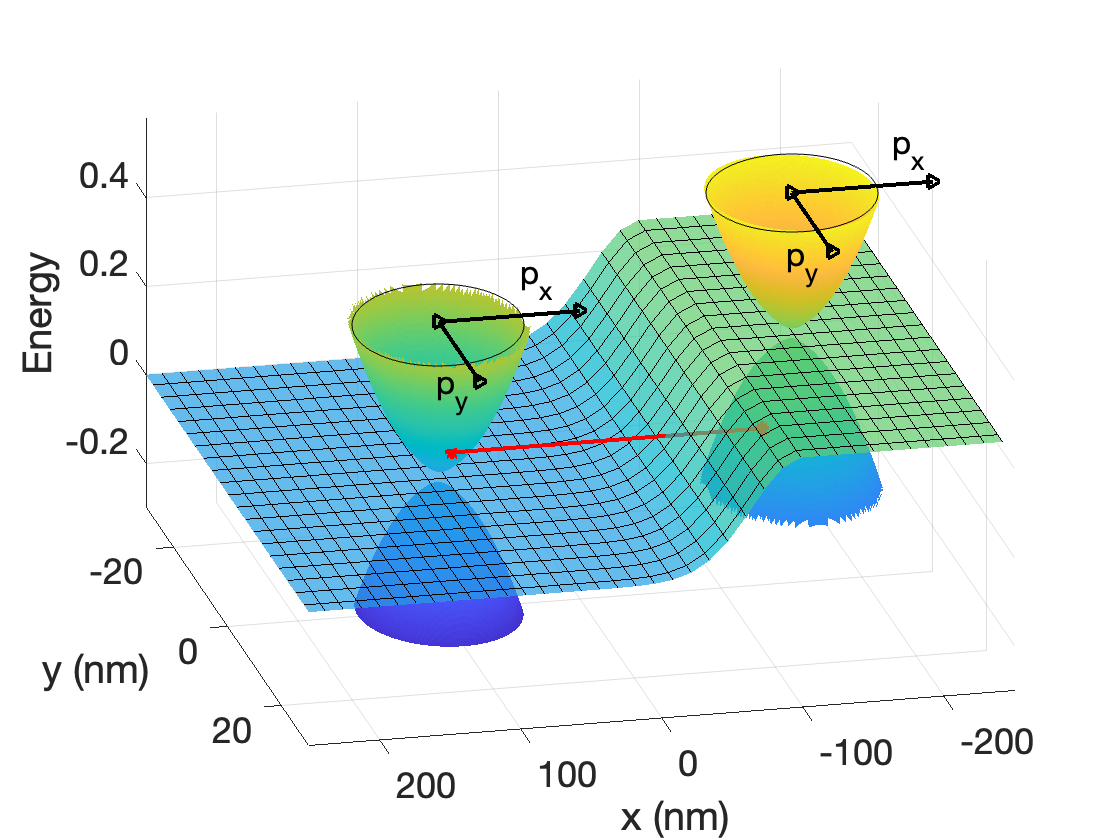} 
		\end{center}\label{fig_Klein_tun} 
		\caption{Illustration of the Klein tunneling mechanism. The meshed surface depicts the external potential $U$. The paraboloids represent the quasiparticle energy at two 
			different positions as a function of the momentum. The horizontal red line connects two momentum states in opposite bands with the same energy.  }
	\end{center}
\end{figure}
%\notaf{code figure in Wig\_4D\_09/work/case\_Bog\_de\_gen\_wall\_pres}
The physics of this phenomenon is illustrated in Fig. \ref{fig_Klein_tun}. We consider a uniform material with a smooth potential step $U$ applied along the $x-$ direction, $
U (x,y)=V_C\; e^{-\frac{(x-x_0)^2}{\zs_U^2}} \Theta(x-x_0) + V_C\;\left(1-\Theta(x-x_0)\right)
$. 
The profile of the applied potential, highlighted by the mesh grid, is represented in the figure. The energy of a quasiparticle is parametrized by its momentum and is represented at two different positions; the external field induces a shift in the quasiparticle energy that depends on the position. The shape of the external potential can be tailored such that states in different bands, located at different spatial positions, share the same energy. This is illustrated by a horizontal red line connecting two isoenegetic states, one in the upper band and one in the lower band. Under these conditions, a particle may tunnel from one topological charge to the opposite, undergoing a Klein-like tunneling process.

\begin{figure}[!h]
	\begin{center}
		\boxed{\includegraphics[width=0.48\textwidth]{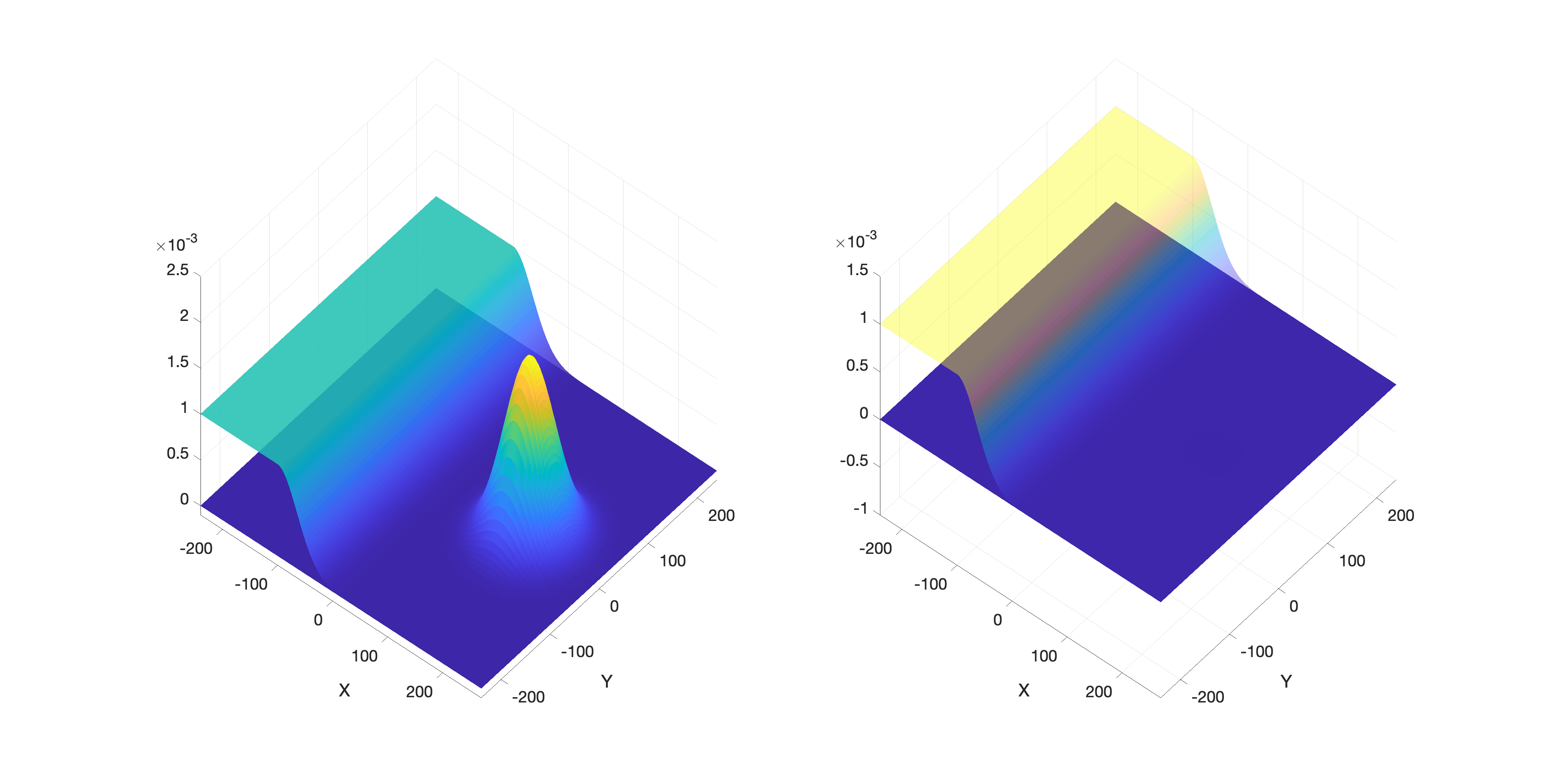}}\;\boxed{\includegraphics[width=0.48\textwidth]{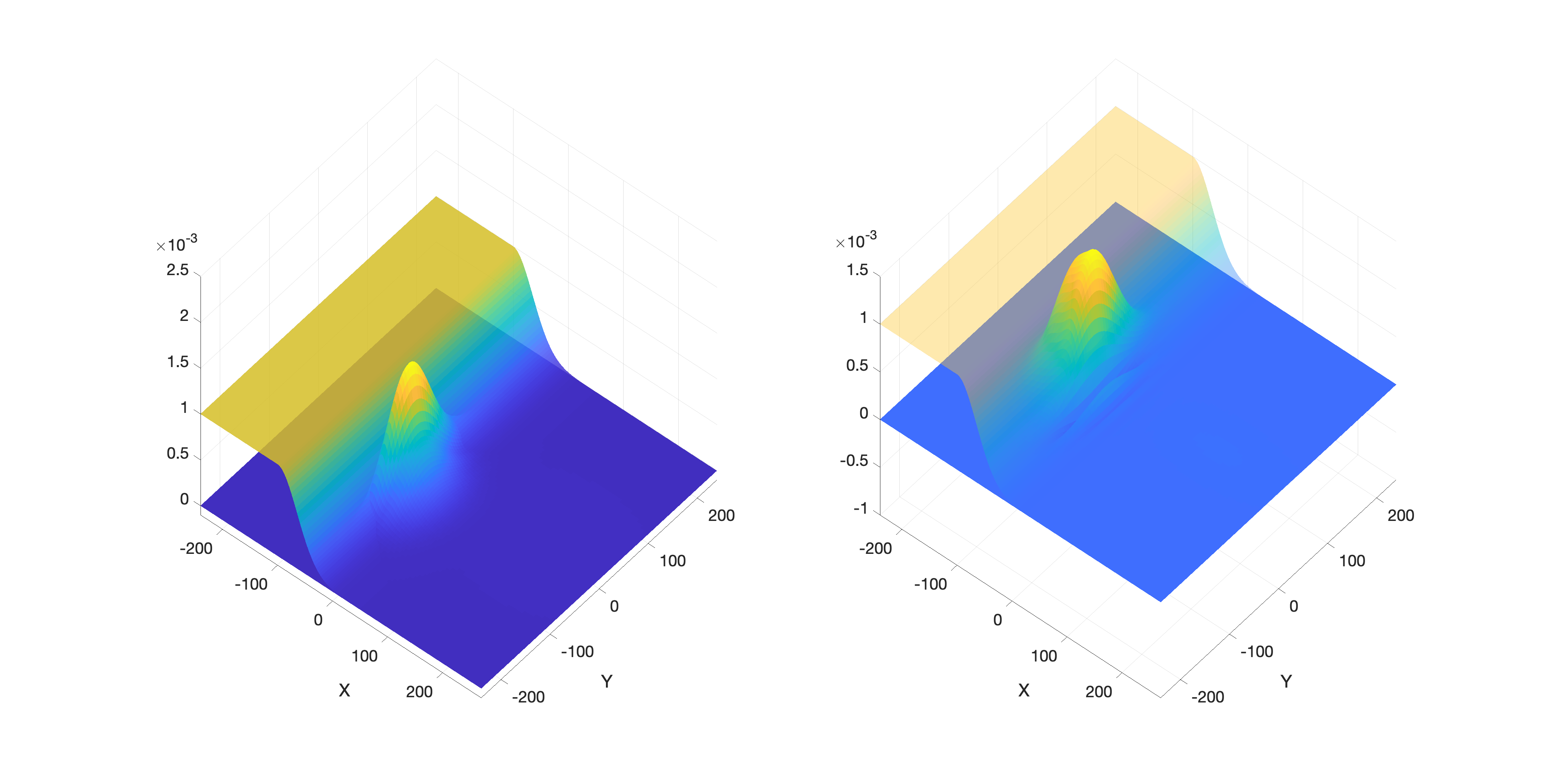}}\\
		\boxed{\includegraphics[width=0.48\textwidth]{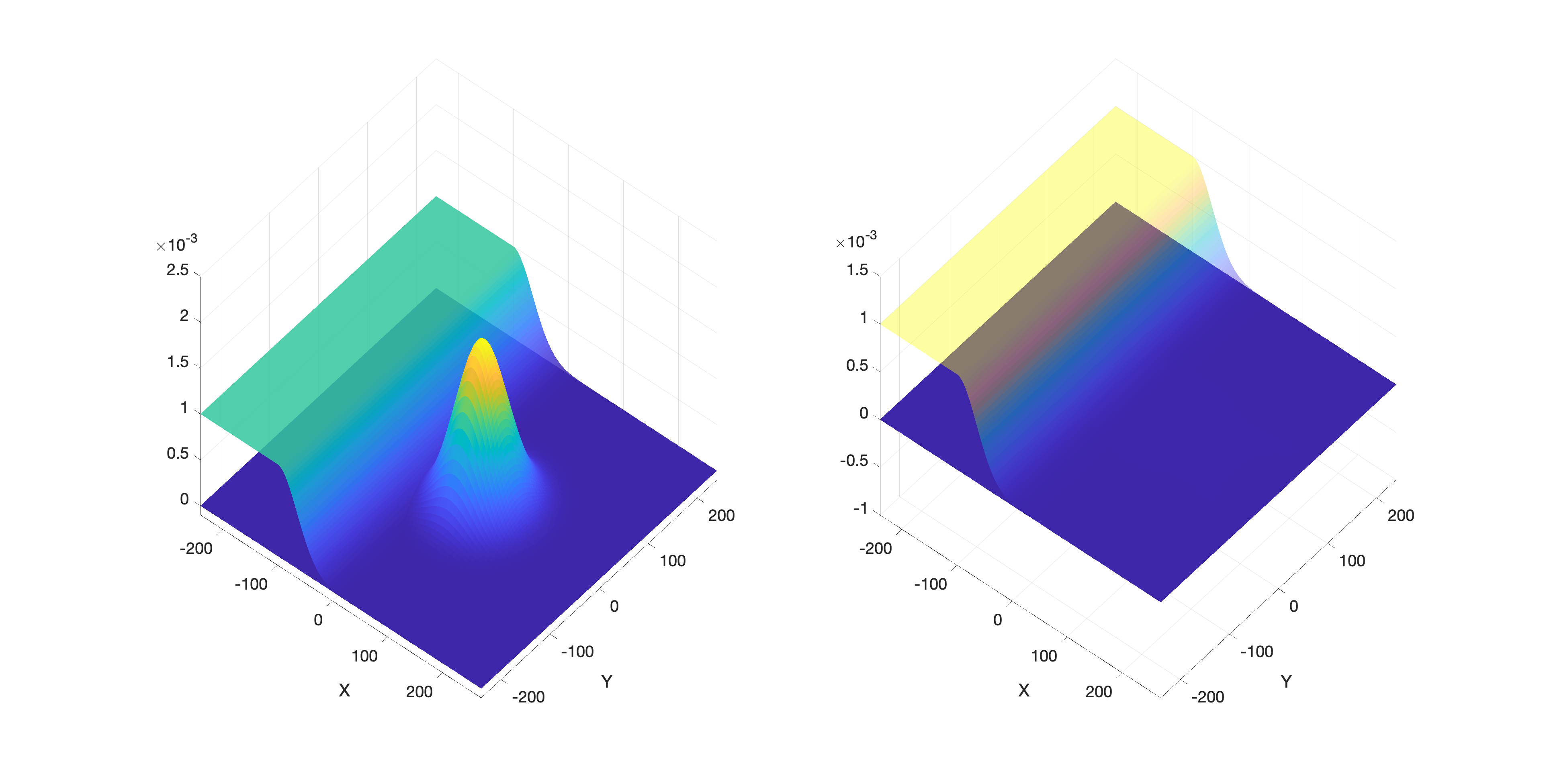}}\;\boxed{\includegraphics[width=0.48\textwidth]{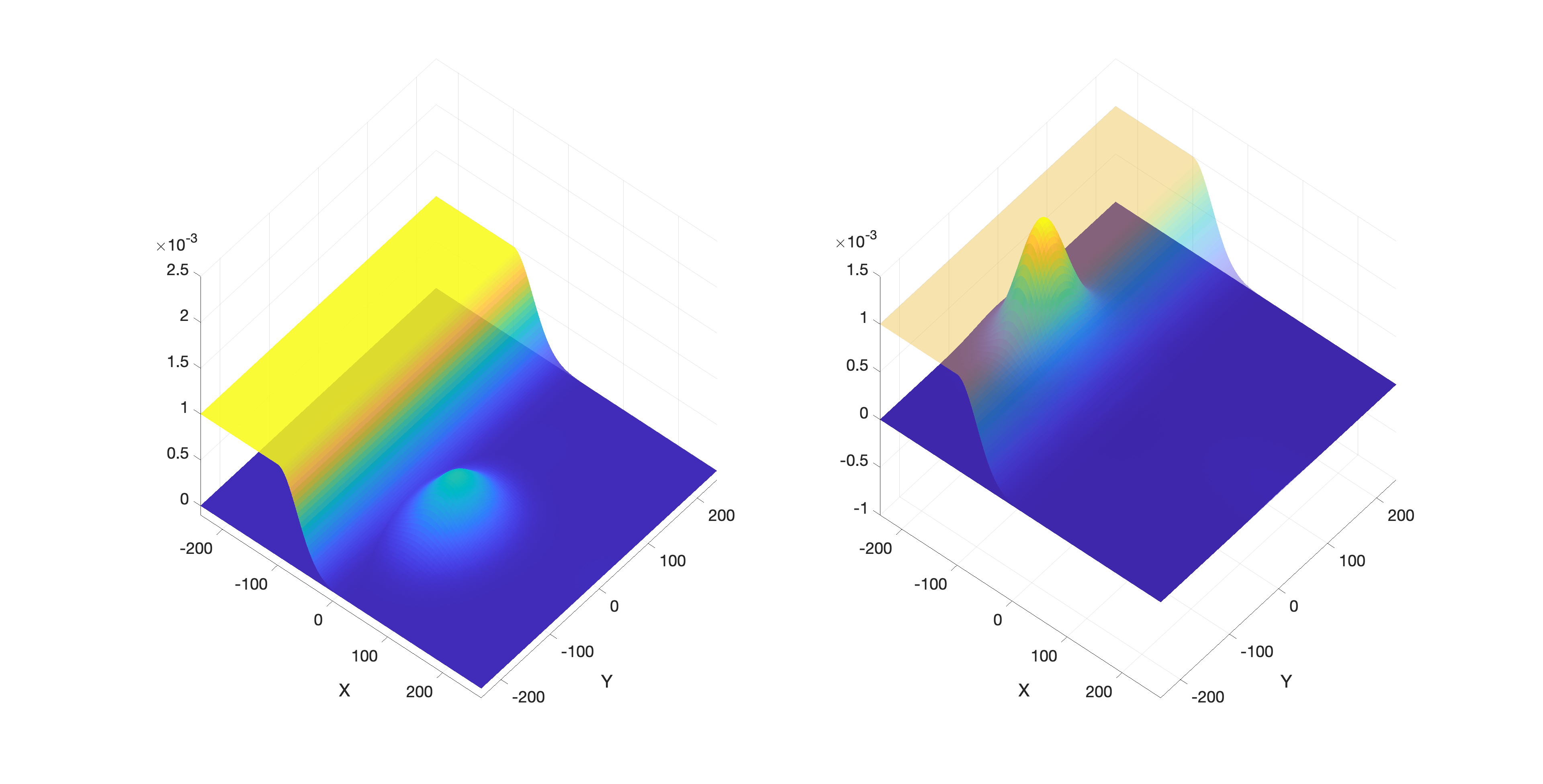}}\\
		\boxed{\includegraphics[width=0.48\textwidth]{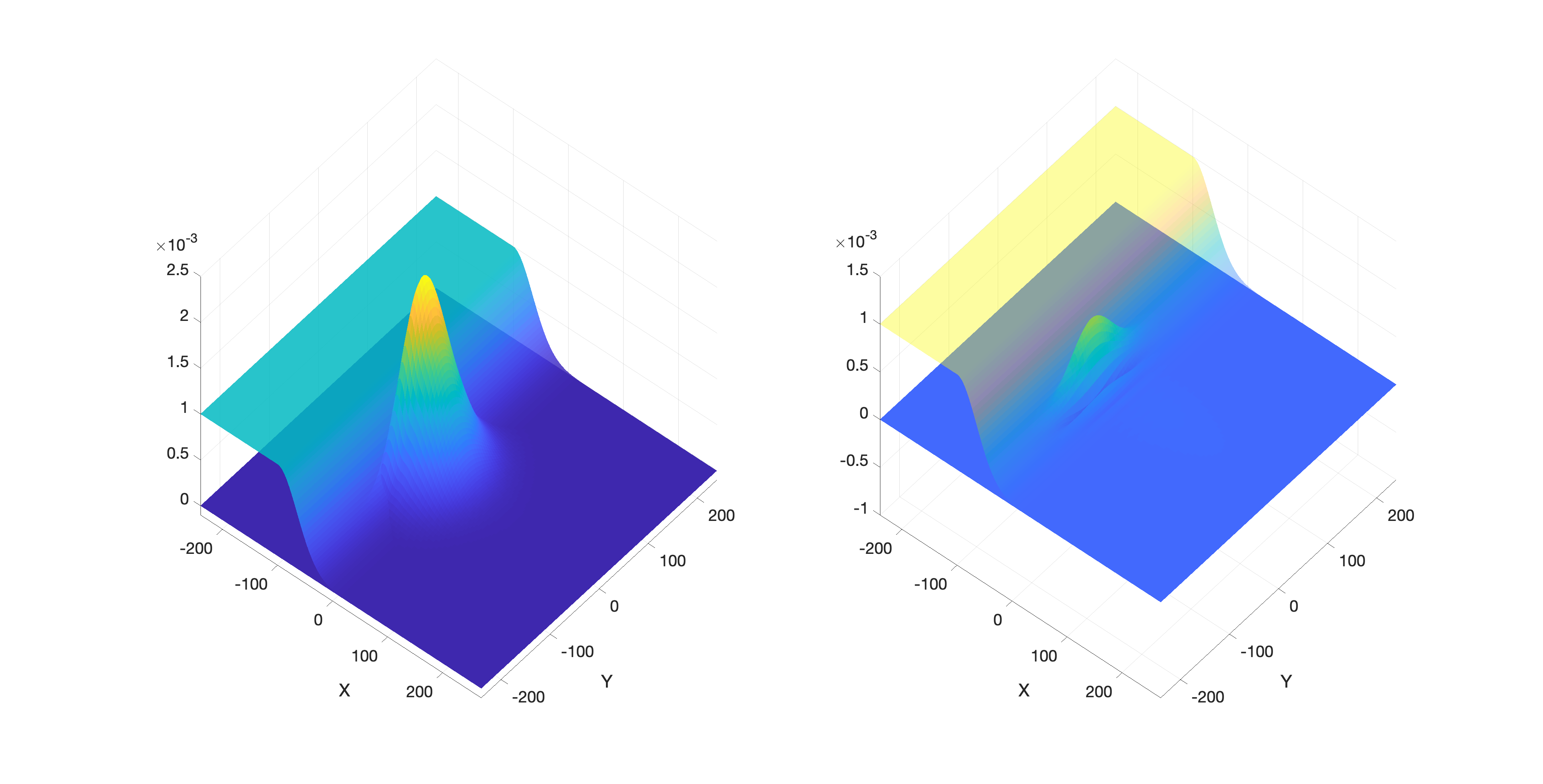}}\;\boxed{\includegraphics[width=0.48\textwidth]{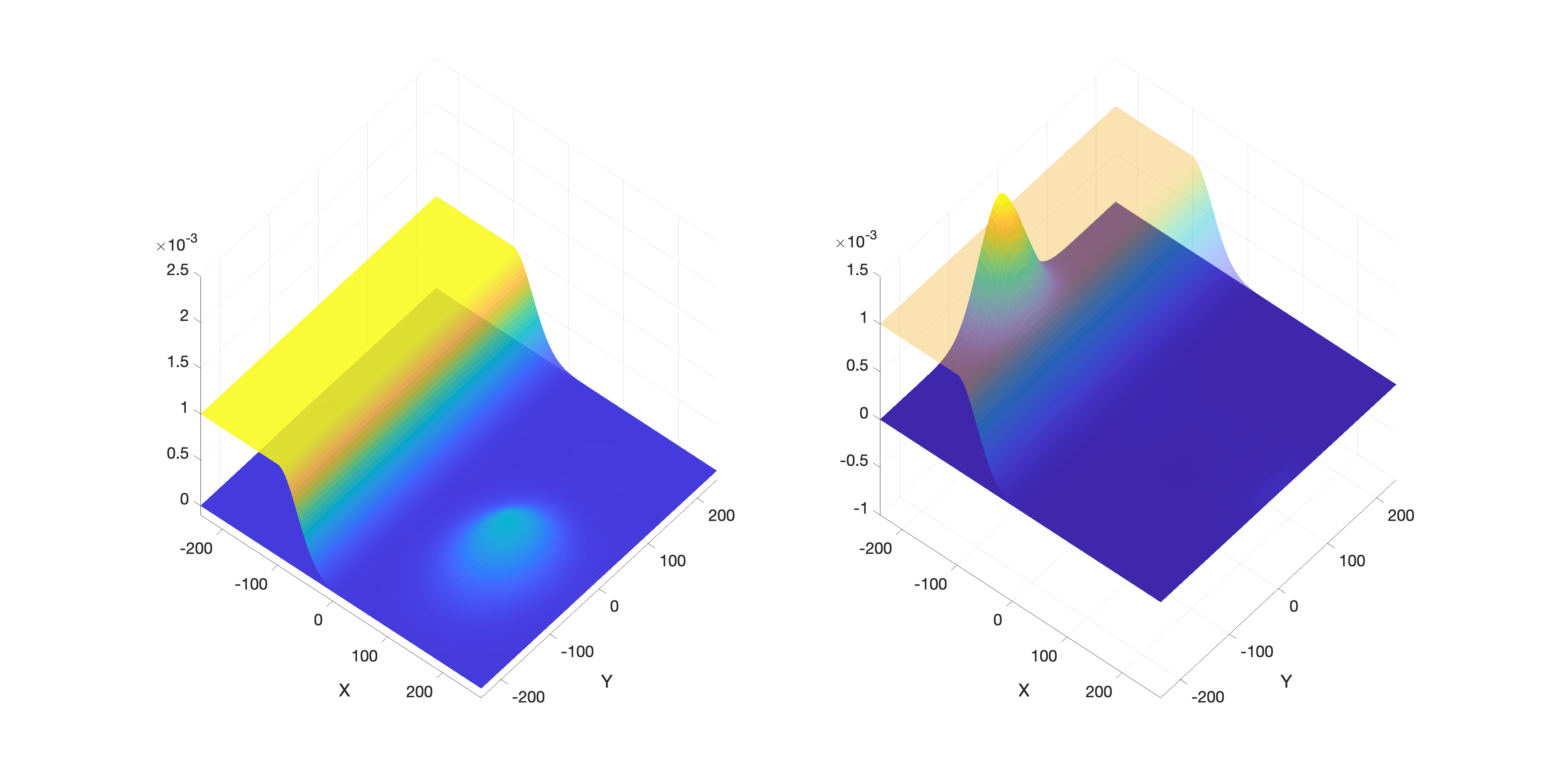}}
		%\\	\href{run:./movie/BdG_wall.mp4}{$\circ$}
		\caption{\href{https://youtu.be/4MfEniosE-s}{Movie (YouTube)}. Evolution of the two band Wigner function. In each panel, the left side represents the $\Pi^+$ component and the right side the  $\Pi^-$ component of the solution. The plots refer to the times $t=30,50,1000,1500,2000,2500\;\qty{}{\femto\second}   $ (from top to bottom, from left to right).}\label{fig_Klein_BdG} 
	\end{center}
\end{figure}
%\notaf{Sim in Wig\_4D\_09/work/case\_Bog\_de\_gen\_wall}

\noindent
In our simulations, we assume that the quasiparticle density is initially localized in the upper band with a Gaussian distribution and travels toward the potential step. This is achieved by setting the initial conditions for the Wigner matrix function as follows
\begin{align*}
 	\left.F\right|_{t=0} =	\frac{1}{(\hbar \pi)^2  } e^{-\frac{\left(x - \overline{x} \right)^2}{\Delta_x^2} -\frac{ y^2}{\Delta_y^2} -\frac{\left(p_x-\overline{p_x}\right)^2}{\Delta_{p_x}^2} -\frac{p_y^2}{\Delta_{p_y}^2}} \;	\Pi^+  \;. 
\end{align*}
We use the following parameters. Band structure:  $\mu=\qty{e-3}{\electronvolt}$,  $m=0.1 \;m_e$, $\Delta = \sqrt{\frac{ \mu q}{ m}}$, where $q$ is the electron charge. External potential: $ x_0 = \qty{-100}{\nano\meter} $,
$\zs_U= \qty{50}{\nano\meter}  $, 
$V_C=\qty{2e-2}{\electronvolt} $. 
Initial condition: $\overline{x}= \qty{150}{\nano\meter}  $,
$\overline{p_x}/\hbar = -\qty{0.15}{\nano\meter^{-1}}$, 
$\Delta_y=\Delta_x= \qty{50}{\nano\meter} $, 
$\Delta_{p_y}/\hbar=\Delta_{p_x}/\hbar =\qty{0.025}{\nano\meter^{-1}}   $.
The simulations show the Gaussian wave packet impinging on the potential step, where the particle density splits into two parts: a major component of the solution is transferred from the upper to the lower energy band, while a smaller contribution is reflected by the potential barrier. Due to the change in the sign of the effective mass in the lower band, the transmitted component is accelerated by the external field, leading to an increase in its associated kinetic energy. This process dynamically manifests a Klein tunneling phenomenon.

\subsection{Electron-hole dynamics in graphene}\label{Sec_ex_graphene}
 
As a final application of our phase-space approach to quantum dynamics, we simulate electron-hole excitation in graphene. The Wigner formalism has been widely employed to investigate electron dynamics in this material \cite{morandi_JPA_11,Stepanov_20,Camiola_21, Figueiredo_22, Bonitz2020, Tarasov2021, Fillion2022, Zhu2023, Tahir2013, Jauho2011, Hansen2016,Zhang_24}. 
Here, we model the excitation of electron-hole pairs induced by the sudden switching of an external potential, representing either an external gate or a laser field. To avoid the singularity at the Dirac cone vertex, we introduce a bandgap $\Delta$ separating the conduction and valence bands. We consider the Hamiltonian  $	\mathcal{H}_{graph.} =v_F\left(\begin{array}{cc} \Delta/2  &  p_x+ i p_y \\
		p_x- i  p_y & -\Delta/2 \end{array}\right) + U (x)\zs_0 $. 
In this context, the higher energy band is denoted as the conduction (electron) band, while the lower energy band is referred to as the valence (hole) band. In intrinsic graphene at zero temperature, the Fermi level lies exactly at the Dirac point (the gapless limit). Consequently, the hole band is fully occupied and the electron band is entirely empty.
As initial condition, we assume a two-band FD distribution at temperature $T$, as defined in Eq. \eqref{F_eq}. For the external potential, we adopt the same 2D Gaussian profile of Eq. \eqref{U_tw_Gauss2D} used in Sec. \ref{Sec_ex_atoms}, with the difference that we now assume that the center of the Gaussian remains fixed at the origin, i.e., $(x(t),y(t))=(0,0)$. The external potential is abruptly switched on at $t=0^+$.  The simulation is performed on a square layer of size $1\times\qty{1}{\micro\meter} $,	with periodic boundary conditions, using the parameters:  $v_F= \qty{e6}{\meter\sec^{-1}}$, $\Delta= \qty{0.01}{\electronvolt}$,  $T= \qty{25}{\kelvin}$, and $U_0= \qty{-0.02}{\electronvolt}$ (see Eq. \eqref{U_tw_Gauss2D}).  
\begin{figure}[!h]
\begin{center}
	\boxed{\includegraphics[width=0.48\textwidth]{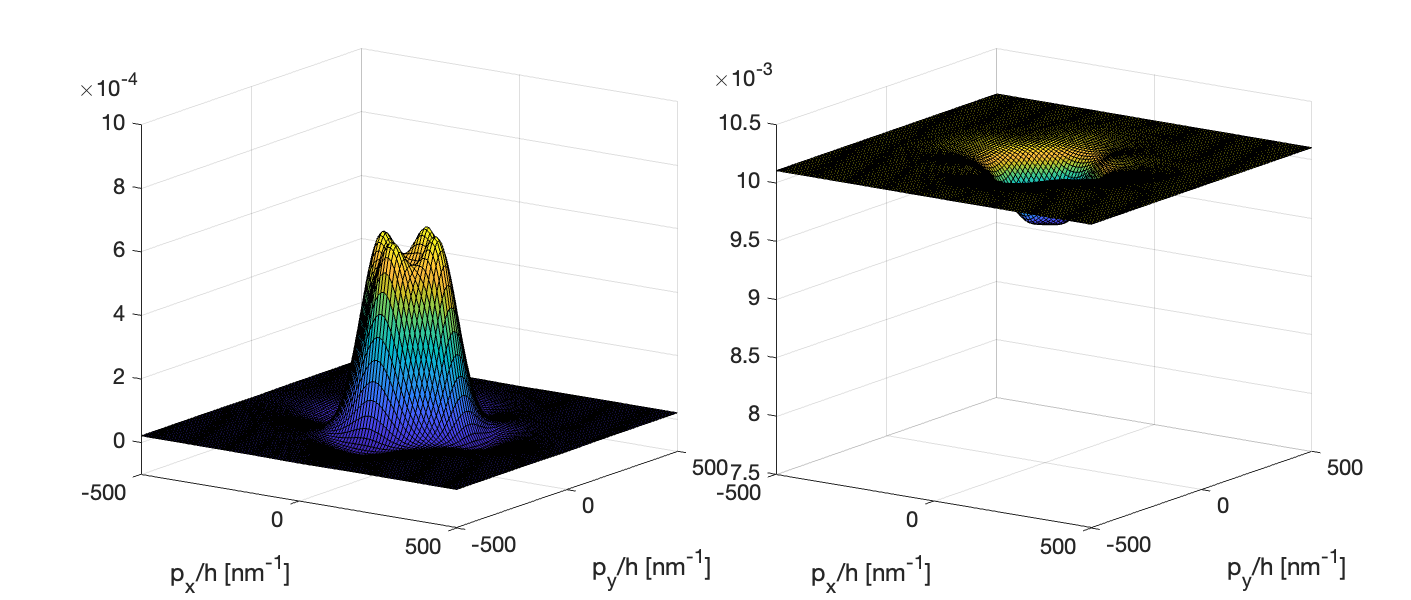}}\\\boxed{\includegraphics[width=0.48\textwidth]{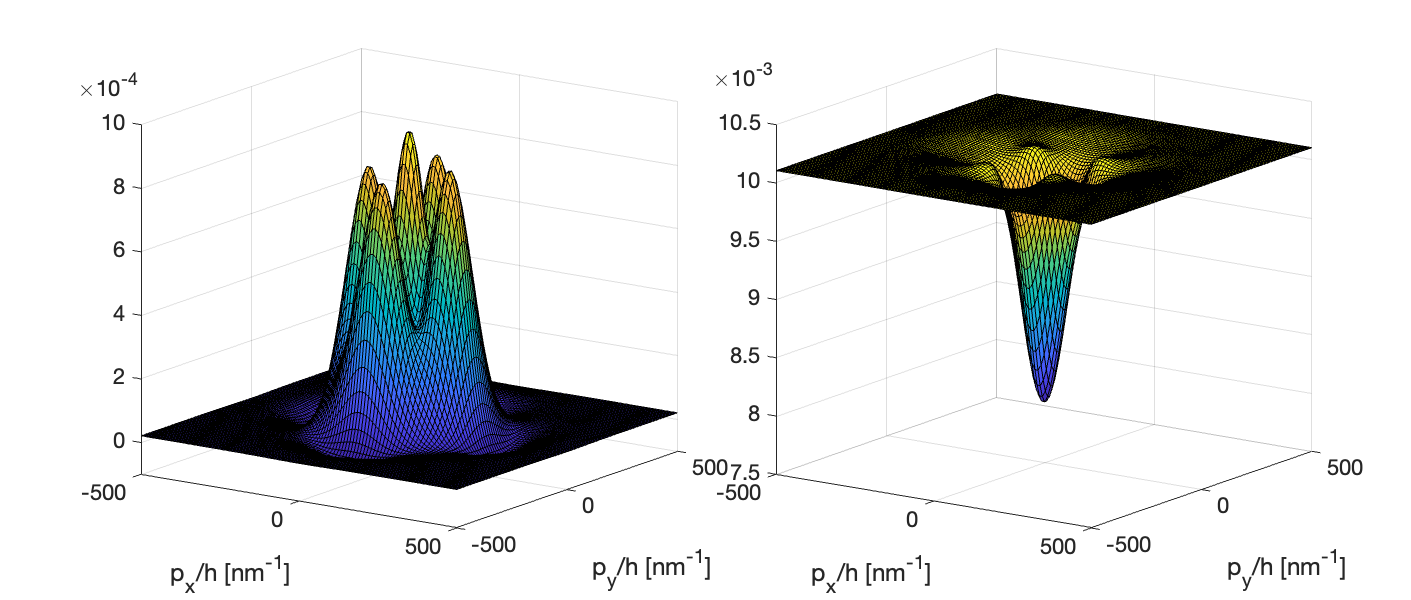}}\\
	\boxed{\includegraphics[width=0.48\textwidth]{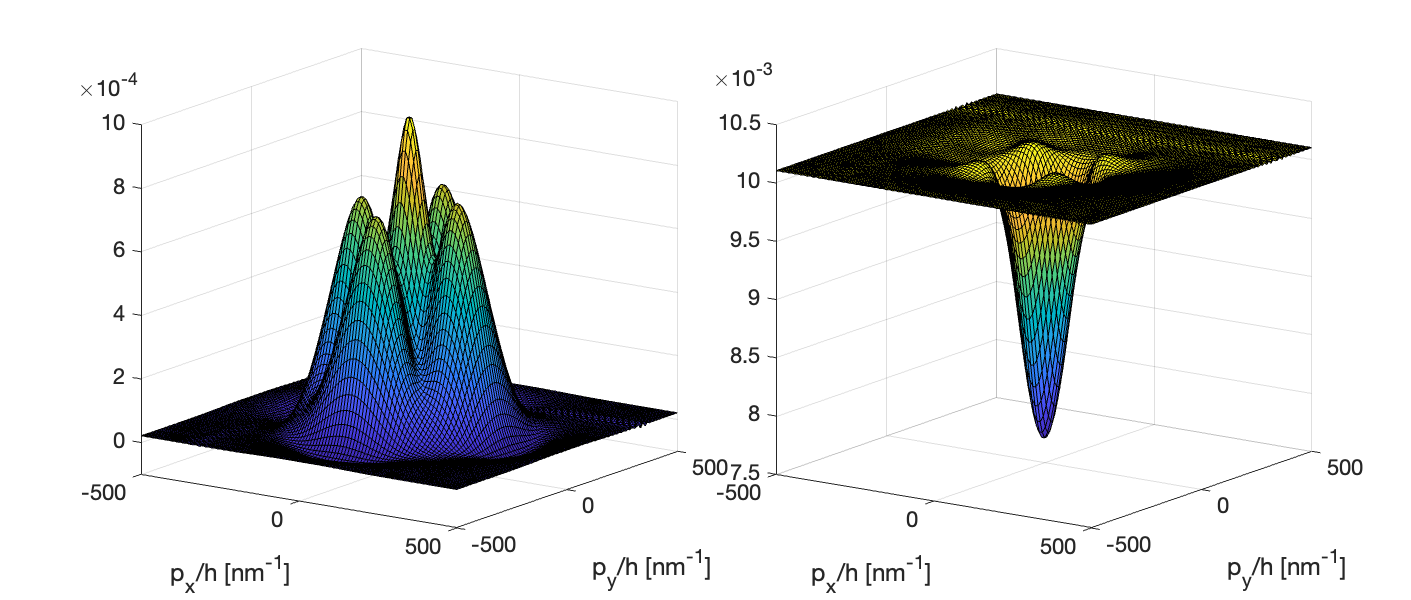}}%\\	\href{run:./movie/movie_res_graph_01.mp4}{$\circ$}
	\caption{\href{https://youtu.be/vvRSrFlSuPI}{Movie (YouTube)}. Evolution of the two-band Wigner function. In each panel, the left side represents the electron band (higher energy band) and the right side the hole band (lower energy band)  component of the solution. The plots refer to the times $t=65,\;130,\;200 \;\qty{}{\femto\second}   $ (from top to bottom).}\label{fig_graph} 
\end{center}
\end{figure}
%\notaf{Sim in Wig\_4D\_10/work/case\_graphene}
The results of the simulations are depicted in Fig. \ref{fig_graph}. In each panel, the left side represents the electron band (higher energy band) and the right side the hole band (lower energy band) component of the solution. Non-equilibrium dynamics are triggered by the external field, which promotes tunneling between the valence and conduction bands. This process results in a net flux of particles from lower to higher energy states, a phenomenon typically described as an interband transition.
\begin{figure}[!h]
	\begin{center}
		\includegraphics[width=0.48\textwidth]{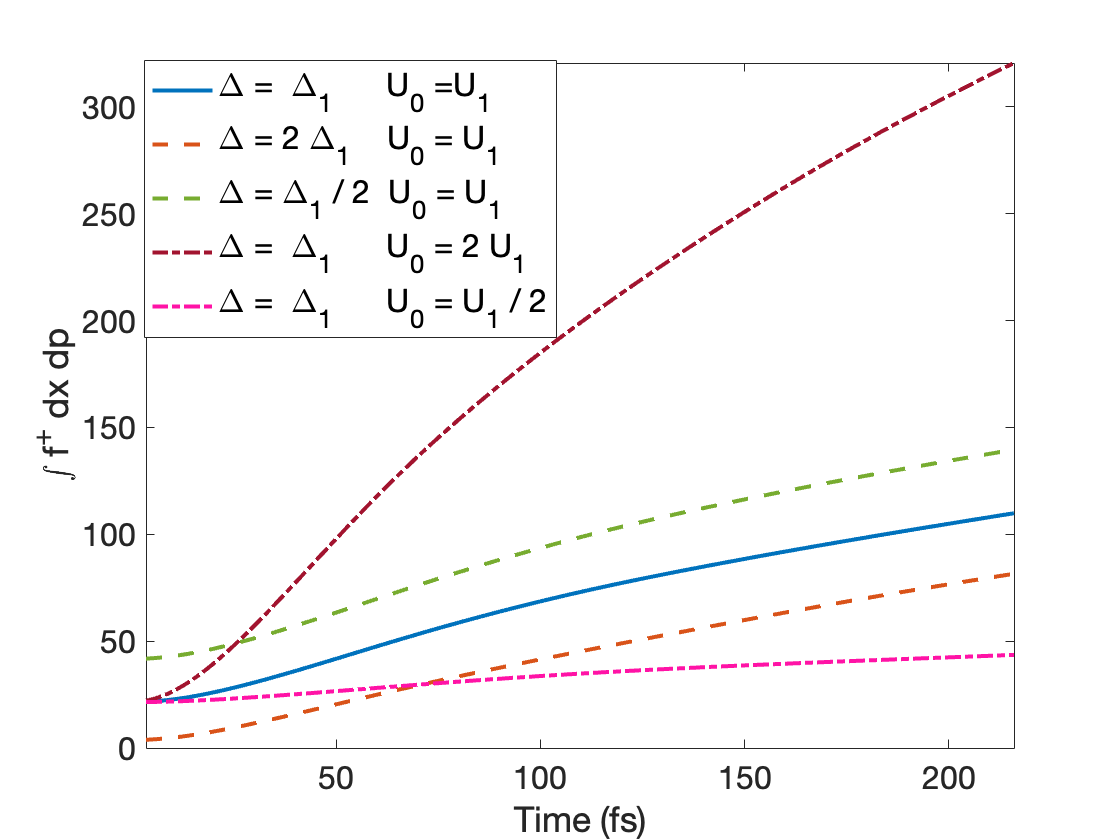} 
		\caption{Comparison of the evolution of the total density in the conduction band, varying the gap $\Delta$ and the potential strength $U_0$. We set $\Delta_1= \qty{0.01}{\electronvolt}$, $U_1= \qty{-0.02}{\electronvolt}$. Continuous blue line refers to the case illustrated in Fig. \ref{fig_graph}.}\label{fig_comp_graph} 
	\end{center}
\end{figure}
%\notaf{Sim in Wig\_4D\_10/work/case\_graphene da sim\_res\_01 a sim\_res\_06}
%
The sensitivity of the solution to the model parameters is illustrated in Fig.  \ref{fig_comp_graph}, which presents a comparison of the total conduction band density evolution for different values of the gap $\Delta$ and the potential strength $U_0$. The continuous blue line corresponds to the reference case previously illustrated in Figure \ref{fig_graph}.
The total electron density in the conduction band is evaluated as the integral of the upper band projection of the Wigner function over the momentum and spatial variables $n^+(t)=	\frac{1}{2} \textrm{Tr} 
\int_{\mathbb{R}^2_x\times\mathbb{R}^2_p}\Pi^+	F  \dif p \dif x$. 
The simulations demonstrate that the interband transition probability is highly sensitive to modifications of the external driving field. Specifically, doubling the excitation strength results in a fourfold increase in the transition rate. Furthermore, varying the gap size leads to a corresponding shift in the total conduction band density. In our simulations, the temperature is maintained at $T = \qty{25}{\kelvin}$ and the chemical potential is set to zero. As the thermal dispersion of the gas increases, the peak particle density decreases, which is consistent with our numerical findings.

\section{Conclusions}

%%%

In this contribution, we have presented a numerical scheme for the simulation of the two-dimensional quantum dynamics in two-level systems. 
The quantum state is represented by a $2 \times 2$ Wigner matrix defined on a 4D phase-space. 
To discretize the phase-space domain, we employ a uniform computational grid for both position and momentum variables, implementing either periodic or transparent open boundary conditions. 
The efficiency of the numerical scheme is enhanced by imposing specific constraints on the discretization steps in both real and Fourier spaces. These constraints reflect the underlying mathematical structure of the Wigner-Weyl integro-differential operators. The versatility of our approach has been demonstrated through several applications, illustrating the potential of the scheme by simulating the quantum evolution of systems relevant to diverse fields of physics, including ultracold neutral atoms, spintronic devices, topological superconductors, and graphene-based devices.

\appendix

\section{Wigner transform: properties and analytical representation of test cases.}\label{App_Wig_from_Sch}
For the reader convenience, we provide the Wigner transform of selected quantum states relevant to physical applications. Consistent with the definition of the Wigner function given in Eq. \eqref{def_wig_tras}, it is convenient to define the Wigner transform of a pairs of wave functions $(\zy_1,\zy_2)$ as 
\begin{align*}
	\mathcal{W}[\zy_{1},\zy_{2}](x,p) \doteq & \frac{1}{(2\pi)^d} \int \zy_{1},\left(x+\frac{\hbar \zh}{2} \right) \overline{\zy_{2},\left(x-\frac{\hbar \zh}{2} \right)}e^{-ip\zh} \dif \zh  \;.
\end{align*}
Denoting by $\zy^{x_0,p_0} \doteq   \zy \left(x-x_0\right)e^{i\frac{x p_0}{\hbar} } $ the shifted wave function, the associated Wigner function is translated in position and momentum by $(x_0,p_0)$, 
$	\mathcal{W}[\zy^{x_0,p_0} ,\zy^{x_0,p_0} ](x,p) =	\mathcal{W}[\zy ,\zy ](x-x_0,p-p_0) 
$. %\cuc
%\subsubsection*{Translation}
%Let $\zy^{x_0,p_0} =  \zy \left(x-x_0\right)e^{i\frac{x p_0}{\hbar} } $ 
%\begin{align*}
%	f^\hbar[\zy^{x_0,p_0}](x,p) =& \frac{1}{(2\pi)^d} \int \zy^{x_0,p_0}\left(x+\frac{\ze \zh}{2} \right) \overline{\zy^{x_0,p_0}\left(x-\frac{\ze \zh}{2} \right)}e^{-ip\zh} \dif \zh \\=& \frac{1}{(2\pi)^d} \int \zy\left(x-x_0+\frac{\hbar \zh}{2} \right) \overline{\zy\left(x-x_0-\frac{\hbar \zh}{2}\right)}e^{-i(p-p_0)\zh} \dif \zh=  f^\hbar[\zy](x-x_0,p-p_0)  
%\end{align*}
Gaussian wave packets in the Schr\"odinger representation correspond to  Gaussian distributions in the quantum phase-space. Defining 
\begin{align*}
	G_{x_0,p_0} \doteq  \frac{2^{d/4}}{\Delta_x^{d/2} \pi^{d/4}} e^{-\left(\frac{x-x_0}{\Delta_x}\right)^2 +\frac{i}{\hbar} p_0 x }\;, 
\end{align*}
we obtain
\begin{align*}
	\mathcal{W}[G_{x_1,p_1},G_{x_2,p_2}](x,p) = &	\frac{1}{(\hbar \pi)^d  } e^{-\frac{2}{\Delta_x^2}\left(x - \frac{x_1+x_2}{2}\right)^2 -\frac{\Delta_x^2}{2\hbar^2} \left(p-\frac{p_1+p_2}{2}\right)^2+\frac{i}{\hbar} \left[x(p_1-p_2)+\left(p-\frac{p_1+p_2}{2}\right) (x_2-x_1)\right]}\;.
\end{align*}
In particular, $	\mathcal{W}[G_{x_0,p_0},G_{x_0,p_0}](x,p) =   	\frac{1}{(\hbar \pi)^d  } e^{-\frac{2}{\Delta_x^2}\left(x - x_0 \right)^2 -\frac{\Delta_x^2}{2\hbar^2} \left(p-p_0\right)^2} 
$. A popular example illustrating quantum interference is represented by the coherent superposition of two Gaussians, $\zy = \za G_{x_1,p_1}+  \zb  G_{x_2,p_2}$, with $\za,\zb\in \mathbb{C}$. In this case, the associated Wigner transform is 
\begin{align*}
	\mathcal{W}[\zy,\zy](x,p) = & |\za|^2	\mathcal{W}[G_{x_1,p_1},G_{x_1,p_1}]+	|\zb|^2 \mathcal{W}[G_{x_2,p_2},G_{x_2,p_2}]+2\textrm{Re} \left(\za \overline{\zb}	\mathcal{W}[G_{x_1,p_1},G_{x_2,p_2}]\right)\;. 
\end{align*}
The interference is most prominent when the two packets overlap. In the case  $x_2 = x_1 =0$, $p_2=-p_1$ and $\za=\zb=1$, we have %\cuc
\begin{align*}
	\mathcal{W}[\zy,\zy](x,p) = & 	\frac{1}{(\hbar \pi)^d  }e^{-\frac{2}{\Delta_x^2} x^2 } \left[e^{ -\frac{\Delta_x^2}{2\hbar^2} \left(p-p_1\right)^2}	+  e^{ -\frac{\Delta_x^2}{2\hbar^2} \left(p+ p_1\right)^2} +2 e^{ -\frac{\Delta_x^2}{2\hbar^2} p^2} \cos\left(\frac{2 x p_1 }{\hbar}\right)	\right]\;.  
\end{align*}
We conclude this section, by few remarks concerning the positivity of the Wigner function in the presence of the spin degrees of freedom. We recall the definition given in Sec. \ref{Sec_Model} concerning the projection of the Wigner function on the upper and lower eigenspaces, $f^\pm = \textrm{tr} \left(\Pi^\pm F \right) $, where $\Pi^\pm = | u^\pm  \rangle \langle u^\pm |$, with $u^\pm \in \mathbb{C}^2$ are the two-component eigenvectors associated to the eigenvalue problem $\Lambda (p) u^\pm (p) = \zl^\pm  u^\pm (p) $. It can be shown that 
\begin{align}
	\int f^{\pm} \dif x \dif y \geq 0\;. \label{poisit_fpm}
\end{align}
This can be verified as follows. Assuming $F_{ij} = \sum_k \zr^k \mathcal{W} (\zy_i,^k\zy_j^k)$ consisting of a convex superposition of pure states, we have $
	\int F_{ij} \dif x \dif y =  \sum_k \zr^k  \hat{\zy_i^k}(p) \overline{\hat{\zy_j^k}} (p)$, and 
\begin{align*}
	\textrm{tr} \left(\Pi^\pm (p)\int F  \dif x \dif y  \right)  = & %\sum_k \zr^k \sum_{i,j} \left(\Pi^\pm_{i,j}(p) \hat{\zy_j^k}(p) \overline{\hat{\zy_i^k}} (p)  \right)  = 
	 \sum_k \zr^k \sum_{i,j} \left(  u^\pm_i \overline{u^\pm_j}  \hat{\zy_j^k}\overline{\hat{\zy_i^k}}   \right)  
	 %\\
	%=&  \sum_k \zr^k\left(  \sum_{i}  u^\pm_i \overline{\hat{\zy_i^k}}  \right) \left(  \sum_{j} \overline{u^\pm_j}  \hat{\zy_j^k} \right)  
	= \sum_k \zr^k\left|  \langle \hat{\zy^k} | u^\pm    \rangle \right|^2\;,
\end{align*}
which satisfies Eq. \eqref{poisit_fpm}. 
Furthermore, concerning the integral over the momentum, we remark that  the non-negativity property holds true only for the trace 
\begin{align*}
	\textrm{tr} \left(\int F  \dif p_x \dif p_y  \right)  \geq 0 \;.
\end{align*}
In general, the integral of the projections $\int f^{\pm} \dif p_x \dif p_y =\int \textrm{tr} \left(\Pi^\pm (p)F  \right) \dif p_x \dif p_y   $ may take negative values. %\notaf{queste considerazioni vanno anche messe nel testo}

\section*{Acknowledgments}
	The work has been developed under the auspices of GNFM (INdAM). %\end{acknowledgments}

\end{document}